\documentclass[11pt]{article}

\usepackage{siunitx}
\usepackage{amsmath,amsfonts,graphicx,epsfig,mathrsfs,amssymb}
\usepackage{subcaption}
\usepackage{bm}
\usepackage{bbm}
\usepackage{svrsymbols}
\usepackage{dsfont}
\usepackage{mathtools}
\usepackage{color}
\usepackage[dvipsnames]{xcolor}
\usepackage{tensor}
\usepackage[colorlinks]{hyperref}
\hypersetup{linktocpage}
\usepackage[titletoc]{appendix}
\usepackage{cite}
\usepackage{ragged2e}
\usepackage{upgreek}
\usepackage{physics}
\usepackage{cancel}

\usepackage{tikz} 
\usetikzlibrary{calc} 

\usepackage{setspace}

\usepackage{amssymb,amsmath}
\usepackage{latexsym}
\usepackage{mathrsfs}
\usepackage{hyperref} 
\usepackage{verbatim}
\usepackage{graphicx}

\newcommand{\x}{{\boldsymbol x}}

\def\k{{\boldsymbol k}}
\newcommand{\q}{{\boldsymbol q}}

\def\bea{\begin{eqnarray}}
\def\eea{\end{eqnarray}}
\def\be{\begin{equation}}
\def\ee{\end{equation}}
\def\ba{\begin{array}}
\def\ea{\end{array}}
\def\nn{\nonumber}

\newcommand{\de}{{\rm d}}

\definecolor{carmine}{rgb}{0.59, 0.0, 0.09}
\definecolor{sgreen}{rgb}{0.0, 0.44, 0.0}
\definecolor{prussianblue}{rgb}{0.0, 0.06, 0.54}
\hypersetup{colorlinks=true, linkcolor=carmine, filecolor=magenta, urlcolor=prussianblue, citecolor=sgreen}

\begin{document}

\renewcommand*{\thefootnote}{\fnsymbol{footnote}}

\renewcommand{\theequation}{\arabic{section}.\arabic{equation}}
\setcounter{page}{1}

\begin{titlepage}

\begin{center}

\vskip 1.5 cm

{\LARGE \bf Non-Gaussian statistics of de Sitter spectators: \\ A perturbative derivation of stochastic dynamics}

\vskip 2.0cm

{\large Gonzalo A. Palma$^a$ and Spyros Sypsas$^{b,c}$ }

\vskip .6cm

$^a${\it Departamento de F\'isica, FCFM, Universidad de Chile,\\ Blanco Encalada 2008, Santiago, Chile}

$^b${\it High Energy Physics Research Unit, Faculty of Science, Chulalongkorn University, Bangkok 10330, Thailand}

$^c${\it National Astronomical Research Institute of Thailand, Don Kaeo, Mae Rim, Chiang Mai 50180, Thailand}

\vskip 2.5cm

\end{center}

\begin{abstract}

Scalar fields interacting with the primordial curvature perturbation during inflation may communicate their statistics to the latter. This situation motivates the study of how the probability density function (PDF) of a light spectator field $\varphi$ in a pure de Sitter space-time, becomes non-Gaussian under the influence of a scalar potential ${\mathcal V(\varphi)}$. One approach to this problem is offered by the stochastic formalism introduced by Starobinsky and Yokoyama. It results in a Fokker-Planck equation for the time-dependent PDF $\rho (\varphi , t)$ describing the statistics of $\varphi$ which, in the limit of equilibrium gives one back the solution $\rho (\varphi) \propto \exp \big[ - \frac{8 \pi^2}{3 H^4} {\mathcal V(\varphi)} \big]$. We study the derivation of $\rho (\varphi , t)$ using quantum field theory tools. Our approach yields an almost Gaussian distribution function, distorted by minor corrections comprised of terms proportional to powers of $\mathcal O_\varphi {\mathcal V(\varphi)}$, where $\mathcal O_\varphi$ stands for a derivative operator acting on ${\mathcal V(\varphi)}$ proportional to $\Delta N$, the number of $e$-folds succeeding the Hubble-horizon crossing of $\varphi$'s wavelengths. This general form is obtained perturbatively and remains valid even with loop corrections. Our solution satisfies a Fokker-Planck equation that receives corrections with respect to the one found within the stochastic approach, allowing us to comment on the validity of the standard equilibrium solution for generic potentials. We posit that higher order corrections to the Fokker-Planck equation may become important towards the equilibrium.

\end{abstract}

\end{titlepage}

\tableofcontents

\section{Introduction}

Upcoming cosmological surveys represent a unique opportunity to probe the statistics of primordial fluctuations, thereby shedding light on the dynamical characteristics of the inflationary universe~\cite{Starobinsky:1980te, Guth:1980zm,Linde:1981mu,Albrecht:1982wi}. Although the simplest models of cosmic inflation (e.g., single-field slow-roll inflation) predict a nearly Gaussian statistics, compelling reasons exist to explore the possibility of non-Gaussian deviations as more of a norm rather than an exception~\cite{Bartolo:2004if, Komatsu:2009kd, Chen:2010xka, Achucarro:2022qrl}. One such class of models leading generically to non-Gaussian statistics is multifield inflation.\footnote{Non-Gaussianity is also generic in single-field inflation~\cite{Cheung:2007st,Chen:2006nt}, especially when one is interested in the tail distribution~\cite{Ezquiaga:2019ftu,Celoria:2021vjw,Hooshangi:2023kss,Cai:2022erk,Cai:2021zsp,Hooshangi:2021ubn,Pi:2022ysn,Vennin:2020kng, Achucarro:2021pdh, Gow:2022jfb}.} 

Indeed, numerous well-grounded scenarios~\cite{Baumann:2014nda} propose the presence of spectator fields interacting with the inflaton~\cite{Gong:2016qmq}. In such models, the primordial curvature perturbation $\zeta$ that seeded our universe's structure, can inherit the statistics of another field, say $\varphi$, thanks to generic interaction terms of the form $\mathcal L_{\rm mix} \propto \alpha \dot \zeta \varphi$, where $\alpha$ is a coupling that may depend on time. In those cases where $\alpha$ remains small, and provided that the spectator field is light, the Fourier space $n$-point correlation functions $\big \langle \zeta(\k_1) \cdots \zeta(\k_N) \big \rangle$ —which determine the statistics of $\zeta$— not only inherit the values of the $n$-point correlation functions of $\varphi$, but they do so in an enhanced way~\cite{Chen:2009we, Chen:2009zp, Chen:2018uul, Chen:2018brw}:
\begin{equation}
\big \langle \zeta(\k_1) \cdots \zeta(\k_N) \big \rangle \propto ( \alpha \Delta N )^n \big \langle \varphi(\k_1) \cdots \varphi(\k_N) \big \rangle .
\end{equation}
Here, $\big \langle \varphi(\k_1) \cdots \varphi(\k_N) \big \rangle$ represents the $n$-point correlator of the spectator field $\varphi(\k, t)$, while $\Delta N$ (with $\de N = H \de t$) indicates the amount of $e$-folds after the horizon crossing of modes in the range of wavelengths relevant for the computation. This straightforward  connection between $n$-point functions of $\zeta$ and $\varphi$ is sufficient to motivate the study of spectator fields' statistics during inflation: Once we know the distribution of $\varphi$, for instance in the form of a probability density function $\rho(\varphi)$, we can infer the statistics of $\zeta$ as sourced by $\varphi$~\cite{Flauger:2016idt, Chen:2018uul, Chen:2018brw, Panagopoulos:2019ail, Palma:2019lpt}, which renders de Sitter spectators phenomenologically relevant. 

To lay the groundwork, let us consider a canonical quantum spectator scalar field $\varphi (t , \x)$ in a de Sitter spacetime characterized by a constant Hubble expansion rate $H$, with the field's self-interactions determined by an arbitrary scalar potential ${\mathcal V(\varphi)}$ (not driving inflation). The action parametrizing this system is given by
\be
S = \int \! \de^3 x \, \de t \, a^3 \bigg[ \frac{1}{2} \dot \varphi^2 - \frac{1}{2 a^2} ( \nabla \varphi )^2 - {\mathcal V(\varphi)} \bigg] , \label{intro:action}
\ee
where $a(t) = e^{H t}$ is the scale factor describing the spatial exponential expansion characteristic of de Sitter spacetime charted with cosmological coordinates. We will assume that the potential satisfies ${\mathcal V(\varphi)} \ll H^4$, irrespective of its shape, even for a field range $|\varphi| \gg H$.\footnote{That is, ${\mathcal V(\varphi)}$ may have a large second derivative at a given field value $\varphi_0$, but this situation could change drastically at a nearby value $\varphi_0 + \sqrt{{\cal V} / {\cal V}^{\, ''}}$. An example of this would be an axionic potential of the form ${\cal V}(\varphi) = \Lambda^4 (1 - \cos (\varphi / f_a))$. In this case, the mass parameter $m$ at $\varphi = 0$ is given by $m = \Lambda^2 / f_a$, but the usual meaning attributed to it changes if we displace the field away by an amount $\Delta \varphi \sim f_a$.} Because the natural amplitude of light spectator fields in de Sitter is of order $\varphi \sim H$ then, if ${\cal V} /H^4 \ll 1$, the statistical distribution of $\varphi$, described by $\rho(\varphi)$, is expected to receive contributions from ${\mathcal V(\varphi)}$. In other words, the probability density function (PDF) of $\varphi$ will reflect the structure of the potential ${\mathcal V(\varphi)}$ via gradients, in such a way that it becomes slightly more likely to measure values of $\varphi$ that coincide with minima (rather than maxima) of ${\mathcal V(\varphi)}$.

A commonly employed method to study the PDF of such a spectator field is the Starobinsky-Yokoyama stochastic formalism~\cite{Starobinsky:1986fx,Starobinsky:1994bd,Gorbenko:2019rza}. This approach models the influence of short-wavelength perturbations on superhorizon modes $\varphi_L$ via a Langevin equation:
\be
 \frac{\partial \varphi_L}{\partial N}  = - \frac{1}{3H^2}\frac{\partial \mathcal V}{\partial \varphi_L} (\varphi_L) +  \hat \xi (t) , \label{langevin-1}
\ee
with $\hat \xi(t)$ being a stochastic noise term encapsulating the effects of short-wavelength perturbations, satisfying 
\be \label{noise}
\langle \xi (t) \xi (t') \rangle = \frac{H}{4\pi^2} \delta(t - t').
\ee 
This approach results in a Fokker-Planck equation of the form
\begin{equation}
\frac{\partial \rho}{\partial N} = \frac{H^2}{8 \pi^2} \frac{\partial^2 \rho}{\partial \varphi^2} + \frac{1}{3H^2} \frac{\partial }{\partial \varphi} \left(\frac{\partial \mathcal V}{\partial \varphi} \rho \right)  , \label{Staro-Yoko}
\end{equation}
where $N$, again, refers to the number of $e$-folds. This equation governs the evolution of the probability density function for a spectator $\varphi$, which is composed of superhorizon Fourier modes. The first term on the right hand side of (\ref{Staro-Yoko}) consists in a diffusion term, informing us that, as time passes, the number of modes entering $\varphi_L$ is increasing (with time, new modes become superhorizon joining those already contained in $\varphi_L$). On the other hand, the second term of (\ref{Staro-Yoko}) is the drift term, which accounts for the deformations of $\rho(\varphi, N)$ induced by the potential ${\mathcal V(\varphi)}$ (for example, the increment of probability at $\varphi$-values corresponding to ${\cal V}$-minima). Taking (\ref{Staro-Yoko}) at face value, after a certain period of time the system is expected to reach equilibrium, at which point $\rho (\varphi , N)$ ceases to evolve and the right-hand side of the equation can be made to vanish. Consequently, the equilibrium solution becomes
\begin{equation}
\rho (\varphi) \propto e^{ - \frac{8 \pi^2}{3 H^4} {\mathcal V(\varphi)}} . \label{sol-Staro-Yoko}
\end{equation}
Although this method is often characterized as nonperturbative, a thorough analysis of the procedure leading to equation~(\ref{Staro-Yoko}) indicates that the latter incurs corrections of order $\sim \rho \times \mathcal V^2$~\cite{Gorbenko:2019rza, Cohen:2021fzf,Cohen:2020php,Green:2022ovz,Cohen:2022clv}. This implies that the solution shown in equation~(\ref{sol-Staro-Yoko}) is valid for small-amplitude potentials ${\mathcal V(\varphi)}$ and/or a restricted range of $\varphi$.  An additional consequence of this limitation is that the relaxation time needed to reach equilibrium, roughly estimated as\footnote{To reach this estimate, we took the Gaussian distribution with $\sigma^2 = \frac{H^2}{4 \pi^2} \Delta N$ (solving the Fokker-Planck equation in the absence of drift) and equated both terms on the right hand side of (\ref{Staro-Yoko}).} 
\be
\Delta N \sim \frac{3 H^2}{2 \mathcal V^{\, ''}}, \label{N=H2V2}
\ee
may exceed the ideal duration of inflation of about 60 $e$-folds, implying that the solution (\ref{sol-Staro-Yoko}) might not be useful to characterize the statistics of $\varphi$ at wavelengths relevant for the description of our universe's structure. Thus, given that the wavelengths of modes relevant for cosmology cannot be excessively long, we might need to consider alternative methods able to accommodate the initial non-Gaussian deviations on $\rho (\varphi , N)$ implied by ${\mathcal V(\varphi)}$. Another feature of the distribution $\rho(\varphi,N)$ governed by the Fokker-Planck equation~(\ref{Staro-Yoko}) is that it describes the statistics of a field $\varphi_L$ defined to include every long wavelength $k/a < H$. Nevertheless, one might be more interested in describing the statistics of a field $\varphi_L$ defined within a more narrow range of scales (for instance, within a window of scales comprising 5 $e$-folds, used by the Planck satellite to analyze the CMB anisotropies).

\subsection{General aim and main results}

The purpose of this article is to provide some understanding of the general structure of $\rho (\varphi , N)$, valid for long-wavelength contributions to $\varphi$, during the initial phase of its evolution, when it retains an almost Gaussian nature — a trait that is, in fact, consistent with most observations. Using purely quantum field theoretical techniques, we will demonstrate that the general solution for $\rho (\varphi , N)$ takes the form
\begin{equation}
\rho (\varphi , N) = \frac{1}{\sqrt{2 \pi} \sigma} \exp \left[ - \frac{\varphi^2}{2 \sigma^2} + \sum_{n=1}^\infty \Delta N^n \mathcal O_n (\partial_\varphi) \mathcal V^n + \cdots
\right], \label{main-result}
\end{equation}
where $\mathcal O_n (\partial_\varphi) \mathcal V^n$ represents a term of order $\mathcal V^n$ with $\varphi$-derivatives acting on $\mathcal V(\varphi)$ with a well-determined structure. Its more precise form is given by equation~(\ref{guess-general-rho-V}). Let us stress here that, contrary to the equilibrium state in equation~\eqref{sol-Staro-Yoko}, the structure of the time-dependent corrections contains gradients of the potential instead of its bare value. Our derivation of (\ref{main-result}) is performed by assuming that $\varphi$ comprises modes within a fixed range of momenta. For this reason, to first order in the potential ${\mathcal V(\varphi)}$, the solution (\ref{main-result}) respects the following diffusionless Fokker-Planck equation:\footnote{Recall our comment following equation~(\ref{Staro-Yoko}) regarding the role of the diffusion term in the Fokker-Planck equation: it reflects the fact that as time passes, the definition of $\varphi_L$ includes more and more modes as they cross the horizon.}
\begin{equation}
\frac{\partial \rho}{\partial N} =  \frac{1}{3H^2} \frac{\partial }{\partial \varphi} \left(\frac{\partial \mathcal V}{\partial \varphi} \rho \right) + \mathcal O \left( \mathcal V^2 \right), \label{our_equation}
\end{equation}
where $\mathcal O \left( \mathcal V^2 \right)$ represents specific corrections of second order with respect to the potential ${\mathcal V(\varphi)}$, to be discussed in Section~\ref{sec:FP}. However, our approach also allows us to consider the case in which the range of momenta grows with time, such that the statistics describes the maximum possible range of momenta for modes that crossed the horizon during the period $\Delta N$. In this case, we recover a Fokker-Planck equation of the form
\bea 
\frac{\partial}{\partial N}\rho (\varphi) &=& \frac{H^2}{8\pi^2} \frac{\partial^2}{\partial \varphi^2} \left[ \rho (\varphi) \left(1 -  \frac{2\Delta N}{3H^2} \mathcal V^{\, ''}  \right)  \right] + 
     \frac{1}{3 H^2} \frac{\partial}{\partial \varphi} \left[ \rho (\varphi)  \mathcal V^{\, '}(\varphi)  \right]  . \label{our_equation_2}
\eea
Note the difference in the diffusion compared to the standard Fokker-Planck equation~(\ref{Staro-Yoko}). In Section~\ref{sec:FP}, we will analyze the emergence of this term within the stochastic formalism. The derivation of (\ref{our_equation}) and (\ref{our_equation_2}) will allow us to comment on the validity of the equilibrium solution (\ref{sol-Staro-Yoko}): We will argue that as the system evolves towards the equilibrium ($\partial \rho / \partial N = 0$) higher order corrections to (\ref{Staro-Yoko}) necessarily become important (at least in some cases). Readers interested exclusively in this aspect may jump directly to Section~\ref{sec:FP}.

To reach equation~(\ref{main-result}) we will first spend some time examining the computation of general $n$-point functions within the Schwinger-Keldysh formalism, which provides a simple diagrammatic language allowing us to write down the analytic form of any $n$-point function to the desired order with respect to the available interactions. By doing so, we will show that only certain classes of diagrams contribute leading order terms to the long-wavelength limit of the $n$-point functions under study. More specifically, we will show that if propagators are split into real and imaginary parts, only diagrams with a minimum number of imaginary propagators ---equal to the number of vertices $V$ of the diagram--- can contribute. Moreover, we show that there can be no loops formed entirely by imaginary propagators. As we shall see, a corollary of this statement (together with the fact that in the limit ${\mathcal V(\varphi)} \to 0$ our free theory corresponds to an exactly massless scalar field) is that a term contributing to $\big \langle \varphi(\k_1) \cdots \varphi(\k_N) \big \rangle$ but coming from a diagram with $V$ vertices, must be proportional to powers of $\ln a(t)$ in the following way:
\be
\big \langle \varphi(\k_1) \cdots \varphi(\k_N) \big \rangle_{V} \propto \left[ \ln  \frac{k}{Ha (t)} \right]^{V} \times \mathcal V^{V} , \label{IR-divergence-n-point}
\ee
where $k$ is a comoving momenta characterizing the set $\k_1, \k_2, \cdots , \k_n$ (for instance, the largest value of the set $k_1 , k_2, \cdots k_n$). On the other hand, $\mathcal V^{V}$ stands for a quantity proportional to $V$ powers of the potential ${\mathcal V(\varphi)}$. 

Such infrared divergences~\cite{Weinberg:2005vy,Weinberg:2006ac} are nothing but welcome, as they signal the evolution of $n$-point correlation functions shortly after horizon crossing. To be precise, given that the full $n$-point function $\big \langle \varphi(\k_1) \cdots \varphi(\k_N) \big \rangle$ is the sum of all $n$-point functions coming from diagrams with an arbitrary number of vertices
\be
\big \langle \varphi(\k_1) \cdots \varphi(\k_N) \big \rangle = \sum_{V} \big \langle \varphi(\k_1) \cdots \varphi(\k_N) \big \rangle_{V} , \label{full-n-point-nV}
\ee
then perturbation theory can only be trusted as long as terms coming from diagrams with large number of vertices remain suppressed with respect to those with a small number of vertices: 
\be
\big| \big \langle \varphi(\k_1) \cdots \varphi(\k_N) \big \rangle_{V + 1} \big| \ll \big| \big \langle \varphi(\k_1) \cdots \varphi(\k_N) \big \rangle_{V} \big|. 
\ee
As we shall see, this condition breaks down precisely at an $e$-folding time $\Delta N$ of order $H^2 / {\cal V}^{\, ''}$ as already expressed in equation~(\ref{N=H2V2}). As soon as (\ref{N=H2V2}) is satisfied, one has to resort to nonperturbative methods in order to resum the full value of $\big \langle \varphi(\k_1) \cdots \varphi(\k_N) \big \rangle$ valid at larger times (or smaller momenta).\footnote{An available example of this statement would be the computation of the two point function in a massive theory with ${\cal V}(\varphi) = \frac{m^2}{2} \varphi^2$, where mode functions can be exactly computed, leading to a resummed version of (\ref{full-n-point-nV}).}

Another interesting issue that we shall face in order to obtain equation~(\ref{main-result}) consists in the resummation of $n$-point functions containing large $n$-factorials. These large $n$-factorials are due to the large number of diagrams contributing to $n$-point functions already at tree-level. This has been studied extensively within the context of the computation of amplitudes in the case of processes involving the decay of a single particle into $n$ soft-particles in theories such as $\lambda \varphi^4$ and variants thereof~\cite{Voloshin:1992qn, Brown:1992ay, Son:1995wz, Libanov:1994ug, Khoze:2017ifq, Ghosh:2016fvm}. It has been claimed that large $n$-factorials make it impossible to resum amplitudes. As we shall see, our scheme to reconstruct the full probability density function $\rho(\varphi)$ out of $n$-point functions with large $n$-factorials remains safe thanks to compensating factorials in our general formula relating $\rho(\varphi)$ to the full set of $n$-point functions. We shall deal with this technical issue in Sections~\ref{sec:PDF_V=1} and \ref{sec:general_structure_pdf} and comment more on this aspect in the conclusions.

\subsection{Outline}

The article is organized as follows: In Section~\ref{sec-1-point} we will establish the connection between the 1-point probability distribution function $\rho (\varphi)$ and general $n$-point functions $\big \langle \varphi(\k_1) \cdots \varphi(\k_n) \big \rangle$ of $\varphi (\k , t)$ computed with the help of standard perturbative techniques within the context of quantum field theory in de Sitter space. The result of this section will help us reconstruct the form of $\rho(\varphi)$ once every $n$-point function is known. Then, in Section~\ref{sec-SK} we will offer an overview of the Schwinger-Keldysh approach to compute cosmological $n$-point correlation functions. In Section~\ref{sec:diagrammatics}, we will derive some general results concerning the computation of $n$-point functions via diagrams. This will allow us to discuss in Section~\ref{sec:long-wavelength-n-points} the computation of the leading contribution to long-wavelength $n$-point functions as determined by a general scalar potential ${\mathcal V(\varphi)}$. Having explicit expressions for $n$-point functions in terms of $\mathcal V (\varphi)$, in Section~\ref{sec:PDF_V=1}, we will be able to review the reconstruction of the general form of the PDF to first order in the potential ${\mathcal V(\varphi)}$. While this reconstruction has been already analyzed in previous works (see for instance~\cite{Palma:2017lww, Chen:2018uul, Chen:2018brw, Palma:2019lpt}) this discussion will allow us to clarify a few aspects about this computation and arrive to an even simpler expression for $\rho (\varphi , N)$ in terms of the potential ${\mathcal V(\varphi)}$. Then, in Section~\ref{sec:general_structure_pdf}, we will analyze the general form of $\rho(\varphi , N)$ as a function of ${\mathcal V(\varphi)}$ to all orders. There, we will offer a more precise expression for (\ref{main-result}). Finally, in Section~\ref{sec:FP} we will derive the Fokker-Planck equation respected by $\rho(\varphi , N)$. This derivation will allow us to compare our approach with that found within the stochastic formalism. In Section~\ref{sec:conclusions} we shall offer our concluding remarks.

\subsection{Conventions}

We will work on a de Sitter background parametrized by the metric written in conformal time $\tau \in (-\infty , 0 )$ as
\be
\de s^2 =  a^2 (\tau) \left( - \de \tau^2 + \de \x^2 \right) , \label{conformal-time-metric}
\ee
where $a(\tau) = - \frac{1}{H \tau}$ is the scale factor and $H$ is the Hubble expansion rate, set to be a constant. A Fourier transform $f(\tau, {\k})$ of a function $f(\tau , {\x}) $ will be written as
\be
f(\tau, {\k}) = \int_{\x} e^{- i {\k} \cdot {\x}} f(\tau , {\x})  ,
\ee
with $\int_{\x} \equiv \int \de^3 x$. The inverse relation is therefore 
\be
f(\tau, {\x}) = \int_{\k} e^{ i {\k} \cdot {\x}} f(\tau , {\k}) ,
\ee
with $\int_{\k} \equiv (2 \pi)^{-3} \int \de^3 k$. In particular, the inverse transform of an $n$-point correlation function $\big \langle \varphi (\x_1) \cdots  \varphi (\x_n) \big \rangle$ will take the form:
\be
\big \langle \varphi^n (\k_1 ,  \cdots  , \k_n) \big \rangle  \equiv \int_{\x_1}  e^{- i \k_1 \cdot \x_1 } \cdots \int_{\x_n}  e^{- i \k_n \cdot \x_n } \big \langle \varphi (\x_1) \cdots  \varphi (\x_n) \big \rangle . \label{Fourier-n-point}
\ee
Units will be such that $c=\hbar = 1$.

\setcounter{equation}{0}
\section{PDFs and $n$-point correlation functions}
\label{sec-1-point}

In this section we will establish a general relation between the 1-point probability density function $\rho (\varphi)$ describing the statistics of $\varphi$ and the connected part of $n$-point correlation functions $\langle \varphi (\x_1) \cdots  \varphi (\x_n) \rangle$. This relation will be valid at any given time, so we will omit the time label until Section~\ref{sec:PDF_V=1}, once it becomes necessary. We are interested in the statistics of $\varphi (\x)$ within a given range of scales restricted to be superhorizon (that is, with wavelengths much larger than the Hubble radius $R_H \equiv H^{-1}$). Typically, $n$-point correlation functions can be computed out of a probability distribution functional $P[\varphi(\x)]$, as follows: 
\be
\langle \varphi (\x_1) \cdots  \varphi (\x_n) \rangle = \int D \varphi(\x)  P[\varphi(\x)]   \varphi (\x_1) \cdots  \varphi (\x_n) .  \label{n-point-P}
\ee
Here, the integral $\int D \varphi(\x)$ denotes a functional integral (a path integral) which may satisfy additional boundary conditions not specified explicitly. The functional $P[\varphi(\x)]$ can be obtained in a top-down approach starting from a particular theory. In the context of inflation, available methods are the Hamiltonian in-in formalism~\cite{Maldacena:2002vr,Weinberg:2005vy}, the wavefunction of the universe approach~\cite{Maldacena:2002vr, Anninos:2014lwa,Arkani-Hamed:2018kmz,Pajer:2020wxk}, and the Schwinger-Keldysh formalism~\cite{Calzetta:1986ey}. In Section~\ref{sec-SK}, we will focus on Schwinger-Keldysh diagrammatics but let us stress immediately that the particular approach adopted is irrelevant to  the objective of this work. 

The computation of (\ref{n-point-P}) out of an arbitrary $P[\varphi(\x)]$ is impractical unless one resorts to perturbation theory. To do so, it is useful to define the partition function $Z[J]$ as
\be
Z[J] = \int D \varphi(\x)  P[\varphi(\x)] e^{ i \int_x \varphi (\x) J(\x)} , \label{intro-partition-Z}
\ee
which is a functional of the source function $J(\x)$. Then, from equation~(\ref{n-point-P}) one sees that $n$-point correlation functions can be written in terms of functional derivatives of $Z[J]$ with respect to sources $J(\x_i)$:
\be
\langle \varphi (\x_1) \cdots  \varphi (\x_n) \rangle = \frac{1}{i} \frac{\delta}{\delta J(\x_1)} \cdots  \frac{1}{i} \frac{\delta}{\delta J(\x_n)} Z[J] \Bigg|_{J=0} .
\ee
In general, $\langle \varphi (\x_1) \cdots  \varphi (\x_n) \rangle$ can be organized as the sum of terms that consist in the multiplication of lower $n$-point correlation functions. It is therefore convenient to deal with the so-called connected $n$-point correlation functions $\langle \varphi (\x_1) \cdots  \varphi (\x_n) \rangle_c $ which, by definition, cannot be expressed as the multiplication of lower $n$-point correlation functions. It is a standard procedure to show that the functional $W[J] \equiv \ln Z[J]$ generates only connected $n$-point functions:
\be
\langle \varphi (\x_1) \cdots  \varphi (\x_n) \rangle_c = \frac{1}{i} \frac{\delta}{\delta J(\x_1)} \cdots  \frac{1}{i} \frac{\delta}{\delta J(\x_n)} W[J] \Bigg|_{J=0} .
\ee
By using (\ref{Fourier-n-point}) one finds that in Fourier space the connected $n$-point functions are proportional to a single Dirac-delta function imposing conservation of momentum:
\be
\big \langle \varphi^n (\k_1,  \cdots  , \k_n) \big \rangle_c  = (2 \pi)^3 \delta^{(3)} (\k_1 + \cdots + \k_n) \big \langle \varphi^n (\k_1, \cdots , \k_n) \big \rangle' . \label{amp-connected}
\ee
In this expression, $\langle \varphi (\k_1) \cdots  \varphi (\k_n) \rangle'$ is the amplitude of the connected $n$-point correlation function which, in general may have singularities, but not as strong as a delta function. 

\subsection{1-point statistics}

If we are interested in $1$-point statistics, we can marginalize over the space-dependent fields $\varphi(\x)$ and define the probability density function $\rho(\varphi)$ out of the functional $P[\varphi (\x)]$ in the following way:
\be
\rho(\varphi)  \equiv  \int D \varphi (\x)  P[\varphi (\x)] \delta(\varphi - \varphi(\x) ) , \label{rel-P-rho}
\ee
where $\delta (\cdot)$ corresponds to a one-dimensional Dirac-delta function. Here, $\varphi$ is just a variable, whereas $ \varphi(\x)$ is the function entering the functional PDF. Furthermore, if our interest is to restrict the statistics to a particular range of scales, we can introduce a smoothed field $\varphi_W(\x)$ by writing $\varphi_W (\x) \equiv \int_{\k}  e^{i \k \cdot \x} W(k) \varphi (\k)$, where $W(k)$ is a window function selecting the relevant scales. Then, the 1-point probability density function would be defined as
\be
\rho(\varphi)  \equiv  \int D \varphi (\x)  P[\varphi (\x)] \delta(\varphi - \varphi_W(\x) ) . \label{rel-P-rho-W}
\ee
For simplicity, we will omit the use of $W(k)$ and come back to it in Section~\ref{sec:PDF_V=1}. 

To continue, having a 1-point probability density function $\rho(\varphi)$ one can now compute $n$-moments $\langle \varphi^n \rangle$. The standard definition reads
\be
\langle \varphi^n \rangle \equiv \int \de \varphi \rho (\varphi) \varphi^n . \label{n-moments-rho}
\ee
Inserting (\ref{rel-P-rho}) back into equation~(\ref{n-moments-rho}) one learns that
\be
\langle \varphi^n \rangle =  \int D \varphi (\x)  \rho[\varphi (\x)] \int \de\varphi  \, \delta(\varphi - \varphi(\x) ) \varphi^n =  \int D \varphi (\x)  \rho[\varphi (\x)]  \varphi^n  (\x) .
\ee
Comparing this result with (\ref{n-point-P}), we see that $n$-moments belonging to the 1-point statistics simply correspond to $n$-point correlation functions $\big \langle \varphi (\x_1) \cdots  \varphi (\x_n) \big \rangle$ with the field evaluated at the coincident limit:
\be
\langle \varphi^n \rangle = \langle \varphi (\x_1) \cdots  \varphi (\x_n) \rangle \bigg|_{\x_1 = \cdots  =  \x_n} . \label{n-moment-n-point}
\ee
Given that the $n$-point correlation functions must be invariant under spatial translations, it follows that the resulting $n$-moment is spatially homogeneous. 

\subsection{Connected $n$-moments (cumulants)}

The definition of the partition function $Z[J]$ established in equation~(\ref{intro-partition-Z}) has a 1-point counterpart $z(J)$, which may be defined in terms of $\rho(\varphi)$ via a Fourier transform:
\be
z(J) \equiv \int \de\varphi \,\rho(\varphi) e^{ i J \varphi} , \label{z-intermsof-rho}
\ee
where, now, $J$ corresponds to a variable dual to $\varphi$. Then, $n$-moments can be computed out of $z(J)$ as 
\be
\langle \varphi^n \rangle =  \frac{1}{i^n}  \frac{\de^n}{\de J^n} z(J)  \Bigg|_{J=0} .
\ee
Moreover, just as with $W[J] = \ln Z[J]$, the function $w (J) = \ln z(J)$ gives us connected $n$-moments  $\langle \varphi^n \rangle_c$ (also known as cumulants) which are simply defined as
\be
\langle \varphi^n \rangle_c =  \frac{1}{i^n}  \frac{\de^n}{\de J^n} w(J)  \Bigg|_{J=0} . \label{n-moment-w}
\ee
From this definition, with the help of (\ref{n-moment-n-point}), it follows that cumulants and $n$-point functions are related via:
\be
\langle \varphi^n \rangle_c = \langle \varphi (\x_1) \cdots  \varphi (\x_n) \rangle_c \bigg|_{\x_1 = \cdots  =  \x_n} .
\ee
Equation (\ref{n-moment-w}) implies that we can expand $w(J)$ as
\be
w(J) =  - \frac{1}{2} \sigma^2 J^2 + \sum_{n=1}^{\infty}  \frac{i^n}{n!}  \langle \varphi^n\rangle_c  J^n .
\ee
Notice that we have split the connected $2$-moment into the variance $\sigma^2$ and a second term $\langle \varphi^2\rangle_c$. With this splitting we are anticipating the fact that we shall work with physical theories whereby at zeroth order in perturbation theory all connected pieces vanish except for the variance. This is exactly what one finds in the case of cosmic inflation and scalar field theories in de Sitter spacetimes (among many other theories). 

To continue, notice that we can obtain $\rho(\varphi)$ in terms of $z(J)$ by inverting (\ref{z-intermsof-rho}) as $\rho(\varphi) =  \int \de J \;z(J) e^{ - i J \varphi}$. Using $z = e^w$ one then obtains:
\be
 \rho(\varphi) =  \frac{1}{2 \pi } \int \de J \,  e^{ - \frac{1}{2} \sigma^2 J^2} e^{-i J \varphi} \, e^{\sum_{n=1}^{\infty}  \frac{i^n}{n!}  \langle \varphi^n\rangle_c  J^n}.
\ee
Expanding the last exponential and integrating with respect to $J$, it is then possible to find a general formula relating $\rho(\varphi)$ and every connected $n$-moment:
\be
\rho (\varphi) =  \frac{e^{- \frac{1}{2} \frac{\varphi^2}{\sigma^2}}}{\sqrt{2 \pi} \sigma}  \sum_{N=0}^{\infty} \frac{1}{N!} \sum_{n_1=0}^{\infty} \cdots \sum_{n_N=0}^{\infty}
 \frac{\langle \varphi^{n_1} \rangle_c }{n_1! \sigma^{n_1}} \cdots \frac{ \langle \varphi^{n_N} \rangle_c }{n_N! \sigma^{n_N}} {\rm He}_{n_1 + \cdots + n_N} (\varphi / \sigma) ,   \label{rho-n-moments}
\ee
where ${\rm He}_n(x) = (-1)^n e^{\frac{x^2}{2}} \frac{\de^n}{\de x^n} e^{-\frac{x^2}{2}}$ correspond to the probabilistic Hermite polynomials. Equation~(\ref{rho-n-moments}) is nothing but the Gram-Charlier series expansion of the PDF (the summation can be reordered over integer partitions in $N$ parts such that the Bell polynomials over cumulants arise). For convenience, we have set $\langle \varphi^0 \rangle_c = 1$. This expression shows how to reconstruct the 1-point PDF $\rho (\varphi)$ out of the full set of connected $n$-point functions. In the context of inflation, the models we will be interested in can be described by processes which are perturbatively close to Gaussian; we will thus assume that the series~\eqref{rho-n-moments} converges. Finally, note that in this formulation, the condition $\int \de\varphi \rho (\varphi) = 1$ is automatically satisfied from the orthogonality of the Hermite polynomials with respect to the Gaussian distribution.

\subsection{Perturbative scheme}

We are interested in theories that allow the computation of $n$-point functions perturbatively. This means that there must be a parameter ---or a collection of parameters--- enabling us to perform an expansion of the form
\be
 \langle \varphi^{n} \rangle_c =  \langle \varphi^{n} \rangle_{1} +  \langle \varphi^{n} \rangle_{2} +  \langle \varphi^{n} \rangle_{3} + \cdots , \label{expansion-n-moments}
\ee
where the label $i$ on the right hand side refers to the order of expansion (recall that we are assuming that at zeroth order the connected contributions vanish). Then, from equation~(\ref{rho-n-moments}) it follows that
\be \label{PDF-schematic-pt}
\rho (\varphi) = \rho_0 (\varphi) + \rho_1 (\varphi) + \rho_2 (\varphi) + \cdots .
\ee
The first term $\rho_0$ corresponds to the Gaussian contribution:
\be
\rho_0 (\varphi) =  \frac{e^{- \frac{1}{2} \frac{\varphi^2}{\sigma^2}}}{\sqrt{2 \pi} \sigma} . \label{rho-0-gaussian}
\ee
On the other hand, the first and second order contributions to the 1-point PDF are found to be given by
\bea
\rho_1 (\varphi) &=&  \frac{e^{- \frac{1}{2} \frac{\varphi^2}{\sigma^2}}}{\sqrt{2 \pi} \sigma}  \sum_{n_1=0}^{\infty} 
 \frac{1}{n_1!}  \frac{\langle \varphi^{n_1} \rangle_{1} }{\sigma^{n_1}} {\rm He}_{n_1} (\varphi / \sigma)  , \label{rho_1-sum} \\
\rho_2 (\varphi)  &=&  \frac{e^{- \frac{1}{2} \frac{\varphi^2}{\sigma^2}}}{\sqrt{2 \pi} \sigma}  \Bigg[  \sum_{n_1=0}^{\infty} 
 \frac{1}{n_1!}  \frac{\langle \varphi^{n_1} \rangle_{2} }{\sigma^{n_1}} {\rm He}_{n_1} (\varphi / \sigma)  \nn \\ 
 && + \frac{1}{2!}  \sum_{n_1=0}^{\infty} \sum_{n_2=0}^{\infty}
 \frac{1}{n_1!}  \frac{1}{n_2!}  \frac{\langle \varphi^{n_1} \rangle_{1} }{\sigma^{n_1}} \frac{ \langle \varphi^{n_2} \rangle_{1} }{\sigma^{n_2}} {\rm He}_{n_1 +  n_2} (\varphi / \sigma)  \Bigg] . \label{rho_2-sum}
\eea
We shall come back to these results after obtaining concrete expressions for the connected $n$-moments. In particular, the resummation of (\ref{rho_2-sum}) will highlight several steps valid in the general resummation of an arbitrary term $\rho_{i}$.

\setcounter{equation}{0}
\section{Schwinger-Keldysh formalism}
\label{sec-SK}

Let us now briefly review the Schwinger-Keldysh formalism~\cite{Calzetta:1986ey} (we refer to~\cite{Chen:2017ryl} for a comprehensive introduction to this technique within the context of inflation). The starting point is to consider a quantum theory for the field operator $\varphi (\x , \tau)$, described by the action
\be
S [\varphi (\x , \tau)] = \int \de^3 x \, \de \tau \, a^4 \mathcal L \big( \varphi (\x , \tau) \big) .
\ee
Here, $\mathcal L \big( \varphi (\x , \tau) \big)$ is the Lagrangian of the system. In this work, we will focus on canonical models with arbitrary interactions, which in the conformal-time metric (\ref{conformal-time-metric}) are described by
\be
\mathcal L  =  \frac{1}{2 a^2} \bigg[  \left( \frac{\de \varphi}{\de\tau} \right)^2  - \left( \nabla \varphi \right) ^2 \bigg] - {\mathcal V(\varphi)}.
\ee
The procedure examined here can be easily extended to more general theories. 

The Schwinger-Keldysh formalism states that the partition function employed to compute $n$-point correlation functions, has the following form:
\be
Z [J_+ , J_- ] = \mathcal N \int_{\varphi_{+} (\tau_f) = \varphi_{-} (\tau_f) = \varphi (\tau_f)} \hspace{-3cm} D \varphi_{+}   D \varphi_{-}  \, \, e^{ i \big( S [\varphi_+ ]  - S [\varphi_- ]  \big) }  e^{ i \int \de^4 x  \big( J_+ \varphi_{+}    - J_-  \varphi_{-}   \big) }  , \label{general-partition-Z}
\ee
where $\mathcal N$ is some normalization constant that will play no relevant role whatsoever. In the previous expression the time integral in the definition of the action $S$ goes from $-\infty$ to $\tau_f$. Notice that in this formalism, the number of fields involved in the path integral doubles; in this particular case, one scalar $\varphi$ corresponds to two fields denoted by $\varphi_+$ and $\varphi_-$. Nonetheless, these two fields are identified with the physical field $\varphi (\tau_f)$ at the time $\tau_f$ at which we are interested in evaluating $n$-point functions. This condition can be fixed by inserting the product of Dirac-delta functions $\prod_{\x} \delta \big(\varphi_+ (\x , \tau_f) - \varphi_- (\x , \tau_f) \big)$ inside the path integral. 

Having $Z [J_+ , J_- ] $ at our disposal, an $n$-point function evaluated at $\tau_f$ can be obtained as follows:
\be
\langle \varphi (\x_1) \cdots  \varphi (\x_n) \rangle = \frac{1}{a_1 i} \frac{\delta}{\delta J_{a_1}(\x_1,\tau_f)} \cdots  \frac{1}{a_n i} \frac{\delta}{\delta J_{a_n}(\x_n,\tau_f)} Z [J_+ , J_- ]  \Bigg|_{J=0} , \label{n-point-Z_pm}
\ee
where the factors $a_i$ take the values $+1$ or $-1$ depending of the type of source $J_{\pm}$. Given that both fields are identified at the temporal surface $\tau_f$ (the boundary), the type of source used to evaluate the previous expression turns out to be irrelevant. 

To compute $n$-point functions perturbatively, we can now split the action into free and interacting contributions. In the present context, the interacting contribution is given by the entire potential ${\mathcal V(\varphi)}$:
\be
S_{\rm int} = - \int \! \de^3 x \, \de \tau \,a^4  {\mathcal V(\varphi)} .
\ee
To deal with ${\mathcal V(\varphi)}$ we simply expand it in Taylor series as
\be
{\mathcal V(\varphi)} = \sum_n \frac{1}{n!} \lambda_n \varphi^n , \label{Taylor-V}
\ee
and treat every $\lambda_n$ as a quantity of order ${\mathcal V(\varphi)}$. Of course, given that the dimension of $\lambda_n$ changes according to the value of $n$, this statement involves additional parameters with mass dimensions ---however, recall our comment in the paragraph after equation~(\ref{intro:action}). Then, the partition function becomes
\be
Z [J_+ , J_- ] = e^{-  i  \sum_n \frac{1}{n!} \lambda_n \int \de^3 x \,\de \tau \, a^4 \left(   \left[ \frac{1}{i} \frac{\delta}{\delta J_+} \right]^n  -  \left[ - \frac{1}{i} \frac{\delta}{\delta J_-} \right]^n   \right) }  Z_0 [J_+ , J_- ] , \label{Z-Z0}
\ee
where $Z_0 [J_+ , J_- ]$ corresponds to the zeroth order partition function (\ref{general-partition-Z}) excluding the interaction term $S_{\rm int}$. 

The functional $Z_0 [J_+ , J_- ]$ requires specifying the vacuum in the infinite past $\tau_{\rm ini} \to - \infty$. This can be achieved with the help of the usual $\epsilon$-prescription and the additional requirement that the state of the universe coincides with the Bunch-Davies vacuum state. This step specifies the mode functions of free propagating fields as
\be
\varphi(k , \tau) = \frac{H}{\sqrt{2 k^3}} (1 + i k \tau) e^{- i k \tau} . \label{BD-field-massless}
\ee 
Notice that (\ref{BD-field-massless}) is the mode function of a massless field, which follows from the fact that our zeroth order theory corresponds to $\mathcal L  =  \frac{1}{2 a^2} \big[  \left( \de \varphi / \de\tau \right)^2 - \left( \nabla \varphi  \right)^2 \big]$. 

With all these elements at sight, the Feynman rules for computing $n$-point functions can be obtained after expanding the exponential in (\ref{Z-Z0}) and inserting the result back in equation~(\ref{n-point-Z_pm}). To write down the rules, it is first convenient to write the basic function involved in the definition of every propagator: 
\be
G (k, \tau_1 , \tau_2) = \varphi(k , \tau_1) \varphi^*(k , \tau_2) = \frac{H^2}{2 k^3} (1 + i k \tau_1) (1 - i k \tau_2)  e^{- i k ( \tau_1 - \tau_2 ) } . \label{G-uu}
\ee
We are now ready to review the rules.

\subsection{Feynman rules}

The Feynman rules tell us how to obtain the amplitude $\big \langle \varphi^n ( \k_1 \cdots , \k_n ) \big \rangle'$ given a certain interaction. Notice that this is a function of the external momenta $\k_1$, $\k_2$, $\cdots$, $\k_n$ and the final time $\tau_f$.

A single term of the expansion (\ref{Taylor-V}), proportional to $\lambda_n$, defines two classes of vertices: one corresponding to the field $\varphi_+$ and another corresponding to the field $\varphi_-$. We will associate black solid circles to the former and white empty circles to the latter. This convention then requires the following assignment for the two types of vertices:
\def\nvertexb{\tikz[baseline=-0.6ex,scale=1.8, every node/.style={scale=1.4}]{
\coordinate (v1) at (0ex,0ex);
\coordinate (phi1) at (-3ex,2ex);
\coordinate (phi2) at (-4ex,0ex);
\coordinate (phi3) at (-3ex,-2ex);
\coordinate (phi4) at (3ex,2ex);
\coordinate (phi5) at (4ex,0ex);
\coordinate (phi6) at (3ex,-2ex);
\draw[thick] (v1) -- (phi1);
\draw[thick] (v1) -- (phi2);
\draw[thick] (v1) -- (phi3);
\draw[thick] (v1) -- (phi4);
\draw[thick] (v1) -- (phi5);
\draw[thick] (v1) -- (phi6);
\filldraw[color=black, fill=black, thick] (v1) circle (0.5ex);
\node[anchor=south] at ($(v1)+(0,-2.5ex)$) {\scriptsize{$\tau$}};
\node[anchor=south] at ($(v1)+(0,+1.0ex)$) {\scriptsize{$\cdots$}};
}
}
\def\nvertexw{\tikz[baseline=-0.6ex,scale=1.8, every node/.style={scale=1.4}]{
\coordinate (v1) at (0ex,0ex);
\coordinate (phi1) at (-3ex,2ex);
\coordinate (phi2) at (-4ex,0ex);
\coordinate (phi3) at (-3ex,-2ex);
\coordinate (phi4) at (3ex,2ex);
\coordinate (phi5) at (4ex,0ex);
\coordinate (phi6) at (3ex,-2ex);
\draw[thick] (v1) -- (phi1);
\draw[thick] (v1) -- (phi2);
\draw[thick] (v1) -- (phi3);
\draw[thick] (v1) -- (phi4);
\draw[thick] (v1) -- (phi5);
\draw[thick] (v1) -- (phi6);
\filldraw[color=black, fill=white, thick] (v1) circle (0.5ex);
\node[anchor=south] at ($(v1)+(0,-2.5ex)$) {\scriptsize{$\tau$}};
\node[anchor=south] at ($(v1)+(0,+1.0ex)$) {\scriptsize{$\cdots$}};
}
}
\bea
    \nvertexb \qquad & \longrightarrow & \qquad - i \lambda_n \int^{\tau_f}_{- \infty} \! \de\tau \, a^4(\tau) \bigg[ \cdots \bigg] , \\
     \nvertexw \qquad & \longrightarrow & \qquad + i \lambda_n \int^{\tau_f}_{- \infty} \! \de\tau \, a^4(\tau) \bigg[ \cdots \bigg] .
\eea
Notice that each vertex is characterized by a time label $\tau$ that must be integrated from $-\infty$ up until the final time $\tau_f$ at which $n$-point correlation functions are evaluated. The square brackets on the right hand side of the previous equations indicate that any function of $\tau$ must be integrated in this way.  Notice that a white-dotted vertex is the complex conjugate of a black-dotted one.

Given that we have two classes of vertices, the theory will contain four types of internal propagators. The corresponding internal propagators are given by
\def\propbb{\tikz[baseline=-0.6ex,scale=1.8, every node/.style={scale=1.4}]{
\coordinate (tau1) at (-4ex,0ex);
\coordinate (tau2) at (4ex,0ex);
\draw[thick] (tau1) -- (tau2);
\filldraw[color=black, fill=black, thick] (tau1) circle (0.5ex);
\node[anchor=south] at ($(tau1)+(0,0.5ex)$) {\scriptsize{$\tau_1$}};
\filldraw[color=black, fill=black, thick] (tau2) circle (0.5ex);
\node[anchor=south] at ($(tau2)+(0,0.5ex)$) {\scriptsize{$\tau_2$}};
}
}
\def\propww{\tikz[baseline=-0.6ex,scale=1.8, every node/.style={scale=1.4}]{
\coordinate (tau1) at (-4ex,0ex);
\coordinate (tau2) at (4ex,0ex);
\draw[thick] (tau1) -- (tau2);
\filldraw[color=black, fill=white, thick] (tau1) circle (0.5ex);
\node[anchor=south] at ($(tau1)+(0,0.5ex)$) {\scriptsize{$\tau_1$}};
\filldraw[color=black, fill=white, thick] (tau2) circle (0.5ex);
\node[anchor=south] at ($(tau2)+(0,0.5ex)$) {\scriptsize{$\tau_2$}};
}
}
\def\propbw{\tikz[baseline=-0.6ex,scale=1.8, every node/.style={scale=1.4}]{
\coordinate (tau1) at (-4ex,0ex);
\coordinate (tau2) at (4ex,0ex);
\draw[thick] (tau1) -- (tau2);
\filldraw[color=black, fill=black, thick] (tau1) circle (0.5ex);
\node[anchor=south] at ($(tau1)+(0,0.5ex)$) {\scriptsize{$\tau_1$}};
\filldraw[color=black, fill=white, thick] (tau2) circle (0.5ex);
\node[anchor=south] at ($(tau2)+(0,0.5ex)$) {\scriptsize{$\tau_2$}};
}
}
\def\propwb{\tikz[baseline=-0.6ex,scale=1.8, every node/.style={scale=1.4}]{
\coordinate (tau1) at (-4ex,0ex);
\coordinate (tau2) at (4ex,0ex);
\draw[thick] (tau1) -- (tau2);
\filldraw[color=black, fill=white, thick] (tau1) circle (0.5ex);
\node[anchor=south] at ($(tau1)+(0,0.5ex)$) {\scriptsize{$\tau_1$}};
\filldraw[color=black, fill=black, thick] (tau2) circle (0.5ex);
\node[anchor=south] at ($(tau2)+(0,0.5ex)$) {\scriptsize{$\tau_2$}};
}
}
\bea
\propbb  \qquad  &\longrightarrow & \qquad G_{++} (k, \tau_1 , \tau_2) , \\
\propww  \qquad &\longrightarrow &\qquad  G_{--} (k, \tau_1 , \tau_2) , \\
\propbw  \qquad &\longrightarrow & \qquad G_{+-} (k, \tau_1 , \tau_2) ,\\
\propwb  \qquad &\longrightarrow & \qquad G_{-+} (k, \tau_1 , \tau_2) .
\eea
The analytical expressions for the quantities appearing on the right hand side of the previous assignments are given in terms of the function $G (k, \tau_1 , \tau_2)$ introduced in equation~(\ref{G-uu}). These are:
\bea
G_{++} (k, \tau_1 , \tau_2)  &=& G (k, \tau_1 , \tau_2) \theta (\tau_1 - \tau_2) +  G^* (k, \tau_1 , \tau_2) \theta (\tau_2 - \tau_1) , \\
G_{--} (k, \tau_1 , \tau_2)  &=& G^* (k, \tau_1 , \tau_2) \theta (\tau_1 - \tau_2) +  G (k, \tau_1 , \tau_2) \theta (\tau_2 - \tau_1) , \\
G_{+-} (k, \tau_1 , \tau_2)  &=& G^* (k, \tau_1 , \tau_2) , \\
G_{-+} (k, \tau_1 , \tau_2)  &=& G (k, \tau_1 , \tau_2) .
\eea
Just as in the case of vertices, it can be verified explicitly that propagators with given colors (black or white) at their ends are the complex conjugate of propagators with the opposite color.

Finally, to obtain an amplitude $\big \langle \varphi (\k_1) \cdots  \varphi (\k_n) \big \rangle'$ the vertices (evaluated at times $\tau_1$, $\tau_2$, $\tau_3$, etc) must be connected to a surface labelled with the time $\tau_f$ through bulk-to-boundary propagators. These receive the following assignments:
\def\propbf{\tikz[baseline=-0.6ex,scale=1.8, every node/.style={scale=1.4}]{
\coordinate (tau) at (-4ex,0ex);
\coordinate (phi) at (4ex,0ex);
\draw[thick] (tau) -- (phi);
\filldraw[color=black, fill=black, thick] (tau) circle (0.5ex);
\node[anchor=south] at ($(tau)+(0,0.5ex)$) {\scriptsize{$\tau$}};
\pgfmathsetmacro{\arista}{0.08}
\filldraw[color=black, fill=white, thick] ($(phi)-(\arista,\arista)$) rectangle ($(phi)+(\arista,\arista)$);
\node[anchor=south] at ($(phi)+(0,0.5ex)$){\scriptsize{$\tau_f$}};
}
}
\def\propwf{\tikz[baseline=-0.6ex,scale=1.8, every node/.style={scale=1.4}]{
\coordinate (tau) at (-4ex,0ex);
\coordinate (phi) at (4ex,0ex);
\draw[thick] (tau) -- (phi);
\filldraw[color=black, fill=white, thick] (tau) circle (0.5ex);
\node[anchor=south] at ($(tau)+(0,0.5ex)$) {\scriptsize{$\tau$}};
\pgfmathsetmacro{\arista}{0.08}
\filldraw[color=black, fill=white, thick] ($(phi)-(\arista,\arista)$) rectangle ($(phi)+(\arista,\arista)$);
\node[anchor=south] at ($(phi)+(0,0.5ex)$){\scriptsize{$\tau_f$}};
}
}
\bea
  \propbf   \qquad  &\longrightarrow & \qquad  G_{+}  (k, \tau) , \\
  \propwf  \qquad  &\longrightarrow & \qquad  G_{-}  (k, \tau) .
\eea
Given that fields $\varphi_{+}$ and $\varphi_{-}$ are identified at the boundary, it is unnecessary to assign colors to the square defining the end point evaluated at $\tau_f$. The analytical expressions for these two bulk-to-boundary propagators are given in terms of the function $G(k,\tau_1,\tau_2)$ by:
\bea
G_{+}  (k, \tau_1) &=&  G^* (k, \tau_1 , \tau_f) ,  \\
G_{-}  (k, \tau_1) &=& G (k, \tau_1 , \tau_f) .
\eea

The rules to compute  $\big \langle \varphi (\k_1) \cdots  \varphi (\k_n) \big \rangle'$ consist in adding all possible diagrams of the desired order which have $n$ legs attached to the boundary. The analytical expression corresponding to each diagram is determined by the assignments already declared, procuring that conservation of momentum is imposed in every vertex. Of course, each diagram must have its symmetry factor computed in the standard way. Notice that the rules ensure that after adding every type of vertex in a single type of diagram, the resulting amplitude will always be real, simply because a black vertex is the conjugate of a white vertex (a rule that is transferred to propagators). In order to simplify the writing of a diagram, it is useful to introduce undotted vertices as the sum of two dotted vertices (a black vertex plus a white vertex) in the following way:
\def\nvertex{\tikz[baseline=-0.6ex,scale=1.8, every node/.style={scale=1.4}]{
\coordinate (v1) at (0ex,0ex);
\coordinate (phi1) at (-3ex,2ex);
\coordinate (phi2) at (-4ex,0ex);
\coordinate (phi3) at (-3ex,-2ex);
\coordinate (phi4) at (3ex,2ex);
\coordinate (phi5) at (4ex,0ex);
\coordinate (phi6) at (3ex,-2ex);
\draw[thick] (v1) -- (phi1);
\draw[thick] (v1) -- (phi2);
\draw[thick] (v1) -- (phi3);
\draw[thick] (v1) -- (phi4);
\draw[thick] (v1) -- (phi5);
\draw[thick] (v1) -- (phi6);
\node[anchor=south] at ($(v1)+(0,-2.5ex)$) {\scriptsize{$\tau$}};
\node[anchor=south] at ($(v1)+(0,+1.0ex)$) {\scriptsize{$\cdots$}};
}
}
\bea
    \nvertex \quad = \qquad \nvertexb \quad + \quad \nvertexw \quad.
\eea
It is to be understood that before evaluating a given diagram, every undotted vertex must first be expanded into dotted vertices. 

To finish this discussion, notice that the parameter defining the perturbative scheme laid out in Section~\ref{sec-SK} is the entire potential ${\mathcal V(\varphi)}$. In practice, the order of the correlation function in ${\cal V}$ is equal to the degree of the correlation function as a polynomial in the couplings $\lambda_n$ introduced in \eqref{Taylor-V}. In what follows we review two examples of first and second order.

\subsection{Example 1: First order correlation functions} \label{sec:basic_example_1}

Let us consider the lowest order contribution to the amplitude $\big \langle \varphi^n ( \k_1 \cdots , \k_n ) \big \rangle'$ offered by a vertex of order $\lambda_n$. This consists in the following 1-vertex tree-level diagram:
\def\diagramexone{\tikz[baseline=-1.4ex,scale=1.8, every node/.style={scale=1.4}]{
\coordinate (k1) at (-3ex,0ex);
\coordinate (k2) at (-1ex,0ex);
\coordinate (kn) at (3ex,0ex);
\coordinate (tau) at (0,-4ex);
\coordinate (dots) at (0.9ex,-1.8ex);
\pgfmathsetmacro{\arista}{0.06}
\draw[thick] (-4ex,0ex) -- (4ex,0ex);
\draw[-,thick] (tau) -- (k1);
\draw[-,thick] (tau) -- (k2);
\draw[-,thick] (tau) -- (kn);
\node[anchor=south] at (dots) {\scriptsize{$\cdots$}};
\node[anchor=south] at ($(tau)+(0,-2.0ex)$) {\scriptsize{$\tau$}};
\filldraw[color=black, fill=white, thick] ($(k1)-(\arista,\arista)$) rectangle ($(k1)+(\arista,\arista)$);
\node[anchor=south] at ($(k1)+(0ex,0.5ex)$) {\scriptsize{$\k_1$}};
\filldraw[color=black, fill=white, thick] ($(k2)-(\arista,\arista)$) rectangle ($(k2)+(\arista,\arista)$);
\node[anchor=south] at ($(k2)+(0ex,0.5ex)$) {\scriptsize{$\k_2$}};
\filldraw[color=black, fill=white, thick] ($(kn)-(\arista,\arista)$) rectangle ($(kn)+(\arista,\arista)$);
\node[anchor=south] at ($(kn)+(0ex,0.5ex)$) {\scriptsize{$\k_{n}$}}
}
}
\def\diagramextwo{\tikz[baseline=-1.4ex,scale=1.8, every node/.style={scale=1.4}]{
\coordinate (k1) at (-3ex,0ex);
\coordinate (k2) at (-1ex,0ex);
\coordinate (kn) at (3ex,0ex);
\coordinate (tau) at (0,-4ex);
\coordinate (dots) at (0.9ex,-1.8ex);
\pgfmathsetmacro{\arista}{0.06}
\draw[thick] (-4ex,0ex) -- (4ex,0ex);
\draw[-,thick] (tau) -- (k1);
\draw[-,thick] (tau) -- (k2);
\draw[-,thick] (tau) -- (kn);
\node[anchor=south] at (dots) {\scriptsize{$\cdots$}};
\filldraw[color=black, fill=black, thick] (tau) circle (0.4ex); 
\node[anchor=south] at ($(tau)+(0,-2.0ex)$) {\scriptsize{$\tau$}};
\filldraw[color=black, fill=white, thick] ($(k1)-(\arista,\arista)$) rectangle ($(k1)+(\arista,\arista)$);
\node[anchor=south] at ($(k1)+(0ex,0.5ex)$) {\scriptsize{$\k_1$}};
\filldraw[color=black, fill=white, thick] ($(k2)-(\arista,\arista)$) rectangle ($(k2)+(\arista,\arista)$);
\node[anchor=south] at ($(k2)+(0ex,0.5ex)$) {\scriptsize{$\k_2$}};
\filldraw[color=black, fill=white, thick] ($(kn)-(\arista,\arista)$) rectangle ($(kn)+(\arista,\arista)$);
\node[anchor=south] at ($(kn)+(0ex,0.5ex)$) {\scriptsize{$\k_{n}$}}
}
}
\def\diagramexthree{\tikz[baseline=-1.4ex,scale=1.8, every node/.style={scale=1.4}]{
\coordinate (k1) at (-3ex,0ex);
\coordinate (k2) at (-1ex,0ex);
\coordinate (kn) at (3ex,0ex);
\coordinate (tau) at (0,-4ex);
\coordinate (dots) at (0.9ex,-1.8ex);
\pgfmathsetmacro{\arista}{0.06}
\draw[thick] (-4ex,0ex) -- (4ex,0ex);
\draw[-,thick] (tau) -- (k1);
\draw[-,thick] (tau) -- (k2);
\draw[-,thick] (tau) -- (kn);
\node[anchor=south] at (dots) {\scriptsize{$\cdots$}};
\filldraw[color=black, fill=white, thick] (tau) circle (0.4ex); 
\node[anchor=south] at ($(tau)+(0,-2.0ex)$) {\scriptsize{$\tau$}};
\filldraw[color=black, fill=white, thick] ($(k1)-(\arista,\arista)$) rectangle ($(k1)+(\arista,\arista)$);
\node[anchor=south] at ($(k1)+(0ex,0.5ex)$) {\scriptsize{$\k_1$}};
\filldraw[color=black, fill=white, thick] ($(k2)-(\arista,\arista)$) rectangle ($(k2)+(\arista,\arista)$);
\node[anchor=south] at ($(k2)+(0ex,0.5ex)$) {\scriptsize{$\k_2$}};
\filldraw[color=black, fill=white, thick] ($(kn)-(\arista,\arista)$) rectangle ($(kn)+(\arista,\arista)$);
\node[anchor=south] at ($(kn)+(0ex,0.5ex)$) {\scriptsize{$\k_{n}$}}
}
}
\be
 \langle \varphi^n ( \k_1 , \cdots , \k_n ) \rangle '  \quad =  \quad  \diagramexone  \nn 
\ee
Recall that undotted diagrams corresponds to the sum of dotted diagrams with every possible combination of dotted-vertices. In this case, we simply have
\be
\diagramexone  \quad =  \quad  \diagramextwo \quad + \quad \diagramexthree \nn
\ee
Because black and white vertices (and every propagator connected to them) are the conjugate of each other, we can conveniently rewrite the previous expression as:
 \be
 \langle \varphi^n ( \k_1 , \cdots , \k_n )\rangle '  \quad  = \quad 2 {\rm Re} \Bigg[ \diagramextwo \Bigg] \,\, \nn
\ee
Then, by translating the diagram into its analytical form, we find that the amplitude of the $n$-point function is given by
\bea \label{npt'}
\langle \varphi^n ( \k_1 , \cdots , \k_n ) \rangle ' &=&   \frac{2 \lambda_n }{H^4} 
\int^{\tau_f}_{- \infty} \frac{\de \tau}{\tau^4} \Im \Big\{ G (k_1, \tau , \tau_f ) \cdots G (k_n, \tau , \tau_f) \Big\} .
\eea
This expression can now be integrated to give the desired answer for the $n$-point function as a function of $\tau_f$ and the momenta $\k_1$, $\k_2$, $\cdots$, $\k_n$. However, as we shall discuss in Section~\ref{sec:long-wavelength-n-points}, to obtain the leading contribution of the $n$-point function in the long wavelength limit, we don't need to integrate the entire domain $-\infty < \tau < \tau_f$. Instead, it will be sufficient to integrate within the domain $\tau_0 < \tau < \tau_f$ where $\tau_0$ is some arbitrary time such that $k_i | \tau_0 | \ll 1$. This will considerably simplify the task of obtaining analytical expressions.

\subsection{Example 2: Second order correlation functions} \label{sec:basic_example_2}

As a second example, let us consider the computation of the tree-level contribution of a second order diagram with respect to the number of vertices. To construct an exchange diagram contributing to an $n$-point function we must connect external momenta with two vertices of type $\lambda_{n_1+1}$ and $\lambda_{n_2+1}$ such that $n = n_1 + n_2$. The desired amplitude can then be expressed as
\def\diagramextwovertices{\tikz[baseline=-1.4ex,scale=1.8, every node/.style={scale=1.4}]{
\coordinate (k11) at (-5ex,0ex);
\coordinate (k1n1) at (-1.4ex,0ex);
\coordinate (k21) at (+1.4ex,0ex);
\coordinate (k2n2) at (+5ex,0ex);
\coordinate (t1) at (-3.2ex,-4ex);
\coordinate (t2) at (+3.2ex,-4ex);
\coordinate (dots1) at (-3.2ex,-1.8ex);
\coordinate (dots2) at (+3.2ex,-1.8ex);
\pgfmathsetmacro{\arista}{0.06}
\draw[thick] (-6ex,0ex) -- (6ex,0ex);
\draw[-,thick] (t1) -- (k11);
\draw[-,thick] (t1) -- (k1n1);
\draw[-,thick] (t2) -- (k21);
\draw[-,thick] (t2) -- (k2n2);
\draw[-,thick] (t1) -- (t2);
\node[anchor=south] at ($(dots1)$) {\scriptsize{$\cdots$}};
\node[anchor=south] at ($(dots2)$) {\scriptsize{$\cdots$}};
\node[anchor=south] at ($(t1)+(0,-2.0ex)$) {\scriptsize{$\tau_1$}};
\node[anchor=south] at ($(t2)+(0,-2.0ex)$) {\scriptsize{$\tau_2$}};
\filldraw[color=black, fill=white, thick] ($(k11)-(\arista,\arista)$) rectangle ($(k11)+(\arista,\arista)$);
\node[anchor=south] at ($(k11)+(0ex,0.5ex)$) {\scriptsize{$\k_{11}$}};
\filldraw[color=black, fill=white, thick] ($(k1n1)-(\arista,\arista)$) rectangle ($(k1n1)+(\arista,\arista)$);
\node[anchor=south] at ($(k1n1)+(0ex,0.5ex)$) {\scriptsize{$\k_{1 n_1}$}};
\filldraw[color=black, fill=white, thick] ($(k21)-(\arista,\arista)$) rectangle ($(k21)+(\arista,\arista)$);
\node[anchor=south] at ($(k21)+(0ex,0.5ex)$) {\scriptsize{$\k_{21}$}};
\filldraw[color=black, fill=white, thick] ($(k2n2)-(\arista,\arista)$) rectangle ($(k2n2)+(\arista,\arista)$);
\node[anchor=south] at ($(k2n2)+(0ex,0.5ex)$) {\scriptsize{$\k_{2 n_2}$}}
}
}
\be
\langle \varphi^n (\k_{11} , \cdots , \k_{1 n_1} , \k_{21} , \cdots , \k_{2n_2}) \rangle '  \quad = \quad \diagramextwovertices + {\rm perms .} \nn
\ee
To draw this diagram we have opted for the notation whereby the external momenta connected to the vertex $\tau_1$ (proportional to $\lambda_{n_1+1}$) are labelled as $\k_{11}, \k_{12}, \cdots \k_{1 n_1}$ whereas the momenta connected to the second vertex $\tau_2$ are labelled as $\k_{21}, \k_{22}, \cdots \k_{2 n_2}$. 
The term ``perms" stands for permutations of momenta, which give us new channels contributing to the same amplitude. That is, the permutation must be among momenta connected to different vertices. In total, there are $(n_1 + n_2)! / n_1 ! n_2 !$ different channels allowed by permutations as long as $n_1 \neq n_2$. If instead $n_1 = n_2$, then there are $(n_1 + n_2)! / 2! n_1 ! n_2 !$ channels, where the factor $2!$ appearing in the denominator is due to the extra symmetry of the diagram when the two vertices have the same number of external legs (we will come back to this issue in Section~\ref{sec:general_structure_pdf}). 

Let us now consider the expansion of the undotted diagram corresponding to the first channel in terms of dotted diagrams. One finds:
\def\diagramextwoverticesone{\tikz[baseline=-1.4ex,scale=1.8, every node/.style={scale=1.4}]{
\coordinate (k11) at (-5ex,0ex);
\coordinate (k1n1) at (-1.4ex,0ex);
\coordinate (k21) at (+1.4ex,0ex);
\coordinate (k2n2) at (+5ex,0ex);
\coordinate (t1) at (-3.2ex,-4ex);
\coordinate (t2) at (+3.2ex,-4ex);
\coordinate (dots1) at (-3.2ex,-1.8ex);
\coordinate (dots2) at (+3.2ex,-1.8ex);
\pgfmathsetmacro{\arista}{0.06}
\draw[thick] (-6ex,0ex) -- (6ex,0ex);
\draw[-,thick] (t1) -- (k11);
\draw[-,thick] (t1) -- (k1n1);
\draw[-,thick] (t2) -- (k21);
\draw[-,thick] (t2) -- (k2n2);
\draw[-,thick] (t1) -- (t2);
\node[anchor=south] at ($(dots1)$) {\scriptsize{$\cdots$}};
\node[anchor=south] at ($(dots2)$) {\scriptsize{$\cdots$}};
\node[anchor=south] at ($(t1)+(0,-2.0ex)$) {\scriptsize{$\tau_1$}};
\node[anchor=south] at ($(t2)+(0,-2.0ex)$) {\scriptsize{$\tau_2$}};
\filldraw[color=black, fill=black, thick] (t1) circle (0.4ex); 
\filldraw[color=black, fill=black, thick] (t2) circle (0.4ex); 
\filldraw[color=black, fill=white, thick] ($(k11)-(\arista,\arista)$) rectangle ($(k11)+(\arista,\arista)$);
\node[anchor=south] at ($(k11)+(0ex,0.5ex)$) {\scriptsize{$\k_{11}$}};
\filldraw[color=black, fill=white, thick] ($(k1n1)-(\arista,\arista)$) rectangle ($(k1n1)+(\arista,\arista)$);
\node[anchor=south] at ($(k1n1)+(0ex,0.5ex)$) {\scriptsize{$\k_{1 n_1}$}};
\filldraw[color=black, fill=white, thick] ($(k21)-(\arista,\arista)$) rectangle ($(k21)+(\arista,\arista)$);
\node[anchor=south] at ($(k21)+(0ex,0.5ex)$) {\scriptsize{$\k_{21}$}};
\filldraw[color=black, fill=white, thick] ($(k2n2)-(\arista,\arista)$) rectangle ($(k2n2)+(\arista,\arista)$);
\node[anchor=south] at ($(k2n2)+(0ex,0.5ex)$) {\scriptsize{$\k_{2 n_2}$}}
}
}
\def\diagramextwoverticestwo{\tikz[baseline=-1.4ex,scale=1.8, every node/.style={scale=1.4}]{
\coordinate (k11) at (-5ex,0ex);
\coordinate (k1n1) at (-1.4ex,0ex);
\coordinate (k21) at (+1.4ex,0ex);
\coordinate (k2n2) at (+5ex,0ex);
\coordinate (t1) at (-3.2ex,-4ex);
\coordinate (t2) at (+3.2ex,-4ex);
\coordinate (dots1) at (-3.2ex,-1.8ex);
\coordinate (dots2) at (+3.2ex,-1.8ex);
\pgfmathsetmacro{\arista}{0.06}
\draw[thick] (-6ex,0ex) -- (6ex,0ex);
\draw[-,thick] (t1) -- (k11);
\draw[-,thick] (t1) -- (k1n1);
\draw[-,thick] (t2) -- (k21);
\draw[-,thick] (t2) -- (k2n2);
\draw[-,thick] (t1) -- (t2);
\node[anchor=south] at ($(dots1)$) {\scriptsize{$\cdots$}};
\node[anchor=south] at ($(dots2)$) {\scriptsize{$\cdots$}};
\node[anchor=south] at ($(t1)+(0,-2.0ex)$) {\scriptsize{$\tau_1$}};
\node[anchor=south] at ($(t2)+(0,-2.0ex)$) {\scriptsize{$\tau_2$}};
\filldraw[color=black, fill=black, thick] (t1) circle (0.4ex); 
\filldraw[color=black, fill=white, thick] (t2) circle (0.4ex); 
\filldraw[color=black, fill=white, thick] ($(k11)-(\arista,\arista)$) rectangle ($(k11)+(\arista,\arista)$);
\node[anchor=south] at ($(k11)+(0ex,0.5ex)$) {\scriptsize{$\k_{11}$}};
\filldraw[color=black, fill=white, thick] ($(k1n1)-(\arista,\arista)$) rectangle ($(k1n1)+(\arista,\arista)$);
\node[anchor=south] at ($(k1n1)+(0ex,0.5ex)$) {\scriptsize{$\k_{1 n_1}$}};
\filldraw[color=black, fill=white, thick] ($(k21)-(\arista,\arista)$) rectangle ($(k21)+(\arista,\arista)$);
\node[anchor=south] at ($(k21)+(0ex,0.5ex)$) {\scriptsize{$\k_{21}$}};
\filldraw[color=black, fill=white, thick] ($(k2n2)-(\arista,\arista)$) rectangle ($(k2n2)+(\arista,\arista)$);
\node[anchor=south] at ($(k2n2)+(0ex,0.5ex)$) {\scriptsize{$\k_{2 n_2}$}}
}
}
\def\diagramextwoverticesthree{\tikz[baseline=-1.4ex,scale=1.8, every node/.style={scale=1.4}]{
\coordinate (k11) at (-5ex,0ex);
\coordinate (k1n1) at (-1.4ex,0ex);
\coordinate (k21) at (+1.4ex,0ex);
\coordinate (k2n2) at (+5ex,0ex);
\coordinate (t1) at (-3.2ex,-4ex);
\coordinate (t2) at (+3.2ex,-4ex);
\coordinate (dots1) at (-3.2ex,-1.8ex);
\coordinate (dots2) at (+3.2ex,-1.8ex);
\pgfmathsetmacro{\arista}{0.06}
\draw[thick] (-6ex,0ex) -- (6ex,0ex);
\draw[-,thick] (t1) -- (k11);
\draw[-,thick] (t1) -- (k1n1);
\draw[-,thick] (t2) -- (k21);
\draw[-,thick] (t2) -- (k2n2);
\draw[-,thick] (t1) -- (t2);
\node[anchor=south] at ($(dots1)$) {\scriptsize{$\cdots$}};
\node[anchor=south] at ($(dots2)$) {\scriptsize{$\cdots$}};
\node[anchor=south] at ($(t1)+(0,-2.0ex)$) {\scriptsize{$\tau_1$}};
\node[anchor=south] at ($(t2)+(0,-2.0ex)$) {\scriptsize{$\tau_2$}};
\filldraw[color=black, fill=white, thick] (t1) circle (0.4ex); 
\filldraw[color=black, fill=black, thick] (t2) circle (0.4ex); 
\filldraw[color=black, fill=white, thick] ($(k11)-(\arista,\arista)$) rectangle ($(k11)+(\arista,\arista)$);
\node[anchor=south] at ($(k11)+(0ex,0.5ex)$) {\scriptsize{$\k_{11}$}};
\filldraw[color=black, fill=white, thick] ($(k1n1)-(\arista,\arista)$) rectangle ($(k1n1)+(\arista,\arista)$);
\node[anchor=south] at ($(k1n1)+(0ex,0.5ex)$) {\scriptsize{$\k_{1 n_1}$}};
\filldraw[color=black, fill=white, thick] ($(k21)-(\arista,\arista)$) rectangle ($(k21)+(\arista,\arista)$);
\node[anchor=south] at ($(k21)+(0ex,0.5ex)$) {\scriptsize{$\k_{21}$}};
\filldraw[color=black, fill=white, thick] ($(k2n2)-(\arista,\arista)$) rectangle ($(k2n2)+(\arista,\arista)$);
\node[anchor=south] at ($(k2n2)+(0ex,0.5ex)$) {\scriptsize{$\k_{2 n_2}$}}
}
}
\def\diagramextwoverticesfour{\tikz[baseline=-1.4ex,scale=1.8, every node/.style={scale=1.4}]{
\coordinate (k11) at (-5ex,0ex);
\coordinate (k1n1) at (-1.4ex,0ex);
\coordinate (k21) at (+1.4ex,0ex);
\coordinate (k2n2) at (+5ex,0ex);
\coordinate (t1) at (-3.2ex,-4ex);
\coordinate (t2) at (+3.2ex,-4ex);
\coordinate (dots1) at (-3.2ex,-1.8ex);
\coordinate (dots2) at (+3.2ex,-1.8ex);
\pgfmathsetmacro{\arista}{0.06}
\draw[thick] (-6ex,0ex) -- (6ex,0ex);
\draw[-,thick] (t1) -- (k11);
\draw[-,thick] (t1) -- (k1n1);
\draw[-,thick] (t2) -- (k21);
\draw[-,thick] (t2) -- (k2n2);
\draw[-,thick] (t1) -- (t2);
\node[anchor=south] at ($(dots1)$) {\scriptsize{$\cdots$}};
\node[anchor=south] at ($(dots2)$) {\scriptsize{$\cdots$}};
\node[anchor=south] at ($(t1)+(0,-2.0ex)$) {\scriptsize{$\tau_1$}};
\node[anchor=south] at ($(t2)+(0,-2.0ex)$) {\scriptsize{$\tau_2$}};
\filldraw[color=black, fill=white, thick] (t1) circle (0.4ex); 
\filldraw[color=black, fill=white, thick] (t2) circle (0.4ex); 
\filldraw[color=black, fill=white, thick] ($(k11)-(\arista,\arista)$) rectangle ($(k11)+(\arista,\arista)$);
\node[anchor=south] at ($(k11)+(0ex,0.5ex)$) {\scriptsize{$\k_{11}$}};
\filldraw[color=black, fill=white, thick] ($(k1n1)-(\arista,\arista)$) rectangle ($(k1n1)+(\arista,\arista)$);
\node[anchor=south] at ($(k1n1)+(0ex,0.5ex)$) {\scriptsize{$\k_{1 n_1}$}};
\filldraw[color=black, fill=white, thick] ($(k21)-(\arista,\arista)$) rectangle ($(k21)+(\arista,\arista)$);
\node[anchor=south] at ($(k21)+(0ex,0.5ex)$) {\scriptsize{$\k_{21}$}};
\filldraw[color=black, fill=white, thick] ($(k2n2)-(\arista,\arista)$) rectangle ($(k2n2)+(\arista,\arista)$);
\node[anchor=south] at ($(k2n2)+(0ex,0.5ex)$) {\scriptsize{$\k_{2 n_2}$}}
}
}
\bea
\diagramextwovertices &=& \diagramextwoverticesone  +  \diagramextwoverticestwo \nn \\
&+&  \diagramextwoverticesthree  +  \diagramextwoverticesfour \nn 
\eea
As in the previous example, given that black and white vertices (together with all the propagators attached to them) are conjugate to each other, we can rewrite the previous expression as
\be
\diagramextwovertices = 2 {\rm Re} \Bigg[ \, \diagramextwoverticesone  +  \diagramextwoverticestwo \, \Bigg]  \nn
\ee
Then, according to the Feynman rules, the contribution to this particular channel becomes
\bea
\langle \varphi^n  (\k_{11} , \cdots , \k_{2n_2}) \rangle '  &=& 2 \frac{\lambda_{n_1+1} \lambda_{n_2+1}}{H^8}   \! \int^{\tau_f}_{-\infty}  \frac{\de \tau_1}{\tau_1^4} \! \int^{\tau_f}_{-\infty}  \frac{\de \tau_2}{\tau_2^4} {\rm Re} \bigg[ \nn \\
&& \hspace{-3cm} -  G_+ (k_{11} , \tau_1) \cdots G_+ (k_{1n_1} , \tau_1) G_{++} (q_{12} , \tau_1, \tau_2) G_+ (k_{21} , \tau_2) \cdots G_+ (k_{2n_2} , \tau_2) \nn \\
&& \hspace{-3cm} +  G_+ (k_{11} , \tau_1) \cdots G_+ (k_{1n_1} , \tau_1) G_{+-} (q_{12} , \tau_1, \tau_2) G_- (k_{21} , \tau_2) \cdots G_- (k_{2n_2} , \tau_2) \bigg]  ,  \qquad
\eea
where $q_{12} \equiv | \k_{11} + \cdots +  \k_{1 n_1} | =  | \k_{21} + \cdots  + \k_{2 n_2} |$ is the total momentum flowing through the internal propagator connecting $\tau_1$ and $\tau_2$. This expression can be further simplified by explicitly introducing the function $G$ but we postpone this step until Section~\ref{sec:second_order_n-point}, where we examine the long-wavelength limit of arbitrary diagrams.

\setcounter{equation}{0}
\section{Diagrammatics}
\label{sec:diagrammatics}

The purpose of this section is to prepare the stage for the infrared behaviour of correlators, into which we delve in the next section. In order to extract results in a clear way, it is convenient to distinguish between the real and imaginary parts of the propagator $G (k , \tau_1 , \tau_2)$, which we denote as $G_R (k , \tau_1 , \tau_2)$ and $G_I (k , \tau_1 , \tau_2)$, respectively:
\be
G (k , \tau_1 , \tau_2) = G_R (k , \tau_1 , \tau_2) + i G_I (k , \tau_1 , \tau_2) .
\ee
One can readily verify that 
\bea
G_R (k , \tau_1 , \tau_2) \!\! &=& \!\! \frac{H^2}{2 k^3} \bigg( k ( \tau_1 - \tau_2) \sin \Big[ k (\tau_1 - \tau_2) \Big]  + (1 + k^2 \tau_1 \tau_2) \cos \Big[ k (\tau_1 - \tau_2) \Big]  \bigg), \qquad \\ \label{GI}
G_I (k , \tau_1 , \tau_2) \!\! &=&  \!\! \frac{H^2}{2 k^3} \bigg( k ( \tau_1 - \tau_2) \cos \Big[ k (\tau_1 - \tau_2) \Big]  - (1 + k^2 \tau_1 \tau_2) \sin \Big[ k (\tau_1 - \tau_2) \Big]  \bigg). \qquad
\eea
Notice that $G_R (k , \tau_1 , \tau_2)$ is an even function under the interchange of $\tau_1$ and $\tau_2$ whereas $G_I (k , \tau_1 , \tau_2)$ is found to be odd. One can now split the various propagators of the theory into real and imaginary contributions. These take the form:
\bea
G_{++} (k, \tau_1 , \tau_2)  &=& G_R (k , \tau_1 , \tau_2) + i G_I (k , \tau_1 , \tau_2) I ( \tau_1 , \tau_2)  , \\
G_{--} (k, \tau_1 , \tau_2)  &=& G_R (k , \tau_1 , \tau_2) - i G_I (k , \tau_1 , \tau_2) I ( \tau_1 , \tau_2)  , \\
G_{+-} ( k, \tau_1 , \tau_2)  &=& G_R (k , \tau_1 , \tau_2) -  i G_I (k , \tau_1 , \tau_2)  , \\
G_{-+} (k, \tau_1 , \tau_2)  &=& G_R (k , \tau_1 , \tau_2) + i G_I (k , \tau_1 , \tau_2)  ,
\eea
where we have defined
\be
I ( \tau_1 , \tau_2)  \equiv  \theta (\tau_1 - \tau_2) - \theta (\tau_2 - \tau_1)  . \label{def-I}
\ee
Notice that $I (k , \tau_1 , \tau_2)$ is an odd function under the interchange of $\tau_1$ and $\tau_2$. With these definitions in mind, let us decompose the propagators as
\def\imbb{\tikz[baseline=-0.6ex,scale=1.8, every node/.style={scale=1.4}]{
\coordinate (tau1) at (-4ex,0ex);
\coordinate (tau2) at (4ex,0ex);
\draw[thick, dashed] (tau1) -- (tau2);
\filldraw[color=black, fill=black, thick] (tau1) circle (0.5ex);
\node[anchor=south] at ($(tau1)+(0,0.5ex)$) {\scriptsize{$\tau_1$}};
\filldraw[color=black, fill=black, thick] (tau2) circle (0.5ex);
\node[anchor=south] at ($(tau2)+(0,0.5ex)$) {\scriptsize{$\tau_2$}};
}
}
\def\rebb{\tikz[baseline=-0.6ex,scale=1.8, every node/.style={scale=1.4}]{
\coordinate (tau1) at (-4ex,0ex);
\coordinate (tau2) at (4ex,0ex);
\draw[thick, double] (tau1) -- (tau2);
\filldraw[color=black, fill=black, thick] (tau1) circle (0.5ex);
\node[anchor=south] at ($(tau1)+(0,0.5ex)$) {\scriptsize{$\tau_1$}};
\filldraw[color=black, fill=black, thick] (tau2) circle (0.5ex);
\node[anchor=south] at ($(tau2)+(0,0.5ex)$) {\scriptsize{$\tau_2$}};
}
}
\def\imww{\tikz[baseline=-0.6ex,scale=1.8, every node/.style={scale=1.4}]{
\coordinate (tau1) at (-4ex,0ex);
\coordinate (tau2) at (4ex,0ex);
\draw[thick, dashed] (tau1) -- (tau2);
\filldraw[color=black, fill=white, thick] (tau1) circle (0.5ex);
\node[anchor=south] at ($(tau1)+(0,0.5ex)$) {\scriptsize{$\tau_1$}};
\filldraw[color=black, fill=white, thick] (tau2) circle (0.5ex);
\node[anchor=south] at ($(tau2)+(0,0.5ex)$) {\scriptsize{$\tau_2$}};
}
}
\def\reww{\tikz[baseline=-0.6ex,scale=1.8, every node/.style={scale=1.4}]{
\coordinate (tau1) at (-4ex,0ex);
\coordinate (tau2) at (4ex,0ex);
\draw[thick, double] (tau1) -- (tau2);
\filldraw[color=black, fill=white, thick] (tau1) circle (0.5ex);
\node[anchor=south] at ($(tau1)+(0,0.5ex)$) {\scriptsize{$\tau_1$}};
\filldraw[color=black, fill=white, thick] (tau2) circle (0.5ex);
\node[anchor=south] at ($(tau2)+(0,0.5ex)$) {\scriptsize{$\tau_2$}};
}
}
\def\imbw{\tikz[baseline=-0.6ex,scale=1.8, every node/.style={scale=1.4}]{
\coordinate (tau1) at (-4ex,0ex);
\coordinate (tau2) at (4ex,0ex);
\draw[thick, dashed] (tau1) -- (tau2);
\filldraw[color=black, fill=black, thick] (tau1) circle (0.5ex);
\node[anchor=south] at ($(tau1)+(0,0.5ex)$) {\scriptsize{$\tau_1$}};
\filldraw[color=black, fill=white, thick] (tau2) circle (0.5ex);
\node[anchor=south] at ($(tau2)+(0,0.5ex)$) {\scriptsize{$\tau_2$}};
}
}
\def\rebw{\tikz[baseline=-0.6ex,scale=1.8, every node/.style={scale=1.4}]{
\coordinate (tau1) at (-4ex,0ex);
\coordinate (tau2) at (4ex,0ex);
\draw[thick, double] (tau1) -- (tau2);
\filldraw[color=black, fill=black, thick] (tau1) circle (0.5ex);
\node[anchor=south] at ($(tau1)+(0,0.5ex)$) {\scriptsize{$\tau_1$}};
\filldraw[color=black, fill=white, thick] (tau2) circle (0.5ex);
\node[anchor=south] at ($(tau2)+(0,0.5ex)$) {\scriptsize{$\tau_2$}};
}
}
\def\imwb{\tikz[baseline=-0.6ex,scale=1.8, every node/.style={scale=1.4}]{
\coordinate (tau1) at (-4ex,0ex);
\coordinate (tau2) at (4ex,0ex);
\draw[thick, dashed] (tau1) -- (tau2);
\filldraw[color=black, fill=white, thick] (tau1) circle (0.5ex);
\node[anchor=south] at ($(tau1)+(0,0.5ex)$) {\scriptsize{$\tau_1$}};
\filldraw[color=black, fill=black, thick] (tau2) circle (0.5ex);
\node[anchor=south] at ($(tau2)+(0,0.5ex)$) {\scriptsize{$\tau_2$}};
}
}
\def\rewb{\tikz[baseline=-0.6ex,scale=1.8, every node/.style={scale=1.4}]{
\coordinate (tau1) at (-4ex,0ex);
\coordinate (tau2) at (4ex,0ex);
\draw[thick, double] (tau1) -- (tau2);
\filldraw[color=black, fill=white, thick] (tau1) circle (0.5ex);
\node[anchor=south] at ($(tau1)+(0,0.5ex)$) {\scriptsize{$\tau_1$}};
\filldraw[color=black, fill=black, thick] (tau2) circle (0.5ex);
\node[anchor=south] at ($(tau2)+(0,0.5ex)$) {\scriptsize{$\tau_2$}};
}
}
\bea
\propbb &=& \rebb + \imbb , \nn \\
\propww &=& \reww + \imww , \nn \\
\propbw &=& \rebw + \imbw , \nn \\
\propwb &=& \rewb + \imwb , \nn
\eea
where double lines stand for the real part, whereas dashed lines denote their imaginary part. Of course, this splitting is also to be performed for bulk-to-boundary propagators. 

It will become important to notice that real propagators (represented by double lines) are all identical, independently of the type of vertex to which they are attached, whereas the imaginary propagators (dashed lines) change according to the type of vertices to which they are attached. 

\subsection{Prevalence of imaginary propagators} \label{sec:imaginary_props}

We can now draw an arbitrary diagram with every propagator, internal or external, decomposed into real and imaginary parts. It should be clear that a diagram with only real propagators (double lines) will necessarily vanish. Indeed, in this specific case, since the real propagators are vertex-independent, the only difference between the diagrams comes from the vertices; but given that for every black vertex there is a white one which contributes with the opposite sign, the sum of all diagrams will thus vanish. 

The previous reasoning can be made more precise without the need of restricting our attention only to diagrams where every propagator is real. Consider an arbitrary diagram with a given arbitrary vertex labeled by $\tau$:
\def\arbdiagram
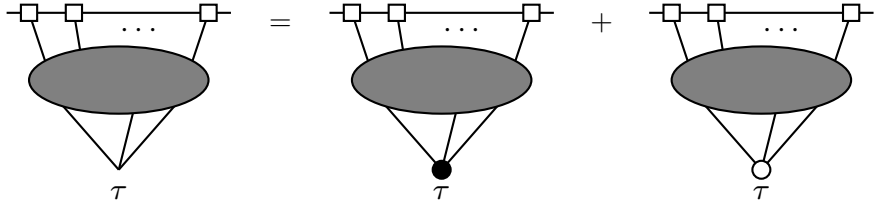
{\tikz[baseline=-1.4ex,scale=1.8, every node/.style={scale=1.4}]{
\coordinate (k1) at (-4ex,0ex);
\coordinate (k2) at (-2ex,0ex);
\coordinate (kn) at (4ex,0ex);
\coordinate (tau) at (0,-7ex);
\coordinate (dots1) at (1ex,-1.7ex);
\pgfmathsetmacro{\arista}{0.06}
\draw[thick] (-5ex,0ex) -- (5ex,0ex);
\draw[-,thick] (-3.0ex,-3ex) -- (k1);
\draw[-,thick] (-1.5ex,-3ex) -- (k2);
\draw[-,thick] (3.0ex,-3ex) -- (kn);
\draw[-,thick] (-3.5ex,-3ex) -- (tau);
\draw[-,thick] (1ex,-3ex) -- (tau);
\draw[-,thick] (3.5ex,-3ex) -- (tau);
\node[anchor=south] at ($(dots1)$) {\scriptsize{$\cdots$}};
\filldraw[color=black, fill=gray, thick] (0.0ex,-3.0ex) ellipse (4ex and 1.5ex); 
\node[anchor=north] at ($(tau)+(0.0ex,-0.1ex)$) {\scriptsize{$\tau$}};
\filldraw[color=black, fill=white, thick] ($(k1)-(\arista,\arista)$) rectangle ($(k1)+(\arista,\arista)$);
\filldraw[color=black, fill=white, thick] ($(k2)-(\arista,\arista)$) rectangle ($(k2)+(\arista,\arista)$);
\filldraw[color=black, fill=white, thick] ($(kn)-(\arista,\arista)$) rectangle ($(kn)+(\arista,\arista)$);
}
}
\def\arbdiagramb{\tikz[baseline=-1.4ex,scale=1.8, every node/.style={scale=1.4}]{
\coordinate (k1) at (-4ex,0ex);
\coordinate (k2) at (-2ex,0ex);
\coordinate (kn) at (4ex,0ex);
\coordinate (tau) at (0,-7ex);
\coordinate (dots1) at (1ex,-1.7ex);
\pgfmathsetmacro{\arista}{0.06}
\draw[thick] (-5ex,0ex) -- (5ex,0ex);
\draw[-,thick] (-3.0ex,-3ex) -- (k1);
\draw[-,thick] (-1.5ex,-3ex) -- (k2);
\draw[-,thick] (3.0ex,-3ex) -- (kn);
\draw[-,thick] (-3.5ex,-3ex) -- (tau);
\draw[-,thick] (1ex,-3ex) -- (tau);
\draw[-,thick] (3.5ex,-3ex) -- (tau);
\node[anchor=south] at ($(dots1)$) {\scriptsize{$\cdots$}};
\filldraw[color=black, fill=gray, thick] (0.0ex,-3.0ex) ellipse (4ex and 1.5ex); 
\node[anchor=north] at ($(tau)+(0.0ex,-0.1ex)$) {\scriptsize{$\tau$}};
\filldraw[color=black, fill=white, thick] ($(k1)-(\arista,\arista)$) rectangle ($(k1)+(\arista,\arista)$);
\filldraw[color=black, fill=white, thick] ($(k2)-(\arista,\arista)$) rectangle ($(k2)+(\arista,\arista)$);
\filldraw[color=black, fill=white, thick] ($(kn)-(\arista,\arista)$) rectangle ($(kn)+(\arista,\arista)$);
\filldraw[color=black, fill=black, thick] (tau) circle (0.4ex);
}
}
\def\arbdiagramw{\tikz[baseline=-1.4ex,scale=1.8, every node/.style={scale=1.4}]{
\coordinate (k1) at (-4ex,0ex);
\coordinate (k2) at (-2ex,0ex);
\coordinate (kn) at (4ex,0ex);
\coordinate (tau) at (0,-7ex);
\coordinate (dots1) at (1ex,-1.7ex);
\pgfmathsetmacro{\arista}{0.06}
\draw[thick] (-5ex,0ex) -- (5ex,0ex);
\draw[-,thick] (-3.0ex,-3ex) -- (k1);
\draw[-,thick] (-1.5ex,-3ex) -- (k2);
\draw[-,thick] (3.0ex,-3ex) -- (kn);
\draw[-,thick] (-3.5ex,-3ex) -- (tau);
\draw[-,thick] (1ex,-3ex) -- (tau);
\draw[-,thick] (3.5ex,-3ex) -- (tau);
\node[anchor=south] at ($(dots1)$) {\scriptsize{$\cdots$}};
\filldraw[color=black, fill=gray, thick] (0.0ex,-3.0ex) ellipse (4ex and 1.5ex); 
\node[anchor=north] at ($(tau)+(0.0ex,-0.1ex)$) {\scriptsize{$\tau$}};
\filldraw[color=black, fill=white, thick] ($(k1)-(\arista,\arista)$) rectangle ($(k1)+(\arista,\arista)$);
\filldraw[color=black, fill=white, thick] ($(k2)-(\arista,\arista)$) rectangle ($(k2)+(\arista,\arista)$);
\filldraw[color=black, fill=white, thick] ($(kn)-(\arista,\arista)$) rectangle ($(kn)+(\arista,\arista)$);
\filldraw[color=black, fill=white, thick] (tau) circle (0.4ex);
}
}
\be
\arbdiagram \quad =\quad \arbdiagramb \quad + \quad \arbdiagramw \quad \nn
\ee
If we now expand only the propagators attached to the $\tau$-vertex into real and imaginary parts, we see that the contributions that contain only the real parts will necessarily cancel each other, as the following diagrammatic relation shows:
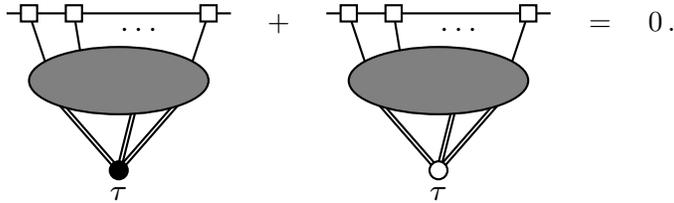
\def\arbdiagrambreal{\tikz[baseline=-1.4ex,scale=1.8, every node/.style={scale=1.4}]{
\coordinate (k1) at (-4ex,0ex);
\coordinate (k2) at (-2ex,0ex);
\coordinate (kn) at (4ex,0ex);
\coordinate (tau) at (0,-7ex);
\coordinate (dots1) at (1ex,-1.7ex);
\pgfmathsetmacro{\arista}{0.06}
\draw[thick] (-5ex,0ex) -- (5ex,0ex);
\draw[-,thick] (-3.0ex,-3ex) -- (k1);
\draw[-,thick] (-1.5ex,-3ex) -- (k2);
\draw[-,thick] (3.0ex,-3ex) -- (kn);
\draw[-,thick,double] (-3.5ex,-3ex) -- (tau);
\draw[-,thick,double] (1ex,-3ex) -- (tau);
\draw[-,thick,double] (3.5ex,-3ex) -- (tau);
\node[anchor=south] at ($(dots1)$) {\scriptsize{$\cdots$}};
\filldraw[color=black, fill=gray, thick] (0.0ex,-3.0ex) ellipse (4ex and 1.5ex); 
\node[anchor=north] at ($(tau)+(0.0ex,-0.1ex)$) {\scriptsize{$\tau$}};
\filldraw[color=black, fill=white, thick] ($(k1)-(\arista,\arista)$) rectangle ($(k1)+(\arista,\arista)$);
\filldraw[color=black, fill=white, thick] ($(k2)-(\arista,\arista)$) rectangle ($(k2)+(\arista,\arista)$);
\filldraw[color=black, fill=white, thick] ($(kn)-(\arista,\arista)$) rectangle ($(kn)+(\arista,\arista)$);
\filldraw[color=black, fill=black, thick] (tau) circle (0.4ex);
}
}
\def\arbdiagramwreal{\tikz[baseline=-1.4ex,scale=1.8, every node/.style={scale=1.4}]{
\coordinate (k1) at (-4ex,0ex);
\coordinate (k2) at (-2ex,0ex);
\coordinate (kn) at (4ex,0ex);
\coordinate (tau) at (0,-7ex);
\coordinate (dots1) at (1ex,-1.7ex);
\pgfmathsetmacro{\arista}{0.06}
\draw[thick] (-5ex,0ex) -- (5ex,0ex);
\draw[-,thick] (-3.0ex,-3ex) -- (k1);
\draw[-,thick] (-1.5ex,-3ex) -- (k2);
\draw[-,thick] (3.0ex,-3ex) -- (kn);
\draw[-,thick,double] (-3.5ex,-3ex) -- (tau);
\draw[-,thick,double] (1ex,-3ex) -- (tau);
\draw[-,thick,double] (3.5ex,-3ex) -- (tau);
\node[anchor=south] at ($(dots1)$) {\scriptsize{$\cdots$}};
\filldraw[color=black, fill=gray, thick] (0.0ex,-3.0ex) ellipse (4ex and 1.5ex); 
\node[anchor=north] at ($(tau)+(0.0ex,-0.1ex)$) {\scriptsize{$\tau$}};
\filldraw[color=black, fill=white, thick] ($(k1)-(\arista,\arista)$) rectangle ($(k1)+(\arista,\arista)$);
\filldraw[color=black, fill=white, thick] ($(k2)-(\arista,\arista)$) rectangle ($(k2)+(\arista,\arista)$);
\filldraw[color=black, fill=white, thick] ($(kn)-(\arista,\arista)$) rectangle ($(kn)+(\arista,\arista)$);
\filldraw[color=black, fill=white, thick] (tau) circle (0.4ex);
}
}
\be
\arbdiagrambreal \quad + \quad \arbdiagramwreal \quad = \quad  0 \,. \nn  
\ee
This is because the only difference between the two diagrams of the previous sum is given by the opposite sign of the two vertices. 
This will not be the case if at least one imaginary propagator remains attached to the vertex $\tau$. This result allows us to conclude that there must be at least one imaginary propagator attached to every vertex for a given diagram not to vanish. 

Taken on face value, this statement implies that for a diagram with $V$ vertices, the minimum number of imaginary (bulk-to-bulk) propagators appears to be $V-1$, which would ensure that every vertex has at least one imaginary propagator attached to it. However, a single diagram constructed with $V$ vertices and $V-1$ imaginary propagators is necessarily imaginary, from where it follows that the sum of subdiagrams constructed in this way must vanish. This finally implies that, for a diagram with $V$ vertices, the minimum number of imaginary propagators must be $V$.

\subsection{Imaginary bulk-to-boundary propagators}

As a final result, we now show that at least one of the imaginary propagators must always correspond to an external leg. If this was not the case, then for a diagram with $V$ vertices and $V$ imaginary propagators, there would necessarily be a closed loop formed entirely by imaginary propagators. Such a loop formed by only imaginary propagators must vanish. To see this, consider an arbitrary undotted diagram $D_n$ with a subdiagram consisting of a loop formed only by imaginary internal propagators (denoted by dashed lines), with $n$ vertices attached to it. The rest of the diagram is constructed with full propagators (solid lines). The undotted diagram can then be written as the sum of diagrams with a dotted imaginary loop (that is, with every other vertex undotted, except for those attached to the loop):
\def\loopy{\tikz[baseline=-1.4ex,scale=1.8, every node/.style={scale=1.4}]{
\coordinate (k1) at (-4ex,0ex);
\coordinate (k2) at (-2ex,0ex);
\coordinate (kn) at (4ex,0ex);
\coordinate (dots1) at (1ex,-1.7ex);
\coordinate (loop) at (0ex,-9ex);
\coordinate (t1) at (-3ex,0ex);
\coordinate (t2) at (-2.121ex,2.121ex);
\coordinate (t3) at (0ex,3ex);
\coordinate (t4) at (2.121ex,2.121ex);
\coordinate (t5) at (3ex,0ex);
\pgfmathsetmacro{\arista}{0.06}
\draw[thick] (-5ex,0ex) -- (5ex,0ex);
\draw[-,thick] (-3.0ex,-3ex) -- (k1);
\draw[-,thick] (-1.5ex,-3ex) -- (k2);
\draw[-,thick] (3.0ex,-3ex) -- (kn);
\draw[-,thick] (-4.5ex,-3ex) -- ($(loop) + (t1)$);
\draw[-,thick] (-3.75ex,-3ex) -- ($(loop) + (t1)$);
\draw[-,thick] (-3.0ex,-3ex) -- ($(loop) + (t2)$);
\draw[-,thick] (-2.0ex,-3ex) -- ($(loop) + (t2)$);
\draw[-,thick] (-1.0ex,-3ex) -- ($(loop) + (t2)$);
\draw[-,thick] (-0.5ex,-3ex) -- ($(loop) + (t3)$);
\draw[-,thick] (+0.5ex,-3ex) -- ($(loop) + (t3)$);
\draw[-,thick] (+2.5ex,-3ex) -- ($(loop) + (t4)$);
\draw[-,thick] (+3.5ex,-3ex) -- ($(loop) + (t4)$);
\draw[-,thick] (4.5ex,-3ex) -- ($(loop) + (t5)$);
\node[anchor=south] at ($(dots1)$) {\scriptsize{$\cdots$}};
\filldraw[color=black, fill=gray, thick] (0.0ex,-3.0ex) ellipse (4.7ex and 1.0ex); 
\filldraw[color=black, fill=white, thick] ($(k1)-(\arista,\arista)$) rectangle ($(k1)+(\arista,\arista)$);
\filldraw[color=black, fill=white, thick] ($(k2)-(\arista,\arista)$) rectangle ($(k2)+(\arista,\arista)$);
\filldraw[color=black, fill=white, thick] ($(kn)-(\arista,\arista)$) rectangle ($(kn)+(\arista,\arista)$);
\draw[thick, fill=white, dashed](loop) circle (3.0ex);
\node[anchor=west] at ($(loop) + (t1) + (-0.2ex,-0.7ex)$) {\scriptsize{$\tau_n$}};
\node[anchor=west] at ($(loop) + (t2) + (-0.7ex,-1.0ex)$) {\scriptsize{$\tau_1$}};
\node[anchor=north] at ($(loop) + (t3) + (0.0ex,0.0ex)$) {\scriptsize{$\tau_2$}};
\node[anchor=east] at ($(loop) + (t4) + (+0.7ex,-1.0ex)$) {\scriptsize{$\tau_3$}};
\node[anchor=east] at ($(loop) + (t5) + (+0.2ex,-0.7ex)$) {\scriptsize{$\tau_4$}};
\filldraw[color=black, fill=black, thick] (-0.7ex,-11.4ex) circle (0.035ex);
\filldraw[color=black, fill=black, thick] (0ex,-11.5ex) circle (0.035ex);
\filldraw[color=black, fill=black, thick] (0.7ex,-11.4ex) circle (0.035ex);
}
}
\def\loopone{\tikz[baseline=-1.4ex,scale=1.8, every node/.style={scale=1.4}]{
\coordinate (k1) at (-4ex,0ex);
\coordinate (k2) at (-2ex,0ex);
\coordinate (kn) at (4ex,0ex);
\coordinate (dots1) at (1ex,-1.7ex);
\coordinate (loop) at (0ex,-9ex);
\coordinate (t1) at (-3ex,0ex);
\coordinate (t2) at (-2.121ex,2.121ex);
\coordinate (t3) at (0ex,3ex);
\coordinate (t4) at (2.121ex,2.121ex);
\coordinate (t5) at (3ex,0ex);
\pgfmathsetmacro{\arista}{0.06}
\draw[thick] (-5ex,0ex) -- (5ex,0ex);
\draw[-,thick] (-3.0ex,-3ex) -- (k1);
\draw[-,thick] (-1.5ex,-3ex) -- (k2);
\draw[-,thick] (3.0ex,-3ex) -- (kn);
\draw[-,thick] (-4.5ex,-3ex) -- ($(loop) + (t1)$);
\draw[-,thick] (-3.75ex,-3ex) -- ($(loop) + (t1)$);
\draw[-,thick] (-3.0ex,-3ex) -- ($(loop) + (t2)$);
\draw[-,thick] (-2.0ex,-3ex) -- ($(loop) + (t2)$);
\draw[-,thick] (-1.0ex,-3ex) -- ($(loop) + (t2)$);
\draw[-,thick] (-0.5ex,-3ex) -- ($(loop) + (t3)$);
\draw[-,thick] (+0.5ex,-3ex) -- ($(loop) + (t3)$);
\draw[-,thick] (+2.5ex,-3ex) -- ($(loop) + (t4)$);
\draw[-,thick] (+3.5ex,-3ex) -- ($(loop) + (t4)$);
\draw[-,thick] (4.5ex,-3ex) -- ($(loop) + (t5)$);
\node[anchor=south] at ($(dots1)$) {\scriptsize{$\cdots$}};
\filldraw[color=black, fill=gray, thick] (0.0ex,-3.0ex) ellipse (4.7ex and 1.0ex); 
\filldraw[color=black, fill=white, thick] ($(k1)-(\arista,\arista)$) rectangle ($(k1)+(\arista,\arista)$);
\filldraw[color=black, fill=white, thick] ($(k2)-(\arista,\arista)$) rectangle ($(k2)+(\arista,\arista)$);
\filldraw[color=black, fill=white, thick] ($(kn)-(\arista,\arista)$) rectangle ($(kn)+(\arista,\arista)$);
\draw[thick, fill=white, dashed](loop) circle (3.0ex);
\filldraw[color=black, fill=black, thick] ($(loop) + (t1)$) circle (0.4ex);
\filldraw[color=black, fill=black, thick] ($(loop) + (t2)$) circle (0.4ex);
\filldraw[color=black, fill=black, thick] ($(loop) + (t3)$) circle (0.4ex);
\filldraw[color=black, fill=black, thick] ($(loop) + (t4)$) circle (0.4ex);
\filldraw[color=black, fill=black, thick] ($(loop) + (t5)$) circle (0.4ex);
\node[anchor=west] at ($(loop) + (t1) + (-0.2ex,-0.7ex)$) {\scriptsize{$\tau_n$}};
\node[anchor=west] at ($(loop) + (t2) + (-0.7ex,-1.0ex)$) {\scriptsize{$\tau_1$}};
\node[anchor=north] at ($(loop) + (t3) + (0.0ex,0.0ex)$) {\scriptsize{$\tau_2$}};
\node[anchor=east] at ($(loop) + (t4) + (+0.7ex,-1.0ex)$) {\scriptsize{$\tau_3$}};
\node[anchor=east] at ($(loop) + (t5) + (+0.2ex,-0.7ex)$) {\scriptsize{$\tau_4$}};
\filldraw[color=black, fill=black, thick] (-0.7ex,-11.4ex) circle (0.035ex);
\filldraw[color=black, fill=black, thick] (0ex,-11.5ex) circle (0.035ex);
\filldraw[color=black, fill=black, thick] (0.7ex,-11.4ex) circle (0.035ex);
}
}
\def\looptwo{\tikz[baseline=-1.4ex,scale=1.8, every node/.style={scale=1.4}]{
\coordinate (k1) at (-4ex,0ex);
\coordinate (k2) at (-2ex,0ex);
\coordinate (kn) at (4ex,0ex);
\coordinate (dots1) at (1ex,-1.7ex);
\coordinate (loop) at (0ex,-9ex);
\coordinate (t1) at (-3ex,0ex);
\coordinate (t2) at (-2.121ex,2.121ex);
\coordinate (t3) at (0ex,3ex);
\coordinate (t4) at (2.121ex,2.121ex);
\coordinate (t5) at (3ex,0ex);
\pgfmathsetmacro{\arista}{0.06}
\draw[thick] (-5ex,0ex) -- (5ex,0ex);
\draw[-,thick] (-3.0ex,-3ex) -- (k1);
\draw[-,thick] (-1.5ex,-3ex) -- (k2);
\draw[-,thick] (3.0ex,-3ex) -- (kn);
\draw[-,thick] (-4.5ex,-3ex) -- ($(loop) + (t1)$);
\draw[-,thick] (-3.75ex,-3ex) -- ($(loop) + (t1)$);
\draw[-,thick] (-3.0ex,-3ex) -- ($(loop) + (t2)$);
\draw[-,thick] (-2.0ex,-3ex) -- ($(loop) + (t2)$);
\draw[-,thick] (-1.0ex,-3ex) -- ($(loop) + (t2)$);
\draw[-,thick] (-0.5ex,-3ex) -- ($(loop) + (t3)$);
\draw[-,thick] (+0.5ex,-3ex) -- ($(loop) + (t3)$);
\draw[-,thick] (+2.5ex,-3ex) -- ($(loop) + (t4)$);
\draw[-,thick] (+3.5ex,-3ex) -- ($(loop) + (t4)$);
\draw[-,thick] (4.5ex,-3ex) -- ($(loop) + (t5)$);
\node[anchor=south] at ($(dots1)$) {\scriptsize{$\cdots$}};
\filldraw[color=black, fill=gray, thick] (0.0ex,-3.0ex) ellipse (4.7ex and 1.0ex); 
\filldraw[color=black, fill=white, thick] ($(k1)-(\arista,\arista)$) rectangle ($(k1)+(\arista,\arista)$);
\filldraw[color=black, fill=white, thick] ($(k2)-(\arista,\arista)$) rectangle ($(k2)+(\arista,\arista)$);
\filldraw[color=black, fill=white, thick] ($(kn)-(\arista,\arista)$) rectangle ($(kn)+(\arista,\arista)$);
\draw[thick, fill=white, dashed](loop) circle (3.0ex);
\filldraw[color=black, fill=black, thick] ($(loop) + (t1)$) circle (0.4ex);
\filldraw[color=black, fill=white, thick] ($(loop) + (t2)$) circle (0.4ex);
\filldraw[color=black, fill=black, thick] ($(loop) + (t3)$) circle (0.4ex);
\filldraw[color=black, fill=black, thick] ($(loop) + (t4)$) circle (0.4ex);
\filldraw[color=black, fill=black, thick] ($(loop) + (t5)$) circle (0.4ex);
\node[anchor=west] at ($(loop) + (t1) + (-0.2ex,-0.7ex)$) {\scriptsize{$\tau_n$}};
\node[anchor=west] at ($(loop) + (t2) + (-0.7ex,-1.0ex)$) {\scriptsize{$\tau_1$}};
\node[anchor=north] at ($(loop) + (t3) + (0.0ex,0.0ex)$) {\scriptsize{$\tau_2$}};
\node[anchor=east] at ($(loop) + (t4) + (+0.7ex,-1.0ex)$) {\scriptsize{$\tau_3$}};
\node[anchor=east] at ($(loop) + (t5) + (+0.2ex,-0.7ex)$) {\scriptsize{$\tau_4$}};
\filldraw[color=black, fill=black, thick] (-0.7ex,-11.4ex) circle (0.035ex);
\filldraw[color=black, fill=black, thick] (0ex,-11.5ex) circle (0.035ex);
\filldraw[color=black, fill=black, thick] (0.7ex,-11.4ex) circle (0.035ex);
}
}
\def\loopthree{\tikz[baseline=-1.4ex,scale=1.8, every node/.style={scale=1.4}]{
\coordinate (k1) at (-4ex,0ex);
\coordinate (k2) at (-2ex,0ex);
\coordinate (kn) at (4ex,0ex);
\coordinate (dots1) at (1ex,-1.7ex);
\coordinate (loop) at (0ex,-9ex);
\coordinate (t1) at (-3ex,0ex);
\coordinate (t2) at (-2.121ex,2.121ex);
\coordinate (t3) at (0ex,3ex);
\coordinate (t4) at (2.121ex,2.121ex);
\coordinate (t5) at (3ex,0ex);
\pgfmathsetmacro{\arista}{0.06}
\draw[thick] (-5ex,0ex) -- (5ex,0ex);
\draw[-,thick] (-3.0ex,-3ex) -- (k1);
\draw[-,thick] (-1.5ex,-3ex) -- (k2);
\draw[-,thick] (3.0ex,-3ex) -- (kn);
\draw[-,thick] (-4.5ex,-3ex) -- ($(loop) + (t1)$);
\draw[-,thick] (-3.75ex,-3ex) -- ($(loop) + (t1)$);
\draw[-,thick] (-3.0ex,-3ex) -- ($(loop) + (t2)$);
\draw[-,thick] (-2.0ex,-3ex) -- ($(loop) + (t2)$);
\draw[-,thick] (-1.0ex,-3ex) -- ($(loop) + (t2)$);
\draw[-,thick] (-0.5ex,-3ex) -- ($(loop) + (t3)$);
\draw[-,thick] (+0.5ex,-3ex) -- ($(loop) + (t3)$);
\draw[-,thick] (+2.5ex,-3ex) -- ($(loop) + (t4)$);
\draw[-,thick] (+3.5ex,-3ex) -- ($(loop) + (t4)$);
\draw[-,thick] (4.5ex,-3ex) -- ($(loop) + (t5)$);
\node[anchor=south] at ($(dots1)$) {\scriptsize{$\cdots$}};
\filldraw[color=black, fill=gray, thick] (0.0ex,-3.0ex) ellipse (4.7ex and 1.0ex); 
\filldraw[color=black, fill=white, thick] ($(k1)-(\arista,\arista)$) rectangle ($(k1)+(\arista,\arista)$);
\filldraw[color=black, fill=white, thick] ($(k2)-(\arista,\arista)$) rectangle ($(k2)+(\arista,\arista)$);
\filldraw[color=black, fill=white, thick] ($(kn)-(\arista,\arista)$) rectangle ($(kn)+(\arista,\arista)$);
\draw[thick, fill=white, dashed](loop) circle (3.0ex);
\filldraw[color=black, fill=black, thick] ($(loop) + (t1)$) circle (0.4ex);
\filldraw[color=black, fill=black, thick] ($(loop) + (t2)$) circle (0.4ex);
\filldraw[color=black, fill=white, thick] ($(loop) + (t3)$) circle (0.4ex);
\filldraw[color=black, fill=black, thick] ($(loop) + (t4)$) circle (0.4ex);
\filldraw[color=black, fill=black, thick] ($(loop) + (t5)$) circle (0.4ex);
\node[anchor=west] at ($(loop) + (t1) + (-0.2ex,-0.7ex)$) {\scriptsize{$\tau_n$}};
\node[anchor=west] at ($(loop) + (t2) + (-0.7ex,-1.0ex)$) {\scriptsize{$\tau_1$}};
\node[anchor=north] at ($(loop) + (t3) + (0.0ex,0.0ex)$) {\scriptsize{$\tau_2$}};
\node[anchor=east] at ($(loop) + (t4) + (+0.7ex,-1.0ex)$) {\scriptsize{$\tau_3$}};
\node[anchor=east] at ($(loop) + (t5) + (+0.2ex,-0.7ex)$) {\scriptsize{$\tau_4$}};
\filldraw[color=black, fill=black, thick] (-0.7ex,-11.4ex) circle (0.035ex);
\filldraw[color=black, fill=black, thick] (0ex,-11.5ex) circle (0.035ex);
\filldraw[color=black, fill=black, thick] (0.7ex,-11.4ex) circle (0.035ex);
}
}
\def\loopfour{\tikz[baseline=-1.4ex,scale=1.8, every node/.style={scale=1.4}]{
\coordinate (k1) at (-4ex,0ex);
\coordinate (k2) at (-2ex,0ex);
\coordinate (kn) at (4ex,0ex);
\coordinate (dots1) at (1ex,-1.7ex);
\coordinate (loop) at (0ex,-9ex);
\coordinate (t1) at (-3ex,0ex);
\coordinate (t2) at (-2.121ex,2.121ex);
\coordinate (t3) at (0ex,3ex);
\coordinate (t4) at (2.121ex,2.121ex);
\coordinate (t5) at (3ex,0ex);
\pgfmathsetmacro{\arista}{0.06}
\draw[thick] (-5ex,0ex) -- (5ex,0ex);
\draw[-,thick] (-3.0ex,-3ex) -- (k1);
\draw[-,thick] (-1.5ex,-3ex) -- (k2);
\draw[-,thick] (3.0ex,-3ex) -- (kn);
\draw[-,thick] (-4.5ex,-3ex) -- ($(loop) + (t1)$);
\draw[-,thick] (-3.75ex,-3ex) -- ($(loop) + (t1)$);
\draw[-,thick] (-3.0ex,-3ex) -- ($(loop) + (t2)$);
\draw[-,thick] (-2.0ex,-3ex) -- ($(loop) + (t2)$);
\draw[-,thick] (-1.0ex,-3ex) -- ($(loop) + (t2)$);
\draw[-,thick] (-0.5ex,-3ex) -- ($(loop) + (t3)$);
\draw[-,thick] (+0.5ex,-3ex) -- ($(loop) + (t3)$);
\draw[-,thick] (+2.5ex,-3ex) -- ($(loop) + (t4)$);
\draw[-,thick] (+3.5ex,-3ex) -- ($(loop) + (t4)$);
\draw[-,thick] (4.5ex,-3ex) -- ($(loop) + (t5)$);
\node[anchor=south] at ($(dots1)$) {\scriptsize{$\cdots$}};
\filldraw[color=black, fill=gray, thick] (0.0ex,-3.0ex) ellipse (4.7ex and 1.0ex); 
\filldraw[color=black, fill=white, thick] ($(k1)-(\arista,\arista)$) rectangle ($(k1)+(\arista,\arista)$);
\filldraw[color=black, fill=white, thick] ($(k2)-(\arista,\arista)$) rectangle ($(k2)+(\arista,\arista)$);
\filldraw[color=black, fill=white, thick] ($(kn)-(\arista,\arista)$) rectangle ($(kn)+(\arista,\arista)$);
\draw[thick, fill=white, dashed](loop) circle (3.0ex);
\filldraw[color=black, fill=black, thick] ($(loop) + (t1)$) circle (0.4ex);
\filldraw[color=black, fill=white, thick] ($(loop) + (t2)$) circle (0.4ex);
\filldraw[color=black, fill=white, thick] ($(loop) + (t3)$) circle (0.4ex);
\filldraw[color=black, fill=black, thick] ($(loop) + (t4)$) circle (0.4ex);
\filldraw[color=black, fill=black, thick] ($(loop) + (t5)$) circle (0.4ex);
\node[anchor=west] at ($(loop) + (t1) + (-0.2ex,-0.7ex)$) {\scriptsize{$\tau_n$}};
\node[anchor=west] at ($(loop) + (t2) + (-0.7ex,-1.0ex)$) {\scriptsize{$\tau_1$}};
\node[anchor=north] at ($(loop) + (t3) + (0.0ex,0.0ex)$) {\scriptsize{$\tau_2$}};
\node[anchor=east] at ($(loop) + (t4) + (+0.7ex,-1.0ex)$) {\scriptsize{$\tau_3$}};
\node[anchor=east] at ($(loop) + (t5) + (+0.2ex,-0.7ex)$) {\scriptsize{$\tau_4$}};
\filldraw[color=black, fill=black, thick] (-0.7ex,-11.4ex) circle (0.035ex);
\filldraw[color=black, fill=black, thick] (0ex,-11.5ex) circle (0.035ex);
\filldraw[color=black, fill=black, thick] (0.7ex,-11.4ex) circle (0.035ex);
}
}
\def\loopfive{\tikz[baseline=-1.4ex,scale=1.8, every node/.style={scale=1.4}]{
\coordinate (k1) at (-4ex,0ex);
\coordinate (k2) at (-2ex,0ex);
\coordinate (kn) at (4ex,0ex);
\coordinate (dots1) at (1ex,-1.7ex);
\coordinate (loop) at (0ex,-9ex);
\coordinate (t1) at (-3ex,0ex);
\coordinate (t2) at (-2.121ex,2.121ex);
\coordinate (t3) at (0ex,3ex);
\coordinate (t4) at (2.121ex,2.121ex);
\coordinate (t5) at (3ex,0ex);
\pgfmathsetmacro{\arista}{0.06}
\draw[thick] (-5ex,0ex) -- (5ex,0ex);
\draw[-,thick] (-3.0ex,-3ex) -- (k1);
\draw[-,thick] (-1.5ex,-3ex) -- (k2);
\draw[-,thick] (3.0ex,-3ex) -- (kn);
\draw[-,thick] (-4.5ex,-3ex) -- ($(loop) + (t1)$);
\draw[-,thick] (-3.75ex,-3ex) -- ($(loop) + (t1)$);
\draw[-,thick] (-3.0ex,-3ex) -- ($(loop) + (t2)$);
\draw[-,thick] (-2.0ex,-3ex) -- ($(loop) + (t2)$);
\draw[-,thick] (-1.0ex,-3ex) -- ($(loop) + (t2)$);
\draw[-,thick] (-0.5ex,-3ex) -- ($(loop) + (t3)$);
\draw[-,thick] (+0.5ex,-3ex) -- ($(loop) + (t3)$);
\draw[-,thick] (+2.5ex,-3ex) -- ($(loop) + (t4)$);
\draw[-,thick] (+3.5ex,-3ex) -- ($(loop) + (t4)$);
\draw[-,thick] (4.5ex,-3ex) -- ($(loop) + (t5)$);
\node[anchor=south] at ($(dots1)$) {\scriptsize{$\cdots$}};
\filldraw[color=black, fill=gray, thick] (0.0ex,-3.0ex) ellipse (4.7ex and 1.0ex); 
\filldraw[color=black, fill=white, thick] ($(k1)-(\arista,\arista)$) rectangle ($(k1)+(\arista,\arista)$);
\filldraw[color=black, fill=white, thick] ($(k2)-(\arista,\arista)$) rectangle ($(k2)+(\arista,\arista)$);
\filldraw[color=black, fill=white, thick] ($(kn)-(\arista,\arista)$) rectangle ($(kn)+(\arista,\arista)$);
\draw[thick, fill=white, dashed](loop) circle (3.0ex);
\filldraw[color=black, fill=black, thick] ($(loop) + (t1)$) circle (0.4ex);
\filldraw[color=black, fill=black, thick] ($(loop) + (t2)$) circle (0.4ex);
\filldraw[color=black, fill=white, thick] ($(loop) + (t3)$) circle (0.4ex);
\filldraw[color=black, fill=white, thick] ($(loop) + (t4)$) circle (0.4ex);
\filldraw[color=black, fill=black, thick] ($(loop) + (t5)$) circle (0.4ex);
\node[anchor=west] at ($(loop) + (t1) + (-0.2ex,-0.7ex)$) {\scriptsize{$\tau_n$}};
\node[anchor=west] at ($(loop) + (t2) + (-0.7ex,-1.0ex)$) {\scriptsize{$\tau_1$}};
\node[anchor=north] at ($(loop) + (t3) + (0.0ex,0.0ex)$) {\scriptsize{$\tau_2$}};
\node[anchor=east] at ($(loop) + (t4) + (+0.7ex,-1.0ex)$) {\scriptsize{$\tau_3$}};
\node[anchor=east] at ($(loop) + (t5) + (+0.2ex,-0.7ex)$) {\scriptsize{$\tau_4$}};
\filldraw[color=black, fill=black, thick] (-0.7ex,-11.4ex) circle (0.035ex);
\filldraw[color=black, fill=black, thick] (0ex,-11.5ex) circle (0.035ex);
\filldraw[color=black, fill=black, thick] (0.7ex,-11.4ex) circle (0.035ex);
}
}
\bea
\loopy \,\, &=& \,\, 2 {\rm Re} \Bigg[ \quad \loopone \quad \Bigg] \nn \\
&& + 2 {\rm Re} \Bigg[ \quad \looptwo \quad + \quad \loopthree \quad + \quad \cdots \quad \Bigg] \nn \\
&& + 2 {\rm Re} \Bigg[ \quad \loopfour \quad + \quad \loopfive \quad + \quad \cdots \quad \Bigg] \nn \\
&& + \cdots \nn
\eea
In translating the sum of these diagrams into Feynman rules, one will find that the imaginary loop will contribute to the entire sum as an overall factor equal to the sum of products of the function $I(\tau_i , \tau_j)$ defined in equation~(\ref{def-I}). 

For instance, in the particular case whereby the number $n$ of vertices attached to the loop is even, the first line in the previous diagrammatic relation would be proportional to the product $I(\tau_1, \tau_2) I(\tau_2, \tau_3) \cdots I(\tau_n, \tau_1)$. On the other hand, the second line, which contains diagrams with all vertices black except for one, is found to be proportional to the product $I(\tau_2, \tau_3) I(\tau_3, \tau_4) \cdots I(\tau_{n-1}, \tau_n)$ plus terms obtained by means of cyclic permutations. Similarly, the third line, which contains diagrams with two consecutive white vertices and every other vertex black, would be proportional to the product $I(\tau_1, \tau_2) I(\tau_3, \tau_4) I(\tau_4, \tau_5) \cdots I(\tau_{n-1}, \tau_n)$ plus terms obtained by means of cyclic permutations.

More generally, the number of functions $I(\tau_i , \tau_j)$ participating in these products will depend on the number of consecutive vertices of the same color. It is then straightforward to find that if $n$ is even, then the diagram under examination must be proportional to the following combination:
\bea
D_n &\propto& 1  + I (\tau_1 , \tau_2) I (\tau_2 , \tau_3)  + {\rm perms .} \nn \\
&& + I (\tau_1 , \tau_2) \times \cdots \times I (\tau_4 , \tau_5 ) + {\rm perms .} \nn \\
&&
+ \cdots  \nn \\
&& + I (\tau_1 , \tau_2) \times \cdots \times I ( \tau_n , \tau_1 ) .
\eea
On the other hand, if $n$ is odd, then one obtains
\bea
D_n &\propto& I (\tau_1 , \tau_2) + I (\tau_2 , \tau_3) + \cdots + I (\tau_n , \tau_1) \nn \\
&& + I (\tau_1 , \tau_2) I (\tau_2 , \tau_3) I (\tau_3 , \tau_4) + {\rm perms .} \nn \\
&& + I (\tau_1 , \tau_2) \times \cdots \times I (\tau_5 , \tau_6 ) + {\rm perms .} \nn \\
&&
+ \cdots  \nn \\
&& + I (\tau_1 , \tau_2) \times \cdots \times I (\tau_n , \tau_1 ).
\eea
One can readily verify that in both cases the total sum adds up to zero, rendering the above diagram null. To see this, it is enough to evaluate the sum for the particular ordered sequence $\tau_1 > \tau_2 > \cdots > \tau_n$, which is easily verified to vanish. Then, one can interchange the order of any pair of times (for example $\tau_1$ and $\tau_2$) and check that a change of sign in any one term is always compensated by a change of sing of another term (which originally had the opposite sign), leaving the conclusion unchanged. 

All in all, we arrive at the following general result: In splitting propagators into real and imaginary parts, the lowest number of imaginary propagators that a diagram of $V$ vertices can have is precisely $V$. Such a diagram cannot have loops formed by imaginary propagators and hence at least one external line must be an imaginary propagator. The following diagram offers an example of this situation:
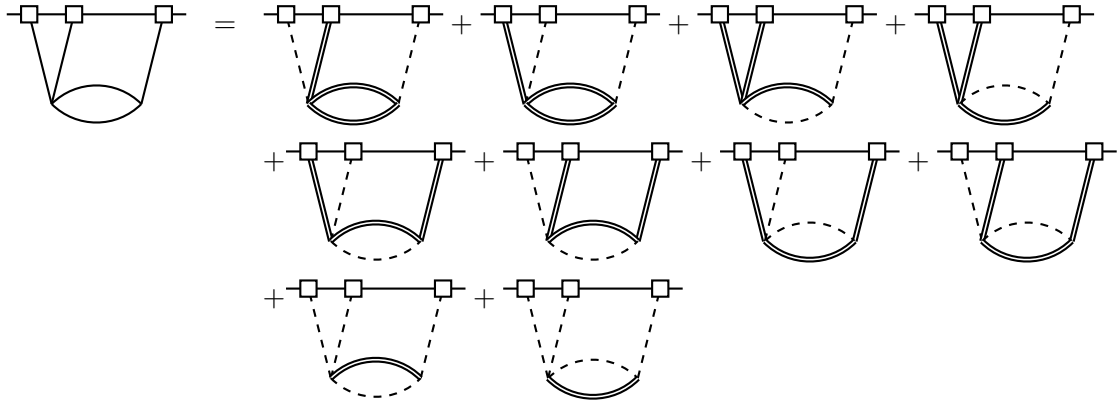
\def\diagramextwoverticesNoLabels{\tikz[baseline=-1.4ex,scale=1.8, every node/.style={scale=1.4}]{
\coordinate (k1) at (-3ex,0ex);
\coordinate (k2) at (-1.ex,0ex);
\coordinate (k3) at (+3.ex,0ex);
\coordinate (t1) at (-2ex,-4ex);
\coordinate (t2) at (+2ex,-4ex);
\pgfmathsetmacro{\arista}{0.06}
\draw[thick] (-4ex,0ex) -- (4ex,0ex);
\draw[-,thick] (t1) -- (k1);
\draw[-,thick] (t1) -- (k2);
\draw[-,thick] (t2) -- (k3);
\draw[thick,scale=2] (t1) to [bend left=45] (t2);
\draw[thick,scale=2] (t1) to [bend left=-45] (t2);
\filldraw[color=black, fill=white, thick] ($(k1)-(\arista,\arista)$) rectangle ($(k1)+(\arista,\arista)$);
\filldraw[color=black, fill=white, thick] ($(k2)-(\arista,\arista)$) rectangle ($(k2)+(\arista,\arista)$);
\filldraw[color=black, fill=white, thick] ($(k3)-(\arista,\arista)$) rectangle ($(k3)+(\arista,\arista)$);
}
}
\def\diagramextwoverticesReImSplit{\tikz[baseline=-1.4ex,scale=1.8, every node/.style={scale=1.4}]{
\coordinate (k1) at (-3ex,0ex);
\coordinate (k2) at (-1.ex,0ex);
\coordinate (k3) at (+3.ex,0ex);
\coordinate (t1) at (-2ex,-4ex);
\coordinate (t2) at (+2ex,-4ex);
\pgfmathsetmacro{\arista}{0.06}
\draw[thick] (-4ex,0ex) -- (4ex,0ex);
\draw[-,thick,dashed] (t1) -- (k1);
\draw[-,thick,double] (t1) -- (k2);
\draw[-,thick,dashed] (t2) -- (k3);
\draw[thick,scale=2,double] (t1) to [bend left=45] (t2);
\draw[thick,scale=2,double] (t1) to [bend left=-45] (t2);
\filldraw[color=black, fill=white, thick] ($(k1)-(\arista,\arista)$) rectangle ($(k1)+(\arista,\arista)$);
\filldraw[color=black, fill=white, thick] ($(k2)-(\arista,\arista)$) rectangle ($(k2)+(\arista,\arista)$);
\filldraw[color=black, fill=white, thick] ($(k3)-(\arista,\arista)$) rectangle ($(k3)+(\arista,\arista)$);
}
}
\def\diagramextwoverticesTwoImExt{\tikz[baseline=-1.4ex,scale=1.8, every node/.style={scale=1.4}]{
\coordinate (k1) at (-3ex,0ex);
\coordinate (k2) at (-1.ex,0ex);
\coordinate (k3) at (+3.ex,0ex);
\coordinate (t1) at (-2ex,-4ex);
\coordinate (t2) at (+2ex,-4ex);
\pgfmathsetmacro{\arista}{0.06}
\draw[thick] (-4ex,0ex) -- (4ex,0ex);
\draw[-,thick,double] (t1) -- (k1);
\draw[-,thick,dashed] (t1) -- (k2);
\draw[-,thick,dashed] (t2) -- (k3);
\draw[thick,scale=2,double] (t1) to [bend left=45] (t2);
\draw[thick,scale=2,double] (t1) to [bend left=-45] (t2);
\filldraw[color=black, fill=white, thick] ($(k1)-(\arista,\arista)$) rectangle ($(k1)+(\arista,\arista)$);
\filldraw[color=black, fill=white, thick] ($(k2)-(\arista,\arista)$) rectangle ($(k2)+(\arista,\arista)$);
\filldraw[color=black, fill=white, thick] ($(k3)-(\arista,\arista)$) rectangle ($(k3)+(\arista,\arista)$);
}
}
\def\diagramextwoverticesThreeIm{\tikz[baseline=-1.4ex,scale=1.8, every node/.style={scale=1.4}]{
\coordinate (k1) at (-3ex,0ex);
\coordinate (k2) at (-1.ex,0ex);
\coordinate (k3) at (+3.ex,0ex);
\coordinate (t1) at (-2ex,-4ex);
\coordinate (t2) at (+2ex,-4ex);
\pgfmathsetmacro{\arista}{0.06}
\draw[thick] (-4ex,0ex) -- (4ex,0ex);
\draw[-,thick,double] (t1) -- (k1);
\draw[-,thick,double] (t1) -- (k2);
\draw[-,thick,dashed] (t2) -- (k3);
\draw[thick,scale=2,double] (t1) to [bend left=45] (t2);
\draw[thick,scale=2,dashed] (t1) to [bend left=-45] (t2);
\filldraw[color=black, fill=white, thick] ($(k1)-(\arista,\arista)$) rectangle ($(k1)+(\arista,\arista)$);
\filldraw[color=black, fill=white, thick] ($(k2)-(\arista,\arista)$) rectangle ($(k2)+(\arista,\arista)$);
\filldraw[color=black, fill=white, thick] ($(k3)-(\arista,\arista)$) rectangle ($(k3)+(\arista,\arista)$);
}
}
\def\diagramextwoverticesFourIm{\tikz[baseline=-1.4ex,scale=1.8, every node/.style={scale=1.4}]{
\coordinate (k1) at (-3ex,0ex);
\coordinate (k2) at (-1.ex,0ex);
\coordinate (k3) at (+3.ex,0ex);
\coordinate (t1) at (-2ex,-4ex);
\coordinate (t2) at (+2ex,-4ex);
\pgfmathsetmacro{\arista}{0.06}
\draw[thick] (-4ex,0ex) -- (4ex,0ex);
\draw[-,thick,double] (t1) -- (k1);
\draw[-,thick,double] (t1) -- (k2);
\draw[-,thick,dashed] (t2) -- (k3);
\draw[thick,scale=2,dashed] (t1) to [bend left=45] (t2);
\draw[thick,scale=2,double] (t1) to [bend left=-45] (t2);
\filldraw[color=black, fill=white, thick] ($(k1)-(\arista,\arista)$) rectangle ($(k1)+(\arista,\arista)$);
\filldraw[color=black, fill=white, thick] ($(k2)-(\arista,\arista)$) rectangle ($(k2)+(\arista,\arista)$);
\filldraw[color=black, fill=white, thick] ($(k3)-(\arista,\arista)$) rectangle ($(k3)+(\arista,\arista)$);
}
}
\def\diagramextwoverticesFiveIm{\tikz[baseline=-1.4ex,scale=1.8, every node/.style={scale=1.4}]{
\coordinate (k1) at (-3ex,0ex);
\coordinate (k2) at (-1.ex,0ex);
\coordinate (k3) at (+3.ex,0ex);
\coordinate (t1) at (-2ex,-4ex);
\coordinate (t2) at (+2ex,-4ex);
\pgfmathsetmacro{\arista}{0.06}
\draw[thick] (-4ex,0ex) -- (4ex,0ex);
\draw[-,thick,double] (t1) -- (k1);
\draw[-,thick,dashed] (t1) -- (k2);
\draw[-,thick,double] (t2) -- (k3);
\draw[thick,scale=2,double] (t1) to [bend left=45] (t2);
\draw[thick,scale=2,dashed] (t1) to [bend left=-45] (t2);
\filldraw[color=black, fill=white, thick] ($(k1)-(\arista,\arista)$) rectangle ($(k1)+(\arista,\arista)$);
\filldraw[color=black, fill=white, thick] ($(k2)-(\arista,\arista)$) rectangle ($(k2)+(\arista,\arista)$);
\filldraw[color=black, fill=white, thick] ($(k3)-(\arista,\arista)$) rectangle ($(k3)+(\arista,\arista)$);
}
}
\def\diagramextwoverticesSixIm{\tikz[baseline=-1.4ex,scale=1.8, every node/.style={scale=1.4}]{
\coordinate (k1) at (-3ex,0ex);
\coordinate (k2) at (-1.ex,0ex);
\coordinate (k3) at (+3.ex,0ex);
\coordinate (t1) at (-2ex,-4ex);
\coordinate (t2) at (+2ex,-4ex);
\pgfmathsetmacro{\arista}{0.06}
\draw[thick] (-4ex,0ex) -- (4ex,0ex);
\draw[-,thick,dashed] (t1) -- (k1);
\draw[-,thick,double] (t1) -- (k2);
\draw[-,thick,double] (t2) -- (k3);
\draw[thick,scale=2,double] (t1) to [bend left=45] (t2);
\draw[thick,scale=2,dashed] (t1) to [bend left=-45] (t2);
\filldraw[color=black, fill=white, thick] ($(k1)-(\arista,\arista)$) rectangle ($(k1)+(\arista,\arista)$);
\filldraw[color=black, fill=white, thick] ($(k2)-(\arista,\arista)$) rectangle ($(k2)+(\arista,\arista)$);
\filldraw[color=black, fill=white, thick] ($(k3)-(\arista,\arista)$) rectangle ($(k3)+(\arista,\arista)$);
}
}
\def\diagramextwoverticesSevenIm{\tikz[baseline=-1.4ex,scale=1.8, every node/.style={scale=1.4}]{
\coordinate (k1) at (-3ex,0ex);
\coordinate (k2) at (-1.ex,0ex);
\coordinate (k3) at (+3.ex,0ex);
\coordinate (t1) at (-2ex,-4ex);
\coordinate (t2) at (+2ex,-4ex);
\pgfmathsetmacro{\arista}{0.06}
\draw[thick] (-4ex,0ex) -- (4ex,0ex);
\draw[-,thick,double] (t1) -- (k1);
\draw[-,thick,dashed] (t1) -- (k2);
\draw[-,thick,double] (t2) -- (k3);
\draw[thick,scale=2,dashed] (t1) to [bend left=45] (t2);
\draw[thick,scale=2,double] (t1) to [bend left=-45] (t2);
\filldraw[color=black, fill=white, thick] ($(k1)-(\arista,\arista)$) rectangle ($(k1)+(\arista,\arista)$);
\filldraw[color=black, fill=white, thick] ($(k2)-(\arista,\arista)$) rectangle ($(k2)+(\arista,\arista)$);
\filldraw[color=black, fill=white, thick] ($(k3)-(\arista,\arista)$) rectangle ($(k3)+(\arista,\arista)$);
}
}
\def\diagramextwoverticesEightIm{\tikz[baseline=-1.4ex,scale=1.8, every node/.style={scale=1.4}]{
\coordinate (k1) at (-3ex,0ex);
\coordinate (k2) at (-1.ex,0ex);
\coordinate (k3) at (+3.ex,0ex);
\coordinate (t1) at (-2ex,-4ex);
\coordinate (t2) at (+2ex,-4ex);
\pgfmathsetmacro{\arista}{0.06}
\draw[thick] (-4ex,0ex) -- (4ex,0ex);
\draw[-,thick,dashed] (t1) -- (k1);
\draw[-,thick,double] (t1) -- (k2);
\draw[-,thick,double] (t2) -- (k3);
\draw[thick,scale=2,dashed] (t1) to [bend left=45] (t2);
\draw[thick,scale=2,double] (t1) to [bend left=-45] (t2);
\filldraw[color=black, fill=white, thick] ($(k1)-(\arista,\arista)$) rectangle ($(k1)+(\arista,\arista)$);
\filldraw[color=black, fill=white, thick] ($(k2)-(\arista,\arista)$) rectangle ($(k2)+(\arista,\arista)$);
\filldraw[color=black, fill=white, thick] ($(k3)-(\arista,\arista)$) rectangle ($(k3)+(\arista,\arista)$);
}
}
\def\diagramextwoverticesNineIm{\tikz[baseline=-1.4ex,scale=1.8, every node/.style={scale=1.4}]{
\coordinate (k1) at (-3ex,0ex);
\coordinate (k2) at (-1.ex,0ex);
\coordinate (k3) at (+3.ex,0ex);
\coordinate (t1) at (-2ex,-4ex);
\coordinate (t2) at (+2ex,-4ex);
\pgfmathsetmacro{\arista}{0.06}
\draw[thick] (-4ex,0ex) -- (4ex,0ex);
\draw[-,thick,dashed] (t1) -- (k1);
\draw[-,thick,dashed] (t1) -- (k2);
\draw[-,thick,dashed] (t2) -- (k3);
\draw[thick,scale=2,double] (t1) to [bend left=45] (t2);
\draw[thick,scale=2,dashed] (t1) to [bend left=-45] (t2);
\filldraw[color=black, fill=white, thick] ($(k1)-(\arista,\arista)$) rectangle ($(k1)+(\arista,\arista)$);
\filldraw[color=black, fill=white, thick] ($(k2)-(\arista,\arista)$) rectangle ($(k2)+(\arista,\arista)$);
\filldraw[color=black, fill=white, thick] ($(k3)-(\arista,\arista)$) rectangle ($(k3)+(\arista,\arista)$);
}
}
\def\diagramextwoverticesTenIm{\tikz[baseline=-1.4ex,scale=1.8, every node/.style={scale=1.4}]{
\coordinate (k1) at (-3ex,0ex);
\coordinate (k2) at (-1.ex,0ex);
\coordinate (k3) at (+3.ex,0ex);
\coordinate (t1) at (-2ex,-4ex);
\coordinate (t2) at (+2ex,-4ex);
\pgfmathsetmacro{\arista}{0.06}
\draw[thick] (-4ex,0ex) -- (4ex,0ex);
\draw[-,thick,dashed] (t1) -- (k1);
\draw[-,thick,dashed] (t1) -- (k2);
\draw[-,thick,dashed] (t2) -- (k3);
\draw[thick,scale=2,dashed] (t1) to [bend left=45] (t2);
\draw[thick,scale=2,double] (t1) to [bend left=-45] (t2);
\filldraw[color=black, fill=white, thick] ($(k1)-(\arista,\arista)$) rectangle ($(k1)+(\arista,\arista)$);
\filldraw[color=black, fill=white, thick] ($(k2)-(\arista,\arista)$) rectangle ($(k2)+(\arista,\arista)$);
\filldraw[color=black, fill=white, thick] ($(k3)-(\arista,\arista)$) rectangle ($(k3)+(\arista,\arista)$);
}
}
\bea
\diagramextwoverticesNoLabels &=& \diagramextwoverticesReImSplit + \diagramextwoverticesTwoImExt  + \diagramextwoverticesThreeIm   + \diagramextwoverticesFourIm \nn \\ &&  + \diagramextwoverticesFiveIm + \diagramextwoverticesSixIm  + \diagramextwoverticesSevenIm  + \diagramextwoverticesEightIm  \nn \\ && + \diagramextwoverticesNineIm + \diagramextwoverticesTenIm 
\nn
\eea
Given that the full result must be real, diagrams with an odd number of imaginary propagators will vanish. In the next section, we shall see that the leading term in this sum in the infrared limit is provided only by those diagrams with two imaginary propagators.

\setcounter{equation}{0}
\section{Infrared behaviour of $n$-point functions}
\label{sec:long-wavelength-n-points}

Now that we have some understanding of how $G_R$ and $G_I$ appear in a general diagram, we are ready to examine the long-wavelength limit of general $n$-point functions. We are interested in evaluating $n$-point functions at scales $\k_i$ such that $k_i | \tau_0 | \ll 1$ for some time $\tau_0$. The introduction of the physical cutoff $\tau_0$ allows us to split time integrals as:
\be \label{tau0}
\int_{- \infty}^{\tau_{f}} \de \tau = \int_{- \infty}^{\tau_0} \de \tau + \int_{\tau_0}^{\tau_{f}} \de \tau .
\ee
Given that we are assuming $k_i | \tau_0 | \ll 1$, the first integral will necessarily act on oscillatory integrands so, together with the $\epsilon$-prescription, we expect it to yield convergent results. On the other hand, the second integral will act on nonoscillatory integrands and can lead to divergent expressions in the formal limit $\tau_f \to 0$ (an infrared divergence).

Anticipating that this is the case, we may explore the leading divergent contribution to an $n$-point function, coming from a diagram built out of $V$ vertices, by only considering integrals from $\tau_0$ to $\tau_f$:
\bea
\langle \varphi^n (\k_1 ,  \cdots , \k_{n} ) \rangle_{V} ' &=&   \int_{\tau_0}^{\tau_{f}}  \!\! \frac{\de \tau_1}{\tau_1^4} \cdots  \int_{\tau_0}^{\tau_{f}}  \!\! \frac{\de \tau_{n_V}}{\tau_{n_V}^4} \bigg[ \cdots \bigg] + \cdots . \label{divergent-n-point-V}
\eea
Now we must worry about the integrand of the previous expression, which consists of various combinations of real and imaginary propagators $G_R (k , \tau_1 , \tau_2)$ and $i G_I (k , \tau_1 , \tau_2)$. By expanding the real and imaginary contributions of $G (k , \tau_1 , \tau_2)$ in powers of $\tau_1$ and $\tau_2$ one obtains:
\bea
G_R (k , \tau_1 , \tau_2) &=& \frac{H^2}{2 k^3} \left[ 1 + \mathcal O \left( k^2\tau_i^2 \right) \right]  , 
\label{expansion-GR} \\
G_I (k , \tau_1 , \tau_2) &=& - \frac{H^2}{2 k^3} \left[ \frac{1}{3} k^3 \left(\tau_1^3 - \tau_2^3\right)+  \mathcal O \left(k^5\tau_i^5\right)  \right]  . 
\label{expansion-GI} 
\eea
Notice that in the long-wavelength limit, where $k |\tau_1| \ll 1$ and $k |\tau_2| \ll 1$, one finds $G_R (k , \tau_1 , \tau_2) \to \frac{H^2}{2 k^3}$ and $G_I (k , \tau_1 , \tau_2) \to 0$. This reveals that to obtain the leading divergent contribution to an $n$-point function we must keep those diagrams that contain the smallest number of imaginary propagators. As we have already seen, for a diagram built from $V$ vertices, there must be at least $V$ imaginary propagators  $i G_I (k , \tau_1 , \tau_2) \simeq - i \frac{H^2}{6} (\tau_1^3 - \tau_2^3)$. Then, simple power counting of the integrands of (\ref{divergent-n-point-V}) leads, after integration, to the general result:
\be \label{log-statement}
\langle \varphi^n (\k_1 ,  \cdots , \k_n ) \rangle_{V} \propto \lambda^{V} \Big[ \ln (\tau_0 / \tau_f) \Big]^{V} + \cdots .
\ee
A version of this statement has been derived in~\cite{Baumgart:2019clc} using the in-in formalism in the commutator language.

The behaviour shown in equation~(\ref{log-statement}) has been studied for simple theories such as ${\mathcal V(\varphi)} = \frac{\lambda}{4!} \varphi^4$. The presence of such infrared (IR) logarithmic divergences is well-known~\cite{Weinberg:2005vy,Weinberg:2006ac} (see~\cite{Hu:2018nxy} for a review), while its relevance to the stability of de Sitter space~\cite{Ford:1984hs,Antoniadis:1985pj,Polyakov:2012uc} is an open question. Instead of a stability issue, all these works can be interpreted as pointing out the need for resummation~\cite{Burgess:2009bs,Seery:2010kh,Kitamoto:2018dek,Baumgart:2019clc,Honda:2023unh} of IR divergences in order to safely (and meaningfully) asses the nonperturbative regime. 
In this paper, we wish to take a step towards the understanding of the perturbative structure of the PDF in order to get an insight into such a resummation. This should eventually coincide with a nonperturbative version of the stochastic formalism~\cite{Cohen:2020php,Cohen:2021fzf,Green:2022ovz,Cohen:2022clv,Choudhury_2019}, whose leading order dynamics is described by the Fokker-Planck equation found in~\cite{Starobinsky:1986fx}. We will have more to say on the subject in Section~\ref{sec:FP}.

With the result~\eqref{log-statement} at hand, it is now easy to anticipate the leading piece of any undotted diagram with $V$ vertices: It is enough to draw all the contributing diagrams with $V$ vertices according  to the rules found in the previous discussion. In what follows we check two examples.

\subsection{Example 1} \label{sec:div_example_1}
Let us consider again the example analyzed in Section~\ref{sec:basic_example_1}. In this case, the leading divergence to the diagram contributing to the $n$-point function is obtained by having ---in each contributing diagram--- all propagators real except from the one connecting the vertex labelled by $\tau$ to one of the external momenta at time $\tau_f$. This is shown in the following diagrammatic expression:
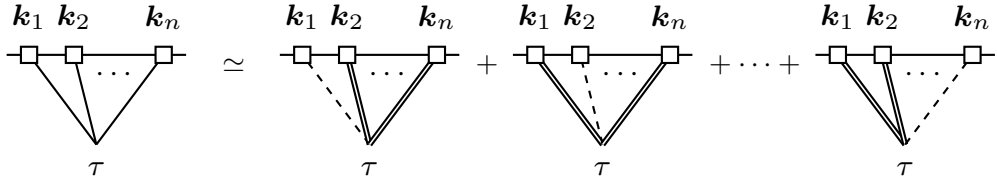
\def\diagramexundottedone{\tikz[baseline=-1.4ex,scale=1.8, every node/.style={scale=1.4}]{
\coordinate (k1) at (-3ex,0ex);
\coordinate (k2) at (-1ex,0ex);
\coordinate (kn) at (3ex,0ex);
\coordinate (tau) at (0,-4ex);
\coordinate (dots) at (0.9ex,-1.8ex);
\pgfmathsetmacro{\arista}{0.06}
\draw[thick] (-4ex,0ex) -- (4ex,0ex);
\draw[-,thick,dashed] (tau) -- (k1);
\draw[-,thick,double] (tau) -- (k2);
\draw[-,thick,double] (tau) -- (kn);
\node[anchor=south] at (dots) {\scriptsize{$\cdots$}};
\node[anchor=south] at ($(tau)+(0,-2.0ex)$) {\scriptsize{$\tau$}};
\filldraw[color=black, fill=white, thick] ($(k1)-(\arista,\arista)$) rectangle ($(k1)+(\arista,\arista)$);
\node[anchor=south] at ($(k1)+(0ex,0.5ex)$) {\scriptsize{$\k_1$}};
\filldraw[color=black, fill=white, thick] ($(k2)-(\arista,\arista)$) rectangle ($(k2)+(\arista,\arista)$);
\node[anchor=south] at ($(k2)+(0ex,0.5ex)$) {\scriptsize{$\k_2$}};
\filldraw[color=black, fill=white, thick] ($(kn)-(\arista,\arista)$) rectangle ($(kn)+(\arista,\arista)$);
\node[anchor=south] at ($(kn)+(0ex,0.5ex)$) {\scriptsize{$\k_{n}$}}
}
}
\def\diagramexundottedtwo{\tikz[baseline=-1.4ex,scale=1.8, every node/.style={scale=1.4}]{
\coordinate (k1) at (-3ex,0ex);
\coordinate (k2) at (-1ex,0ex);
\coordinate (kn) at (3ex,0ex);
\coordinate (tau) at (0,-4ex);
\coordinate (dots) at (0.9ex,-1.8ex);
\pgfmathsetmacro{\arista}{0.06}
\draw[thick] (-4ex,0ex) -- (4ex,0ex);
\draw[-,thick,double] (tau) -- (k1);
\draw[-,thick,dashed] (tau) -- (k2);
\draw[-,thick,double] (tau) -- (kn);
\node[anchor=south] at (dots) {\scriptsize{$\cdots$}};
\node[anchor=south] at ($(tau)+(0,-2.0ex)$) {\scriptsize{$\tau$}};
\filldraw[color=black, fill=white, thick] ($(k1)-(\arista,\arista)$) rectangle ($(k1)+(\arista,\arista)$);
\node[anchor=south] at ($(k1)+(0ex,0.5ex)$) {\scriptsize{$\k_1$}};
\filldraw[color=black, fill=white, thick] ($(k2)-(\arista,\arista)$) rectangle ($(k2)+(\arista,\arista)$);
\node[anchor=south] at ($(k2)+(0ex,0.5ex)$) {\scriptsize{$\k_2$}};
\filldraw[color=black, fill=white, thick] ($(kn)-(\arista,\arista)$) rectangle ($(kn)+(\arista,\arista)$);
\node[anchor=south] at ($(kn)+(0ex,0.5ex)$) {\scriptsize{$\k_{n}$}}
}
}
\def\diagramexundottedthree{\tikz[baseline=-1.4ex,scale=1.8, every node/.style={scale=1.4}]{
\coordinate (k1) at (-3ex,0ex);
\coordinate (k2) at (-1ex,0ex);
\coordinate (kn) at (3ex,0ex);
\coordinate (tau) at (0,-4ex);
\coordinate (dots) at (0.9ex,-1.8ex);
\pgfmathsetmacro{\arista}{0.06}
\draw[thick] (-4ex,0ex) -- (4ex,0ex);
\draw[-,thick,double] (tau) -- (k1);
\draw[-,thick,double] (tau) -- (k2);
\draw[-,thick,dashed] (tau) -- (kn);
\node[anchor=south] at (dots) {\scriptsize{$\cdots$}};
\node[anchor=south] at ($(tau)+(0,-2.0ex)$) {\scriptsize{$\tau$}};
\filldraw[color=black, fill=white, thick] ($(k1)-(\arista,\arista)$) rectangle ($(k1)+(\arista,\arista)$);
\node[anchor=south] at ($(k1)+(0ex,0.5ex)$) {\scriptsize{$\k_1$}};
\filldraw[color=black, fill=white, thick] ($(k2)-(\arista,\arista)$) rectangle ($(k2)+(\arista,\arista)$);
\node[anchor=south] at ($(k2)+(0ex,0.5ex)$) {\scriptsize{$\k_2$}};
\filldraw[color=black, fill=white, thick] ($(kn)-(\arista,\arista)$) rectangle ($(kn)+(\arista,\arista)$);
\node[anchor=south] at ($(kn)+(0ex,0.5ex)$) {\scriptsize{$\k_{n}$}}
}
}
\bea
\diagramexone &\simeq& \diagramexundottedone + \diagramexundottedtwo + \cdots +  \diagramexundottedthree \nn 
\eea
With this, it is now easy to obtain an analytical expression for the leading divergent contribution to the $n$-point function. Before performing the integration, one finds:
\be \label{n-pt-int}
\langle \varphi^n (\k_1 ,  \cdots , \k_n )\rangle ' \simeq  \int_{\tau_0}^{\tau_f}  \frac{\de\tau}{\tau^4}   
\frac{ \lambda_n H^{2 (n-2)}}{3 \times 2^{n-1} }  (\tau^3 - \tau_f^3) \frac{k_1^3 + k_2^3 + \cdots + k_n^3}{k_1^3 k_2^3 \cdots k_n^3} .
\ee
Then, upon integration and keeping the divergent term one is finally left with\footnote{The integral~\eqref{n-pt-int} can be computed exactly without the need of $\tau_0$. The result is~\eqref{n-pt-t0} with the logarithm replaced by $\ln\left(\sum k_i\tau\right)$. However, since we are interested in a finite range of momenta we will keep the restriction $k_i\tau_0\ll1$. The infrared logarithmic dependence is insensitive to this detail: switching to physical momenta, one can write $\ln\left(\sum p_i/H\right)=\ln\left(\sum p_i/\mu\right)+\ln(\mu/H)$, with $\mu$ some arbitrary physical scale. Then~\eqref{n-pt-int} is recovered by setting $\mu=1/(a\tau_0)$ leaving a time-independent remainder. }
\be \label{n-pt-t0}
\langle \varphi^n (\k_1 ,  \cdots , \k_n ) \rangle ' \simeq - \frac{ \lambda_n H^{2 (n-2)}}{3 \times 2^{n-1} }  \frac{k_1^3 + k_2^3 + \cdots + k_n^3}{k_1^3 k_2^3 \cdots k_n^3}  \ln (\tau_0 / \tau_f) .
\ee
Thanks to its dependence on the momenta, the leading contribution to the $n$-point function constitutes a manifestly local term. We will come back to this expression in a moment in order to derive an expression for the probability density distribution $\rho (\varphi)$ to first order in the potential ${\mathcal V(\varphi)}$.

\subsection{Example 2} \label{sec:second_order_n-point}

Let us consider the undotted diagram of the second example analyzed in Section~\ref{sec:basic_example_2}. Here we see that:
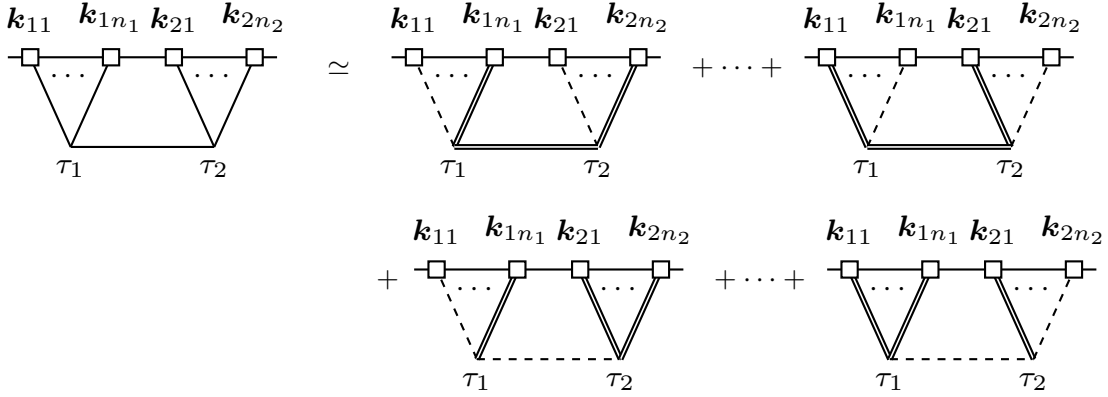
\def\diagramextwoverticesimone{\tikz[baseline=-1.4ex,scale=1.8, every node/.style={scale=1.4}]{
\coordinate (k11) at (-5ex,0ex);
\coordinate (k1n1) at (-1.4ex,0ex);
\coordinate (k21) at (+1.4ex,0ex);
\coordinate (k2n2) at (+5ex,0ex);
\coordinate (t1) at (-3.2ex,-4ex);
\coordinate (t2) at (+3.2ex,-4ex);
\coordinate (dots1) at (-3.2ex,-1.8ex);
\coordinate (dots2) at (+3.2ex,-1.8ex);
\pgfmathsetmacro{\arista}{0.06}
\draw[thick] (-6ex,0ex) -- (6ex,0ex);
\draw[-,thick,dashed] (t1) -- (k11);
\draw[-,thick,double] (t1) -- (k1n1);
\draw[-,thick,dashed] (t2) -- (k21);
\draw[-,thick,double] (t2) -- (k2n2);
\draw[-,thick,double] (t1) -- (t2);
\node[anchor=south] at ($(dots1)$) {\scriptsize{$\cdots$}};
\node[anchor=south] at ($(dots2)$) {\scriptsize{$\cdots$}};
\node[anchor=south] at ($(t1)+(0,-2.0ex)$) {\scriptsize{$\tau_1$}};
\node[anchor=south] at ($(t2)+(0,-2.0ex)$) {\scriptsize{$\tau_2$}};
\filldraw[color=black, fill=white, thick] ($(k11)-(\arista,\arista)$) rectangle ($(k11)+(\arista,\arista)$);
\node[anchor=south] at ($(k11)+(0ex,0.5ex)$) {\scriptsize{$\k_{11}$}};
\filldraw[color=black, fill=white, thick] ($(k1n1)-(\arista,\arista)$) rectangle ($(k1n1)+(\arista,\arista)$);
\node[anchor=south] at ($(k1n1)+(0ex,0.5ex)$) {\scriptsize{$\k_{1 n_1}$}};
\filldraw[color=black, fill=white, thick] ($(k21)-(\arista,\arista)$) rectangle ($(k21)+(\arista,\arista)$);
\node[anchor=south] at ($(k21)+(0ex,0.5ex)$) {\scriptsize{$\k_{21}$}};
\filldraw[color=black, fill=white, thick] ($(k2n2)-(\arista,\arista)$) rectangle ($(k2n2)+(\arista,\arista)$);
\node[anchor=south] at ($(k2n2)+(0ex,0.5ex)$) {\scriptsize{$\k_{2 n_2}$}}
}
}
\def\diagramextwoverticesimtwo{\tikz[baseline=-1.4ex,scale=1.8, every node/.style={scale=1.4}]{
\coordinate (k11) at (-5ex,0ex);
\coordinate (k1n1) at (-1.4ex,0ex);
\coordinate (k21) at (+1.4ex,0ex);
\coordinate (k2n2) at (+5ex,0ex);
\coordinate (t1) at (-3.2ex,-4ex);
\coordinate (t2) at (+3.2ex,-4ex);
\coordinate (dots1) at (-3.2ex,-1.8ex);
\coordinate (dots2) at (+3.2ex,-1.8ex);
\pgfmathsetmacro{\arista}{0.06}
\draw[thick] (-6ex,0ex) -- (6ex,0ex);
\draw[-,thick,double] (t1) -- (k11);
\draw[-,thick,dashed] (t1) -- (k1n1);
\draw[-,thick,double] (t2) -- (k21);
\draw[-,thick,dashed] (t2) -- (k2n2);
\draw[-,thick,double] (t1) -- (t2);
\node[anchor=south] at ($(dots1)$) {\scriptsize{$\cdots$}};
\node[anchor=south] at ($(dots2)$) {\scriptsize{$\cdots$}};
\node[anchor=south] at ($(t1)+(0,-2.0ex)$) {\scriptsize{$\tau_1$}};
\node[anchor=south] at ($(t2)+(0,-2.0ex)$) {\scriptsize{$\tau_2$}};
\filldraw[color=black, fill=white, thick] ($(k11)-(\arista,\arista)$) rectangle ($(k11)+(\arista,\arista)$);
\node[anchor=south] at ($(k11)+(0ex,0.5ex)$) {\scriptsize{$\k_{11}$}};
\filldraw[color=black, fill=white, thick] ($(k1n1)-(\arista,\arista)$) rectangle ($(k1n1)+(\arista,\arista)$);
\node[anchor=south] at ($(k1n1)+(0ex,0.5ex)$) {\scriptsize{$\k_{1 n_1}$}};
\filldraw[color=black, fill=white, thick] ($(k21)-(\arista,\arista)$) rectangle ($(k21)+(\arista,\arista)$);
\node[anchor=south] at ($(k21)+(0ex,0.5ex)$) {\scriptsize{$\k_{21}$}};
\filldraw[color=black, fill=white, thick] ($(k2n2)-(\arista,\arista)$) rectangle ($(k2n2)+(\arista,\arista)$);
\node[anchor=south] at ($(k2n2)+(0ex,0.5ex)$) {\scriptsize{$\k_{2 n_2}$}}
}
}
\def\diagramextwoverticesimthree{\tikz[baseline=-1.4ex,scale=1.8, every node/.style={scale=1.4}]{
\coordinate (k11) at (-5ex,0ex);
\coordinate (k1n1) at (-1.4ex,0ex);
\coordinate (k21) at (+1.4ex,0ex);
\coordinate (k2n2) at (+5ex,0ex);
\coordinate (t1) at (-3.2ex,-4ex);
\coordinate (t2) at (+3.2ex,-4ex);
\coordinate (dots1) at (-3.2ex,-1.8ex);
\coordinate (dots2) at (+3.2ex,-1.8ex);
\pgfmathsetmacro{\arista}{0.06}
\draw[thick] (-6ex,0ex) -- (6ex,0ex);
\draw[-,thick,dashed] (t1) -- (k11);
\draw[-,thick,double] (t1) -- (k1n1);
\draw[-,thick,double] (t2) -- (k21);
\draw[-,thick,double] (t2) -- (k2n2);
\draw[-,thick,dashed] (t1) -- (t2);
\node[anchor=south] at ($(dots1)$) {\scriptsize{$\cdots$}};
\node[anchor=south] at ($(dots2)$) {\scriptsize{$\cdots$}};
\node[anchor=south] at ($(t1)+(0,-2.0ex)$) {\scriptsize{$\tau_1$}};
\node[anchor=south] at ($(t2)+(0,-2.0ex)$) {\scriptsize{$\tau_2$}};
\filldraw[color=black, fill=white, thick] ($(k11)-(\arista,\arista)$) rectangle ($(k11)+(\arista,\arista)$);
\node[anchor=south] at ($(k11)+(0ex,0.5ex)$) {\scriptsize{$\k_{11}$}};
\filldraw[color=black, fill=white, thick] ($(k1n1)-(\arista,\arista)$) rectangle ($(k1n1)+(\arista,\arista)$);
\node[anchor=south] at ($(k1n1)+(0ex,0.5ex)$) {\scriptsize{$\k_{1 n_1}$}};
\filldraw[color=black, fill=white, thick] ($(k21)-(\arista,\arista)$) rectangle ($(k21)+(\arista,\arista)$);
\node[anchor=south] at ($(k21)+(0ex,0.5ex)$) {\scriptsize{$\k_{21}$}};
\filldraw[color=black, fill=white, thick] ($(k2n2)-(\arista,\arista)$) rectangle ($(k2n2)+(\arista,\arista)$);
\node[anchor=south] at ($(k2n2)+(0ex,0.5ex)$) {\scriptsize{$\k_{2 n_2}$}}
}
}
\def\diagramextwoverticesimfour{\tikz[baseline=-1.4ex,scale=1.8, every node/.style={scale=1.4}]{
\coordinate (k11) at (-5ex,0ex);
\coordinate (k1n1) at (-1.4ex,0ex);
\coordinate (k21) at (+1.4ex,0ex);
\coordinate (k2n2) at (+5ex,0ex);
\coordinate (t1) at (-3.2ex,-4ex);
\coordinate (t2) at (+3.2ex,-4ex);
\coordinate (dots1) at (-3.2ex,-1.8ex);
\coordinate (dots2) at (+3.2ex,-1.8ex);
\pgfmathsetmacro{\arista}{0.06}
\draw[thick] (-6ex,0ex) -- (6ex,0ex);
\draw[-,thick,double] (t1) -- (k11);
\draw[-,thick,double] (t1) -- (k1n1);
\draw[-,thick,double] (t2) -- (k21);
\draw[-,thick,dashed] (t2) -- (k2n2);
\draw[-,thick,dashed] (t1) -- (t2);
\node[anchor=south] at ($(dots1)$) {\scriptsize{$\cdots$}};
\node[anchor=south] at ($(dots2)$) {\scriptsize{$\cdots$}};
\node[anchor=south] at ($(t1)+(0,-2.0ex)$) {\scriptsize{$\tau_1$}};
\node[anchor=south] at ($(t2)+(0,-2.0ex)$) {\scriptsize{$\tau_2$}};
\filldraw[color=black, fill=white, thick] ($(k11)-(\arista,\arista)$) rectangle ($(k11)+(\arista,\arista)$);
\node[anchor=south] at ($(k11)+(0ex,0.5ex)$) {\scriptsize{$\k_{11}$}};
\filldraw[color=black, fill=white, thick] ($(k1n1)-(\arista,\arista)$) rectangle ($(k1n1)+(\arista,\arista)$);
\node[anchor=south] at ($(k1n1)+(0ex,0.5ex)$) {\scriptsize{$\k_{1 n_1}$}};
\filldraw[color=black, fill=white, thick] ($(k21)-(\arista,\arista)$) rectangle ($(k21)+(\arista,\arista)$);
\node[anchor=south] at ($(k21)+(0ex,0.5ex)$) {\scriptsize{$\k_{21}$}};
\filldraw[color=black, fill=white, thick] ($(k2n2)-(\arista,\arista)$) rectangle ($(k2n2)+(\arista,\arista)$);
\node[anchor=south] at ($(k2n2)+(0ex,0.5ex)$) {\scriptsize{$\k_{2 n_2}$}}
}
}
\bea
\diagramextwovertices &\simeq& \diagramextwoverticesimone + \cdots + \diagramextwoverticesimtwo \nn \\
&& + \diagramextwoverticesimthree + \cdots + \diagramextwoverticesimfour \nn
\eea
Notice that the expansion consists of two classes of diagrams: those that have two imaginary external legs (one from each vertex) and those that have one imaginary internal propagator and a single imaginary external leg. A given diagram with both vertices connected to the boundary via imaginary propagators, for instance carrying external momenta $\k_{1 i}$ and $\k_{2,j}$, will contribute a term of the form:
\def\diagramextwoverticesparticular{\tikz[baseline=-1.4ex,scale=1.8, every node/.style={scale=1.4}]{
\coordinate (k11) at (-7ex,0ex);
\coordinate (k1i) at (-4.25ex,0ex);
\coordinate (k1n1) at (-1.5ex,0ex);
\coordinate (k21) at (+1.5ex,0ex);
\coordinate (k2j) at (+4.25ex,0ex);
\coordinate (k2n2) at (+7ex,0ex);
\coordinate (t1) at (-4.25ex,-4ex);
\coordinate (t2) at (+4.25ex,-4ex);
\pgfmathsetmacro{\arista}{0.06}
\draw[thick] (-8ex,0ex) -- (8ex,0ex);
\draw[-,thick, double] (t1) -- (k11);
\draw[-,thick, dashed] (t1) -- (k1i);
\draw[-,thick, double] (t1) -- (k1n1);
\draw[-,thick, double] (t2) -- (k21);
\draw[-,thick, dashed] (t2) -- (k2j);
\draw[-,thick, double] (t2) -- (k2n2);
\draw[-,thick, double] (t1) -- (t2);
\node[anchor=south] at ($(t1)+(0,-2.5ex)$) {\scriptsize{$\tau_1$}};
\node[anchor=south] at ($(t2)+(0,-2.5ex)$) {\scriptsize{$\tau_2$}};
\filldraw[color=black, fill=white, thick] ($(k11)-(\arista,\arista)$) rectangle ($(k11)+(\arista,\arista)$);
\node[anchor=south] at ($(k11)+(0ex,0.5ex)$) {\scriptsize{$\k_{11}$}};
\filldraw[color=black, fill=white, thick] ($(k1i)-(\arista,\arista)$) rectangle ($(k1i)+(\arista,\arista)$);
\node[anchor=south] at ($(k1i)+(0ex,0.5ex)$) {\scriptsize{$\k_{1i}$}};
\filldraw[color=black, fill=white, thick] ($(k1n1)-(\arista,\arista)$) rectangle ($(k1n1)+(\arista,\arista)$);
\node[anchor=south] at ($(k1n1)+(0ex,0.5ex)$) {\scriptsize{$\k_{1 n_1}$}};
\filldraw[color=black, fill=white, thick] ($(k21)-(\arista,\arista)$) rectangle ($(k21)+(\arista,\arista)$);
\node[anchor=south] at ($(k21)+(0ex,0.5ex)$) {\scriptsize{$\k_{21}$}};
\filldraw[color=black, fill=white, thick] ($(k2j)-(\arista,\arista)$) rectangle ($(k2j)+(\arista,\arista)$);
\node[anchor=south] at ($(k2j)+(0ex,0.5ex)$) {\scriptsize{$\k_{2 j}$}};
\filldraw[color=black, fill=white, thick] ($(k2n2)-(\arista,\arista)$) rectangle ($(k2n2)+(\arista,\arista)$);
\node[anchor=south] at ($(k2n2)+(0ex,0.5ex)$) {\scriptsize{$\k_{2 n_2}$}};
}
}
\bea
\diagramextwoverticesparticular \!\!\!\!\!\!\! &=& \!\!\! \frac{H^2}{2 q_{12}^3} \!\! \left( 
\frac{\lambda_{n_1+1} H^{2 (n_1-2)}}{3 \times 2^{n_1-1} } \frac{k_{1 i}^3}{ k_{11}^3 \cdots k_{1 n_1}^3} \int_{\tau_0}^{\tau_f} \frac{\de\tau_1}{\tau_1^4} (\tau_1^3 - \tau_f^3) \right) \nn \\[-30pt]
 &&  \times \left(  \frac{\lambda_{n_2+1} H^{2 (n_2-2)}}{3 \times 2^{n_2-1} } \frac{ k_{2 j}^3}{k_{21}^3 \cdots k_{2 n_2}^3} \int_{\tau_0}^{\tau_f} \frac{\de\tau_2}{\tau_2^4} (\tau_2^3 - \tau_f^3) \right) \!\!, 
\eea
where $q_{12} = |\q_{12}|$, with $\q_{12} = \k_{11} + \cdots + \k_{1 n_1} $ the total momentum flowing through the internal line from $\tau_1$ to $\tau_2$. 

On the other hand, a diagram with only one imaginary propagator connecting one vertex to the boundary, say $\tau_1$, and another imaginary propagator connecting both vertices, gives a contribution of the form:
\def\diagramextwoverticesimvvone{\tikz[baseline=-1.4ex,scale=1.8, every node/.style={scale=1.4}]{
\coordinate (k11) at (-7ex,0ex);
\coordinate (k1i) at (-4.25ex,0ex);
\coordinate (k1n1) at (-1.5ex,0ex);
\coordinate (k21) at (+1.5ex,0ex);
\coordinate (k2n2) at (+4.25ex,0ex);
\coordinate (t1) at (-4.25ex,-4ex);
\coordinate (t2) at (+2.875ex,-4ex);
\pgfmathsetmacro{\arista}{0.06}
\draw[thick] (-8ex,0ex) -- (5.25ex,0ex);
\draw[-,thick, double] (t1) -- (k11);
\draw[-,thick, dashed] (t1) -- (k1i);
\draw[-,thick, double] (t1) -- (k1n1);
\draw[-,thick, double] (t2) -- (k21);
\draw[-,thick, double] (t2) -- (k2n2);
\draw[-,thick, dashed] (t1) -- (t2);
\node[anchor=south] at ($(t1)+(0,-2.5ex)$) {\scriptsize{$\tau_1$}};
\node[anchor=south] at ($(t2)+(0,-2.5ex)$) {\scriptsize{$\tau_2$}};
\filldraw[color=black, fill=white, thick] ($(k11)-(\arista,\arista)$) rectangle ($(k11)+(\arista,\arista)$);
\node[anchor=south] at ($(k11)+(0ex,0.5ex)$) {\scriptsize{$\k_{11}$}};
\filldraw[color=black, fill=white, thick] ($(k1i)-(\arista,\arista)$) rectangle ($(k1i)+(\arista,\arista)$);
\node[anchor=south] at ($(k1i)+(0ex,0.5ex)$) {\scriptsize{$\k_{1i}$}};
\filldraw[color=black, fill=white, thick] ($(k1n1)-(\arista,\arista)$) rectangle ($(k1n1)+(\arista,\arista)$);
\node[anchor=south] at ($(k1n1)+(0ex,0.5ex)$) {\scriptsize{$\k_{1 n_1}$}};
\filldraw[color=black, fill=white, thick] ($(k21)-(\arista,\arista)$) rectangle ($(k21)+(\arista,\arista)$);
\node[anchor=south] at ($(k21)+(0ex,0.5ex)$) {\scriptsize{$\k_{21}$}};
\filldraw[color=black, fill=white, thick] ($(k2n2)-(\arista,\arista)$) rectangle ($(k2n2)+(\arista,\arista)$);
\node[anchor=south] at ($(k2n2)+(0ex,0.5ex)$) {\scriptsize{$\k_{2 n_2}$}};
}
}
\bea
\!\!\!\!\!\!\!\!\diagramextwoverticesimvvone \!\!\!\!\!\!\!\!\!\!&=& \!\!\!\!\frac{H^2}{2} \Bigg(  \frac{\lambda_{n_1+1} H^{2 (n_1-2)}}{3 \times 2^{n_1-1} }    \frac{k_{1i}^3}{k_{11}^3 \cdots k_{1 n_1}^3 }     \int_{\tau_0}^{\tau_f}  \frac{\de\tau_1}{\tau_1^4}  (\tau_1^3 - \tau_f^3) \Bigg)
   \nn \\ [-30pt]
\!\!\!\!\!\!\!\!\!\!&& \!\!\!\!\times \Bigg( \frac{\lambda_{n_2+1} H^{2 (n_2-2)}}{3 \times 2^{n_2-1} }  \frac{1}{k_{21}^3 \cdots k_{2 n_2}^3}  \int_{\tau_0}^{\tau_f}  \frac{\de\tau_2}{\tau_2^4}  (\tau_2^3 - \tau_1^3)  \theta (\tau_1 - \tau_2)  \Bigg) . \label{log/2}
\eea
Then, by adding all the diagrams with two imaginary propagators (with at least one of them connected to the boundary) it is now straightforward to see that the analytical form of this channel's contribution to the $n$-point function (with $n = n_1 + n_2$) is
\bea
&& \!\!\!\!\!\!\!\!\! \langle \varphi^{n} ({\k_{11} , \cdots, \k_{2n_2}}) \rangle ' 
\simeq \nn \\
&& \quad \frac{H^2}{2 q_{12}^3} \left( 
\frac{\lambda_{n_1+1} H^{2 (n_1-2)}}{3 \times 2^{n_1-1} } \frac{k_{11}^3 + \cdots + k_{1 n_1}^3}{ k_{11}^3 \cdots k_{1 n_1}^3} \int_{\tau_0}^{\tau_f} \frac{\de\tau_1}{\tau_1^4} (\tau_1^3 - \tau_f^3) \right) \nn \\
&& \quad \times \left(  \frac{\lambda_{n_2+1} H^{2 (n_2-2)}}{3 \times 2^{n_2-1} } \frac{k_{21}^3 +  \cdots + k_{2n_2}^3}{k_{21}^3 \cdots k_{2 n_2}^3} \int_{\tau_0}^{\tau_f} \frac{\de\tau_2}{\tau_2^4} (\tau_2^3 - \tau_f^3) \right)
 \nn \\
&& \quad - \frac{\lambda_{n+1} \lambda_{m+1} H^{2 (n+m-3)}}{9 \times 2^{n+m-1} }   \int_{\tau_0}^{\tau_f}  \frac{\de\tau_1}{\tau_1^4}  \int_{\tau_0}^{\tau_f}  \frac{\de\tau_2}{\tau_2^4}  (\tau_1^3 - \tau_2^3) \\
&&
\quad  \Bigg[   \frac{k_{11}^3 + \cdots + k_{1 n_1}^3}{k_{11}^3 \cdots k_{1 n_1}^3 k_{21}^3 \cdots k_{2 n_2}^3}  (\tau_1^3 - \tau_f^3)  \theta (\tau_1 - \tau_2)   -  \frac{k_{21}^3 + \cdots + k_{2 n_2}^3}{k_{11}^3 \cdots k_{1 n_1}^3 k_{21}^3 \cdots k_{2 n_2}^3} (\tau_2^3 - \tau_f^3) \theta (\tau_2 - \tau_1)
  \Bigg].  \nn 
\label{n-point-V=2}  
\eea
Upon performing the integrals it is easy to show that this diagram diverges as $\ln(\tau_f/\tau_0)^2$ in the IR. We will come back to this result in Section~\ref{sec:general_structure_pdf}.

\setcounter{equation}{0}
\section{PDF to first order in ${\mathcal V(\varphi)}$} \label{sec:PDF_V=1}

In this section we derive an exact result for the PDF $\rho (\varphi , N)$ to first order in ${\mathcal V(\varphi)}$, which corresponds to the resummation of the series written in Eq.~(\ref{rho_1-sum}). We then simplify the expression using the leading behaviour of the $n$-point correlators derived in the previous section. This simplified version coincides with the expression derived in Refs.~\cite{Palma:2017lww, Chen:2018uul} for the particular case of an axionic potential ${\mathcal V(\varphi)} = \Lambda^4 (1 - \cos (\varphi/f_a))$ and in Ref.~\cite{Chen:2018brw} for an arbitrary potential ${\mathcal V(\varphi)}$. In all of these previous works the computation considered the role of loops, obscuring the form of the final result. Here, we shall first examine the computation at tree level, allowing us to obtain a simpler dependence of $\rho (\varphi)$ in terms of ${\mathcal V(\varphi)}$. We shall then move on to review the incorporation of loops and show that they have no observational consequences whatsoever (as expected from loops with no external momenta flowing through them).

\subsection{Zeroth order PDF}

Let us start by noticing that, in the absence of a potential ${\mathcal V(\varphi)}$, the statistics is completely determined by the free theory, and therefore it is exactly Gaussian. Consequently, the only nonvanishing connected moment is the variance $\langle \varphi^2 \rangle  = \sigma_{\rm tot}^2$, which is given by the $2$-point function as 
\bea
\sigma_{\rm tot}^2 &=& \int_{\k} G(k, \tau_f , \tau_f) = \frac{H^2}{4 \pi^2} \int \frac{\de k}{k} \left( 1 + k^2 \tau_f^2 \right) .
\eea
We can rewrite this result in terms of physical momentum $p = k / a(\tau_f) = - H k \tau_f$. This allows us to place physical cutoff scales $\Lambda_{\rm IR}$ and $\Lambda_{\rm UV}$:
\be
\sigma_{\rm tot}^2 = \frac{H^2}{4 \pi^2} \int_{\Lambda_{\rm IR}}^{\Lambda_{\rm UV}} \frac{\de p}{p} \left( 1 + \frac{p^2}{H^2} \right) . \label{variance-0-1}
\ee
This result diverges as $\Lambda_{\rm IR} \to 0$ and $\Lambda_{\rm UV} \to + \infty$ but such a behaviour should not be a concern. Recall that we are interested in long-wavelength modes satisfying the condition $ k_i |\tau_0|\ll 1$. Furthermore, in practice we don't have access to arbitrarily long wavelengths, so we are actually interested in a more restricted range of scales. We can select those scales by introducing a (dimensionless) window function $W(k/a)$ that filters all but a specific range of physical momenta (assumed to be well-within the region $k \ll 1/\tau_0$ at the final time $\tau_f$ of interest). For instance, we may choose $W(k / a)$ to be a unitary tophat function with nonvanishing values for momenta $k/a$ such that ${p_{\rm IR}<k/a<p_L}$, where $p_{\rm IR}$ and $p_{L}$ are physical cutoff scales satisfying $p_{\rm IR} \ll 1/ a \tau_0$ and $p_L \ll 1/a \tau_0$. The details on how we choose $W(k/a)$ will not be particularly relevant for the following discussion. 

With the introduction of this device, we can now define the long-wavelength part of $\varphi (\x , \tau)$ as
\be
\varphi_L (\x , \tau) = \int_{\k}  W(k/a) \varphi(\k, \tau) e^{- i \k \cdot \x} .
\ee
This means that the variance for $\varphi_L$ is not the one found in equation~(\ref{variance-0-1}) but a more restricted version of it given by
\be
\sigma_L^2 = \frac{H^2}{4 \pi^2} \int \frac{\de p}{p} W^2(p) \left(1 + \frac{p^2}{H^2} \right) . \label{variance-0-2}
\ee
Because $W(p)$ selects superhorizon scales $p \ll H$, we may simply write 
\be
\sigma_L^2 \simeq \frac{H^2}{4 \pi^2} \int \frac{\de p}{p} W^2(p) . \label{variance-0-3}
\ee
In this way, the probability density function describing the statistics of $\varphi_L$ is a Gaussian function with a variance given by (\ref{variance-0-3}):
\be \label{rho-0}
\rho_0 (\varphi) = \frac{e^{- \frac{\varphi^2}{2 \sigma_L^2}}}{\sqrt{2 \pi \sigma_L^2} }  .
\ee
It is important to emphasize here that the variance (\ref{variance-0-3}) is not necessarily constant. For instance, consider a window function $W(p)$ corresponding to a tophat function such that $W(p) = 1$ for $p_{\rm IR} < p < p_L$ and zero otherwise. Then, at any given time $N$, the infrared cutoff $p_{\rm IR}$ must coincide with the smallest possible momentum that crossed the horizon around $\tau_0$, which is given by $\sim H e^{-\Delta N}$ where $\Delta N = \ln (\tau_0 / \tau_f)$. Thus, if we fix the upper cutoff momentum $p_L$ to remain constant, then the variance will grow linearly with respect to $\Delta N$. In particular, in the stochastic approach one chooses $p_L$ to be of order $H$, in which case one obtains $\sigma_L^2 = \frac{H^2 }{4\pi^2} \Delta N$. The time dependence of the variance and its relation to diffusion of probability will further concern us in Sections~\ref{sec:variance-t-dep} and~\ref{sec:FP}.

\subsection{Resummation of the cumulant expansion to first order in ${\cal V}(\varphi)$}

Using Eq.~\eqref{npt'} and $(2 \pi)^3 \delta ({\k}) = \int \de^3 x \,e^{- i {\k} \cdot {\x}}$, we may integrate over every momentum after multiplying every leg with the window function. The result for the tree-level, connected $n$-point correlation functions is
\be  \label{fi-n}
 \Big\langle \varphi^n (\tau) \Big\rangle_c   = \frac{8 \pi}{H^4} \, {\rm Im} \bigg\{  \int^{\infty}_0 \de x 
\int^{\tau}_{-\infty}  \de \bar\tau  \, \frac{x^2}{\bar\tau^4} \lambda_n  \Big[ g(x,\bar\tau,\tau) \Big]^n \bigg\} ,
\ee
with
\be  \label{g-def}
g(x,\bar\tau,\tau) = \int_0^\infty \frac{\de k}{k} \,   W(k,\tau) \frac{\sin(kx)}{kx} G(k, \bar \tau , \tau).
\ee
Before proceeding let us refine Eq.~\eqref{fi-n}. Recall Eq.~\eqref{tau0} and the discussion below~\eqref{variance-0-1}. For $\bar\tau < \tau_{0}$, the integrand of $\int^{\tau}_{-\infty }\dd\bar\tau$ is dominated by oscillatory functions that make the integral converge. For $\tau_{0} < \bar\tau < \tau$ the integrand leads to a logarithmic contribution proportional to $\Delta N = \ln \tau_{0}/\tau$. Thus, for $\Delta N \gg 1$ we can write:
\be  \label{fi-n-2}
 \Big\langle \varphi^n (\tau) \Big\rangle_c   \simeq  \frac{8 \pi}{H^4} \, {\rm Im} \bigg\{  \int^{\infty}_0 \de x 
\int^{\tau}_{\tau_0}  \de \bar\tau  \, \frac{x^2}{\bar\tau^4} \lambda_n  \Big[ g(x,\bar\tau,\tau) \Big]^n \bigg\} ,
\ee
with corrections proportional to powers of $\tau / \tau_{0} - 1$.

All we now need to do is plug this expression in the series~(\ref{rho_1-sum}) and resum. To do so, it is convenient to introduce an operator whose eigenfunctions are the Hermite polynomials with integer eigenvalues:
\be \label{Herm-Op}
{\cal O}_\varphi \equiv  \sigma_L^2\frac{\partial^2}{\partial\varphi^2} - \varphi \frac{\partial}{\partial\varphi}\,, \qquad {\cal O}_\varphi \, {\rm He}_n(\varphi / \sigma_L) =  - n \,{\rm He}_n(\varphi / \sigma_L).
\ee 
This operator allows one to replace $\left( g/\sigma_L^2 \right)^{n} \to \left( g/\sigma_L^2 \right)^{- \mathcal O_{\varphi}}$ in Eq.~(\ref{rho_1-sum}). The remaining series may then be identified as the inverse Weierstrass transform of $\ {\mathcal V} (\varphi)$:
 \be \label{Weir-1}
\mathcal U (\varphi) \equiv \sum_{n=0}^{\infty} 
\frac{\lambda_n \sigma_L^n }{n!}   {\rm He}_{n} (\varphi / \sigma_L) = e^{- \frac{\sigma_L^2}{2} \frac{\partial^2}{\partial \varphi^2}} \mathcal V (\varphi) .
\ee
Upon writing the PDF as
\bea
\rho (\varphi) &=& 
\rho_0 (\varphi)  \bigg[ 1 +  \Delta (\varphi) \bigg] ,
\eea
with the zeroth-order part given in~\eqref{rho-0}, the non-Gaussian deformation\footnote{In the notation of Eq.~\eqref{PDF-schematic-pt}, $\Delta=\rho_1/\rho_0$.} reads
\be \label{Delta-1}
\Delta (\varphi,\tau) \simeq \frac{8 \pi}{H^4} {\rm Im} \int^{\infty}_0 \de x 
\int^{\tau}_{\tau_{0}} \de \tau' \,\frac{x^2}{(\tau')^4} \! \left(\frac{g(x, \tau',\tau)}{\sigma^2}  \right)^{ \!-{\cal O}_\varphi}  \mathcal U (\varphi)   .
\ee
\subsubsection{Resummation of loops} \label{loops-V=1}
Having derived the tree-level expression for the PDF at first order, we can now easily tackle the introduction of loops. In the case of diagrams with a single vertex, the only type of loops contributing to the computation of an $n$-point function are daisy loops without external momenta flowing through them, as shown in the following undotted diagram:
\def\diagramloopzero{\tikz[baseline=-1.4ex,scale=1.8, every node/.style={scale=1.4}]{
\coordinate (k1) at (-3ex,0ex);
\coordinate (k2) at (-1ex,0ex);
\coordinate (kn) at (3ex,0ex);
\coordinate (tau) at (0,-4ex);
\coordinate (dots) at (0.9ex,-1.8ex);
\pgfmathsetmacro{\arista}{0.06}
\draw[thick] (-4ex,0ex) -- (4ex,0ex);
\draw[-,thick] (tau) -- (k1);
\draw[-,thick] (tau) -- (k2);
\draw[-,thick] (tau) -- (kn);
\node[anchor=south] at (dots) {\scriptsize{$\cdots$}};
\filldraw[color=black, fill=white, thick] ($(k1)-(\arista,\arista)$) rectangle ($(k1)+(\arista,\arista)$);
\node[anchor=south] at ($(k1)+(0ex,0.5ex)$) {\scriptsize{$\k_1$}};
\filldraw[color=black, fill=white, thick] ($(k2)-(\arista,\arista)$) rectangle ($(k2)+(\arista,\arista)$);
\node[anchor=south] at ($(k2)+(0ex,0.5ex)$) {\scriptsize{$\k_2$}};
\filldraw[color=black, fill=white, thick] ($(kn)-(\arista,\arista)$) rectangle ($(kn)+(\arista,\arista)$);
\node[anchor=south] at ($(kn)+(0ex,0.5ex)$) {\scriptsize{$\k_{n}$}};
}
}
\def\diagramexloopone{\tikz[baseline=-1.4ex,scale=1.8, every node/.style={scale=1.4}]{
\coordinate (k1) at (-3ex,0ex);
\coordinate (k2) at (-1ex,0ex);
\coordinate (kn) at (3ex,0ex);
\coordinate (tau) at (0,-4ex);
\coordinate (dots) at (0.9ex,-1.8ex);
\pgfmathsetmacro{\arista}{0.06}
\draw[thick] (-4ex,0ex) -- (4ex,0ex);
\draw[-,thick] (tau) -- (k1);
\draw[-,thick] (tau) -- (k2);
\draw[-,thick] (tau) -- (kn);
\node[anchor=south] at (dots) {\scriptsize{$\cdots$}};
\filldraw[color=black, fill=white, thick] ($(k1)-(\arista,\arista)$) rectangle ($(k1)+(\arista,\arista)$);
\node[anchor=south] at ($(k1)+(0ex,0.5ex)$) {\scriptsize{$\k_1$}};
\filldraw[color=black, fill=white, thick] ($(k2)-(\arista,\arista)$) rectangle ($(k2)+(\arista,\arista)$);
\node[anchor=south] at ($(k2)+(0ex,0.5ex)$) {\scriptsize{$\k_2$}};
\filldraw[color=black, fill=white, thick] ($(kn)-(\arista,\arista)$) rectangle ($(kn)+(\arista,\arista)$);
\node[anchor=south] at ($(kn)+(0ex,0.5ex)$) {\scriptsize{$\k_{n}$}};
\draw[thick,scale=1.5] (tau)  to[in=-60,out=-120,loop] (tau);
}
}
\def\diagramexlooptwo{\tikz[baseline=-1.4ex,scale=1.8, every node/.style={scale=1.4}]{
\coordinate (k1) at (-3ex,0ex);
\coordinate (k2) at (-1ex,0ex);
\coordinate (kn) at (3ex,0ex);
\coordinate (tau) at (0,-4ex);
\coordinate (dots) at (0.9ex,-1.8ex);
\pgfmathsetmacro{\arista}{0.06}
\draw[thick] (-4ex,0ex) -- (4ex,0ex);
\draw[-,thick] (tau) -- (k1);
\draw[-,thick] (tau) -- (k2);
\draw[-,thick] (tau) -- (kn);
\node[anchor=south] at (dots) {\scriptsize{$\cdots$}};
\filldraw[color=black, fill=white, thick] ($(k1)-(\arista,\arista)$) rectangle ($(k1)+(\arista,\arista)$);
\node[anchor=south] at ($(k1)+(0ex,0.5ex)$) {\scriptsize{$\k_1$}};
\filldraw[color=black, fill=white, thick] ($(k2)-(\arista,\arista)$) rectangle ($(k2)+(\arista,\arista)$);
\node[anchor=south] at ($(k2)+(0ex,0.5ex)$) {\scriptsize{$\k_2$}};
\filldraw[color=black, fill=white, thick] ($(kn)-(\arista,\arista)$) rectangle ($(kn)+(\arista,\arista)$);
\node[anchor=south] at ($(kn)+(0ex,0.5ex)$) {\scriptsize{$\k_{n}$}};
\draw[thick,scale=1.5] (tau)  to[in=-30,out=-90,loop] (tau);
\draw[thick,scale=1.5] (tau)  to[in=-90,out=-150,loop] (tau);
}
}
\be
 \langle \varphi^n (\k_1 ,  \cdots , \k_n ) \rangle '  \quad =  \quad  \diagramloopzero + \diagramexloopone+ \diagramexlooptwo + \cdots . \nn
\ee
Independently of whether the vertex is black or white, a diagram with $L$ loops consists of the original diagram without loops but with the following replacement affecting $\lambda_n$:
\be
\lambda_n \to \frac{\lambda_{n + 2 L}}{L!} \bigg[ \frac{1}{2} \int_{\k}  G_{R} (k , \tau , \tau) \bigg]^L ,
\ee
where the factor $\frac{1}{2}$ comes from the symmetry of the loop insertion (recall that the imaginary propagator is odd under the interchange of its time arguments, implying that $G_{I} (k , \tau , \tau) = 0$). Similarly, the factor $\frac{1}{L!}$ is due to the symmetry of a diagram with $L$ loops under their exchange. The integral can be rewritten as
\bea
\int_{\k}  G_{R} (k , \tau , \tau) &=&  \frac{H^2}{4\pi^2} \int   \frac{\de k}{ k} (1 + k^2 \tau^2) ,  \\
&=&  \frac{H^2}{4\pi^2} \int_{\Lambda_{\rm IR}}^{\Lambda_{\rm UV}}    \frac{\de p}{ p} \left(1 + \frac{p^2}{H^2} \right)  ,
\eea
where in the second line we have made the change of variables from comoving to physical momentum $p = k/ a = - H k \tau$. The result is nothing but the full variance $\sigma_{\rm tot}^2$ encountered in equation~(\ref{variance-0-1}). We can then perform the sum of all diagrams containing loops to find that the result of the entire sum is equivalent to replacing the coupling $\lambda_n$ in the diagram without loops by a new coupling $\lambda_n^{\rm loops}$ given by~\cite{Chen:2018brw}
\be
\lambda_n \to \lambda_n^{\rm loops} =  \sum_{L=0}^{\infty}   \frac{ \lambda_{n+2L} }{L!} \left[ \frac{\sigma_{\rm tot}^2}{2} \right]^L  . \label{lambda-lambda-loops}
\ee
This simply corresponds to a forward Weirstrass transform,
\be \label{Weir-2}
\bar{ \mathcal U} (\varphi) \equiv \exp \bigg[ \frac{\sigma_{\rm tot}^2}{2}   \frac{\de^{2}}{\de \varphi^{2}} \bigg]  \mathcal V (\varphi) ,
\ee
which is a convolution with a Gaussian of variance $\sigma^2_{\rm tot}$.
Combining the two transforms~\eqref{Weir-1} and~\eqref{Weir-2}, we are led to the exact result (valid to linear order in the interaction potential and in the limit $\Delta N \gg1$)
\be \label{D-exact}
\Delta (\varphi,\tau) = \frac{8 \pi}{H^4} {\rm Im} \int^{\infty}_0 \de x 
\int^{\tau}_{\tau_{0}} \de \tau' \,\frac{x^2}{(\tau')^4} \! \left(\frac{g(x, \tau',\tau)}{\sigma^2}  \right)^{ \!-{\cal O}_\varphi}  \mathcal V_{\rm ren} (\varphi)   ,
\ee
where
\be \label{Vren}
\mathcal V_{\rm ren} (\varphi)  =  \exp \bigg[ \frac{\sigma_{\rm tot}^2-\sigma_L^2}{2}   \frac{\de^{2}}{\de \varphi^{2}} \bigg]  \mathcal V (\varphi).
\ee

Putting aside the restriction to long modes, it should be clear that we don't have direct access to the bare Lagrangian, thus the observable potential is just $\bar {\mathcal U}(\varphi) \equiv \exp \big[ \frac{\sigma_{\rm tot}^2}{2}  \frac{\de^{2}}{\de \varphi^{2}} \big]  {\mathcal V} (\varphi)$ (which becomes $\mathcal V_{\rm ren}$ once we restrict the statistics with the window function). Conversely, from the beginning, we could have written the bare potential as $\mathcal V(\varphi) \equiv \exp \big[ - \frac{\sigma_{\rm tot}^2}{2}  \frac{\de^{2}}{\de \varphi^{2}} \big]  \bar {\mathcal U} (\varphi)$, to later on find that the observable potential is given by $\bar{ \mathcal U} (\varphi)$. Either way, the effect of these loops have no physical consequence whatsoever, so we can simply disregard them.  In Section~\ref{higher-loops} we shall comment on the incorporation of loops into higher order diagrams (containing more than one vertex). There, we will arrive at similar conclusions (as long as we focus on the leading IR behavior of $n$-point correlation functions).

In order to simplify this expression for practical purposes, from~\eqref{Delta-1}, we may observe that the temporal arguments of the integrand satisfy $k\tau'<k\tau<1$; we may thus use the asymptotic expansion~\eqref{expansion-GR} and~\eqref{expansion-GI} of the real and imaginary parts of the propagator to deduce that in this region  $|g_I / g_R| \ll 1$, with $g$ defined in~\eqref{g-def}. This allows us to expand 
\be 
{\rm Im} \left \{ \left(\frac{g}{\sigma^2}\right)^{-{\cal O}_\varphi} \right\} \simeq  -{\cal O}_\varphi \,  \frac{g_I}{g_R}  \,  \left( \frac{g_R}{\sigma^2  }\right)^{-{\cal O}_\varphi} ,
\ee 
in~(\ref{Delta-1}), where
\bea
 g_R  &\simeq& \frac{H^2}{4 \pi^2} \int^{k_*(t)}_{k_*(t_{0})} \! \frac{\de k}{k} \frac{\sin (kx)}{kx}   , 
\label{def-g_I} \\
 g_I   &\simeq&  \frac{\tau^3 - (\tau')^3}{3} \,  \frac{H^2}{4 \pi^2} \int^{k_*(t)}_{k_*(t_{0})} \! \frac{\de k}{k} \frac{\sin (kx)}{kx}  k^3 . \qquad \label{def-g_R}
\eea
Next, combining Eqs.~\eqref{Weir-1} and~\eqref{Vren}, we can write ${\cal V}_{\rm ren}(\varphi) = \exp \bigg[ \frac{\sigma_{\rm tot}^2}{2}   \frac{\de^{2}}{\de \varphi^{2}} \bigg]  \mathcal U (\varphi)$ in~\eqref{D-exact} and pull out the exponential. Finally, let us note that the change to physical variables $r=xa$ and $p=k/a$, renders the integrals time independent, enabling us to perform the time integral in~\eqref{D-exact} as $\int^{\tau}_{\tau_{0}} \de \tau'  \, \frac{1}{(\tau')^4}  \,  (\tau^3 - (\tau')^3) = \Delta N + {\cal O}\left( \Delta N^{-2} \right)$. As long as we focus on large enough times $t\gg H^{-1}$, this integral is equal to $\Delta N = H (t - t_{0}) \gg 1$. Putting everything together, we obtain
\be
\Delta (\varphi,\tau) = \frac{4\pi^2}{3H^4} \exp \bigg[ \frac{\sigma_{\rm tot}^2}{2}   \frac{\de^{2}}{\de \varphi^{2}} \bigg] {\cal O}_\varphi \sum_{n=0}^\infty \frac{\lambda_n \sigma^n }{n!} {\rm He}_{n} (\varphi / \sigma) \int^{\infty}_0 \frac{\de r}{r}  \,  K(r,\Delta N) \left( \frac{g_R}{\sigma^2  }\right)^{n}  ,
\ee
with the kernel given by
\be \label{ker-def}
K(r,\Delta N)  = 8 \pi   \frac{\sigma^2}{H^2}  \frac{   \int  \frac{\de p}{p} \,W(p) \sin (pr)    (pr)^2  }{   \int  \frac{\de p}{p} \,W(p)  \frac{\sin (pr)}{pr}   }.
\ee
Upon defining the effective potential as
\be \label{Vobs-def}
{\cal V}_{\rm eff}(\varphi) = \int^{\infty}_0 \frac{\de r}{r}  \,  K(r,\Delta N) \, {\cal V}_{\rm ren}\left( \frac{\varphi g_R}{\sigma^2} \right),
\ee
with ${\cal V}_{\rm ren}$ given in~\eqref{Vren}, the probability density function to first order acquires the following simple form:
\be
\rho (\varphi , N) = 
 \frac{e^{- \frac{\varphi^2}{2 \sigma_L^2}}}{\sqrt{2 \pi \sigma_L^2}}  \Bigg[ 1 +  \frac{4\pi^2 }{3H^4}  {\cal O}_\varphi {\cal V}_{\rm eff}(\varphi) \bigg]  , \label{main-result-first-order}
\ee
with the differential operator defined in~\eqref{Herm-Op}.
This limit coincides with the result of~\cite{Chen:2018brw}. Notice that the particular way in which the effective potential enters the previous expression, that is, as $\sigma_L^2{\mathcal {V}''_{\rm eff}}(\varphi)  - \varphi{\mathcal {V}'_{\rm eff}}(\varphi)$, is exactly what is needed in order to preserve the unitarity condition $\int \de\varphi \rho (\varphi) = 1$ already ensured by the general equation~(\ref{rho-n-moments}).

\subsection{Effective potential}
\label{sec:effective-pot}

Let us briefly examine the form of the effective potential $\mathcal V_{\rm eff} (\varphi)$ defined in equation~(\ref{Vobs-def}). Upon adopting a window function of the form
\be
W(p) = \theta (p - p_{\rm IR})\times \theta( p_L - p) , \label{window-hat}
\ee
that is, $W(p) = 1$ if $p_{\rm IR} < p < p_L$ and $W(p) = 0$ otherwise. Changing variables as $p \to s = p/p_{L}$ and $r \to x = r \,  p_{L}$, the integration kernel~\eqref{ker-def} reads
\be \label{kerx-def}
K_\xi(x)  = 8 \pi\frac{  \sigma_L^2}{H^2}  \frac{   \int^{1}_{\xi^{-1}}  \frac{\de s}{s} \, \sin (sx)    (sx)^2  }{   \int^{1}_{\xi^{-1}}  \frac{\de s}{s} \,   \frac{\sin (sx)}{sx}   },
\ee
with the variable $\xi$ quantifying the range of scales supporting the statistics:
\be \label{xi}
\xi \equiv \frac{p_L}{p_{\rm IR}} .
\ee
It is easy to verify that $K_{\xi} (x)$ is finite for any $x$. Furthermore, notice that in terms of $\xi$, the variance of long modes reads
\be \label{sL}
\sigma_L^2 = \frac{H^2}{4 \pi^2} \ln \xi .
\ee

There is a formal limit that will turn out to be useful for our analysis of Section~\ref{sec:FP}, where we analyze our results within the context of the stochastic formalism. It is possible to show (c.f. Appendix C of Ref.~\cite{Chen:2018uul}) that in the limit $\xi \to \infty$ (for instance, reached by letting $p_{\rm IR} \to 0$), the function $g_R(x)/\sigma^2 \to 1$ for all $x$. Then, in this limit one sees that $\mathcal V (\varphi g_R(x)/\sigma^2) \to \mathcal V (\varphi) $ and the $x$-integral can be performed. Omitting for a moment that $\sigma_L$ also tends to $\infty$ in the same limit, the effective potential now takes the form
\be
 \mathcal V_{\rm eff} (\varphi) = \mathcal   V_{\rm ren} \big(  \varphi \big) .
\ee

\subsection{Regime of validity of our solution}
\label{sec:variance-t-dep}
To finish, let us come back to our result~\eqref{main-result-first-order}, valid to first order in the potential ${\mathcal V(\varphi)}$.
Apart from the requirement that the non-Gaussian correction $\frac{  4\pi^2}{3 H^4 } \mathcal O_\varphi \mathcal V_{\rm eff}$ must remain small (setting a limit to how large $\Delta N$ can be) our approach has a less obvious limitation with regard to how large the range of scales determining the variance $\sigma_L^2$ can be. Recall that the variance entering~\eqref{main-result-first-order} is set by the choice of the window function selecting the scales on which the statistics is defined. If we choose a tophat window function limited by physical cutoff momenta $p_{\rm IR} < p_{L}$, one finds the variance~\eqref{sL} with $\xi$ given by~\eqref{xi}.

Because $p_{\rm IR}$ corresponds to the physical momentum of the mode with the longest wavelength described by $\rho (\varphi, N)$, then it cannot be too small compared to $p_{L}$ (the momentum of the mode with the shortest wavelength). This is because the maximum number of $e$-folds that could have elapsed since the longest wavelength became superhorizon is precisely $\Delta N$. In other words, at a given time of interest $N$ (at which we wish to compute $\rho(\varphi , N)$) the smallest physical momentum $p_{\rm IR}$ must be such that
\be
p_{\rm IR} \sim H e^{- \Delta N} ,
\ee
and the range of momenta $\xi = p_{L} / p_{\rm IR}$ must be limited as 
\be
\ln \xi \leq \Delta N. \label{bound_DeltaN-xi}
\ee
For instance, in Ref.~\cite{Chen:2018brw} the result (\ref{main-result-first-order}) was used to assess the possibility of having non-Gaussianity beyond the bispectrum in the CMB anisotropy map. In that analysis the chosen values for these parameters were $\Delta N \sim 60$ (the estimated number of $e$-folds after horizon characterizing CMB modes) and $\ln \xi \sim 8$ (the range of scales spanning the CMB modes).

\setcounter{equation}{0}
\section{General structure of the PDF}
\label{sec:general_structure_pdf}

We are now ready to embark on the analysis of the general form of the PDF $\rho (\varphi , N)$. We will start this section by first discussing the tree-level case to second order, and then move on to discuss the tree-level case to arbitrary order. We will comment on the inclusion of loops in Section~\ref{higher-loops}.

\subsection{Tree-level case: second order} \label{sec:tree-level-second-order}

Before analysing how a general diagram of order $V$ contributes to the probability density function, let us study the specific case $V=2$. This example will highlight key features that apply to the more general case of an arbitrary diagram of order $V$. Recall that the leading contribution to an $n$-point function to second order in the potential ${\mathcal V(\varphi)}$ is given by equation~(\ref{n-point-V=2}). In obtaining that result we learnt that a given diagram $D_{\tau_i}^{(1)} $ with a vertex labelled by $\tau_i$ connected to the boundary through a single imaginary propagator must contain a factor of the form
\be
\sum D_{\tau_i}^{(1)} 
=
\cdots   \times \Bigg[ \lambda_{n_i + 1} \frac{H^{2 (n_i-2)}}{3 \times 2^{n_i-1}} \frac{k_{i 1}^3 + \cdots + k_{i n_i}^3}{k_{i 1}^3 \cdots  k_{i n_i}^3} \int^{\tau_f}_{\tau_0} \frac{\de\tau_i}{\tau_i^4} (\tau_i^3 - \tau_f^3) \Bigg] \times \cdots . 
\ee
Each term in the sum $k_{i 1}^3 + \cdots + k_{i n_i}^3$ comes from a diagram where the vertex $\tau_i$ connects to the boundary through an imaginary propagator carrying momentum $\k_{ij}$ with $j = 1, \cdots , n_i$. Hence, diagrams where only the first vertex $\tau_1$ is connected to the boundary with imaginary propagators will contribute a term proportional to
\be
\sum D^{(1)}_{\tau_1} \propto \frac{k_{1 1}^3 + \cdots + k_{1 n_1}^3}{k_{1 1}^3  \cdots  k_{1 n_1}^3}  . \label{D1tau1}
\ee
Similarly, diagrams where only the second vertex $\tau_2$ is connected to the boundary with imaginary propagators will add up to contribute a term proportional to
\be
\sum D^{(1)}_{\tau_2} \propto \frac{k_{2 1}^3 + \cdots + k_{2 n_2}^3}{k_{2 1}^3  \cdots  k_{2 n_2}^3} . 
\ee
Finally, diagrams $D^{(2)}_{\tau_1 \tau_2}$ where both vertices $\tau_1$ and $\tau_2$ are connected to the boundary via imaginary propagators will contribute a term proportional to
\be
\sum D^{(2)}_{\tau_1 \tau_2} \propto \frac{k_{1 1}^3 + \cdots + k_{1 n_1}^3}{k_{1 1}^3  \cdots  k_{1 n_1}^3} \frac{k_{2 1}^3 + \cdots + k_{2 n_2}^3}{k_{2 1}^3  \cdots  k_{2 n_2}^3} . \label{D1tau2}
\ee
To extract the $n$-moment out of these diagrams, we must first integrate each external momentum and then add every channel. After that, we must sum all diagrams with $n_1$ and $n_2$ respecting $n_1 + n_2 = n$. Given that all momenta appear on an equal footing, a consequence of the first step (integrating over all external momenta) will be the appearance of a factor $n_i$ in each diagram where the vertex $\tau_i$ is connected to the boundary through imaginary propagators. For instance, in the case of (\ref{D1tau1}) one gets:
\be
\int_{\k_{11}} \!\!\! \cdots \int_{\k_{2n_2}} \sum D^{(1)}_{\tau_1} \propto \int_{\k_{11}}  \!\!\! \cdots \int_{\k_{2n_2}} \frac{k_{1 1}^3 + \cdots + k_{1 n_1}^3}{k_{1 1}^3  \cdots  k_{1 n_1}^3} =  n_1 \int_{\k_{11}} \!\!\! \cdots \int_{\k_{2n_2}} \frac{k_{1 1}^3}{k_{1 1}^3  \cdots  k_{1 n_1}^3} .
\ee
Similarly, in the case of (\ref{D1tau2}) one obtains
\be
\int_{\k_{11}} \!\!\! \cdots \int_{\k_{2n_2}}  \sum D^{(2)}_{\tau_1 \tau_2} 
\propto n_1 n_2
\int_{\k_{11}} \!\!\! \cdots \int_{\k_{2n_2}}
\frac{k_{1 1}^3}{k_{1 1}^3  \cdots  k_{1 n_1}^3} \frac{k_{2 1}^3 }{k_{2 1}^3  \cdots  k_{2 n_2}^3} .
\ee
Notice that after integrating all momenta, diagrams representing different channels yield the same result. Thus, to perform the second step (adding every channel) we can simply take the result obtained by integrating the diagram of the first channel and multiply if by the total number of channels. As already discussed in Section~\ref{sec:basic_example_2}, in the case where $n_1 \neq n_2$ the total number of different channels is
\be
N_{\rm ch} = \frac{(n_1 + n_2)!}{n_1! n_2!} = \frac{n!}{n_1! n_2!}  . 
\ee
On the other hand, if $n_1 = n_2$, the number of channels becomes 
\be
N_{\rm ch} = \frac{1}{2!}\frac{(n_1 + n_2)!}{n_1! n_2!} = \frac{1}{2!}\frac{n!}{n_1! n_2!} .
\ee
The factor $1/2!$ comes from the symmetry of the diagram when both vertices have an equal number of external legs attached to the boundary. The third and last step allowing us to obtain the $n$-moment consists in adding up every group of diagrams for which $n_1 + n_2 = n$. To not over-count diagrams contributing to the desired $n$-moment, we may add every combination respecting $n_1 + n_2 = n$ by keeping the restriction $n_1 < n_2$ in the case where $n_1 \neq n_2$. Then, the total sum containing the cases $n_1 \neq n_2$ and $n_1 = n_2$ can be combined into an unrestricted sum as
\be
\sum_{n_1 < n_2}^{n_1 + n_2 = n} (\cdots ) +  \sum_{n_1 = n_2}^{n_1 + n_2 = n} \frac{1}{2!}  (\cdots ) = \frac{1}{2!} \sum_{n_1 n_2}^{n_1 + n_2 = n}  (\cdots ) ,
\ee 
where the sum of the right hand side goes over all partitions of $n$ into two integers. Notice that the sum with $n_1 = n_2$ is only possible when $n=n_1 + n_2$ is even. This last step gives us back the desired $n$-moment. The contribution to $\langle \varphi^{n} \rangle_{2}$ coming from diagrams with only one imaginary leg connected to the boundary is given by
\bea \label{fi^n_2^1}
\langle \varphi^{n} \rangle_{2}^{(1)} &=& \frac{2}{9 H^6}  \left[ \ln \frac{\tau_0 }{ \tau_f } \right]^2 \frac{n!}{2!} \sum_{n_1 n_2}^{n_1 + n_2 = n} \frac{1}{n_1! n_2!} 
\left( \frac{ H^{2}}{ 2 } \right)^{n_1} \left( \frac{ H^{2}}{ 2 } \right)^{n_2} \nn \\
&& 
\times \frac{1}{2} \bigg[ ( n_1 \lambda_{n_1 + 1} )  \lambda_{n_2 + 1} I^{(1)}_{n_1  n_2} + \lambda_{n_1 + 1} ( n_2 \lambda_{n_2 + 1} ) I^{(1)}_{n_2  n_1} \bigg]  .
\eea
Note that the factor $1/2$ in the second line of this expression is not a symmetry factor; it comes from the time-integration of the diagram~\eqref{log/2}.
The contribution to $\langle \varphi^{n} \rangle_{2}$ coming from those diagrams with two imaginary legs connected to the boundary is found to be given by
\bea \label{fi^n_2^2}
\langle \varphi^{n} \rangle_{2}^{(2)} &=&  \frac{2}{9 H^6} \Big[ \ln \frac{\tau_0 }{ \tau_f } \Big]^2  \frac{n!}{2!}    \sum_{n_1 n_2}^{n_1 + n_2 = n} \frac{1}{n_1! n_2!}  
\left( \frac{ H^{2}}{ 2 } \right)^{n_1} \left( \frac{ H^{2}}{ 2 } \right)^{n_2} \nn \\
&&
\times ( n_1 \lambda_{n_1 + 1} ) ( n_2 \lambda_{n_2 + 1} ) I^{(2)}_{n_1  n_2} .
\eea
In the previous two expressions, both $I^{(2)}_{n_1 n_2}$ and $I^{(1)}_{n_1 n_2}$ are integrals where $n_1$ and $n_2$ only enter as powers. The upper labels of $I^{(2)}_{n_1 n_2}$ and $I^{(1)}_{n_1 n_2}$ indicate the number of imaginary legs that connect vertices to the boundary in those diagrams contributing to their final form. To be concrete, they are given by
\bea
I^{(2)}_{n_1 n_2} &=& \frac{1}{n_1 n_2} \int_{\k_{11}} \cdots \int_{\k_{2 n_2}} \!\!\! (2 \pi)^3 \delta^{(3)} \left(\k_{11} + \cdots + \k_{2 n_2} \right) W(k_{11}) \cdots W(k_{2 n_2}) \nn \\
&& \times \frac{1}{q_{12}^3} \times  \frac{k_{11}^3 + \cdots + k_{1n_1}^3  }{ k_{11}^3 \cdots k_{1 n_1}^3}    \times \frac{k_{21}^3 + \cdots + k_{2n_2}^3 }{k_{21}^3 \cdots k_{2 n_2}^3} ,
\eea
and
\bea
 I^{(1)}_{n_1 n_2} &=& \frac{1}{n_1} \int_{\k_{11}} \cdots \int_{\k_{2 n_2}} (2 \pi)^3 \delta^{(3)} \left(\k_{11} + \cdots + \k_{2 n_2} \right) W(k_{11}) \cdots W(k_{2 n_2}) 
\nn \\
&&
\times  \frac{k_{11}^3 + \cdots + k_{1n_1}^3}{k_{11}^3 \cdots k_{1 n_1}^3 k_{21}^3 \cdots k_{2 n_2}^3} .
\eea
Recall that $q_{12} = |\k_{11} + \cdots  + \k_{1n_1}|= |\k_{21} + \cdots  + \k_{2n_2}|$ is the total momentum running through the internal line connecting $\tau_1$ to $\tau_2$. For the sake of argument, let us disregard the specific way in which integrals appear in their definitions and focus only on their dependence on $n_1$ and $n_2$ by writing
\be
I_{n_1 n_2} \sim A^2 \times B^{n_1} B^{n_2} ,
\ee
independently of their class (two or one vertex connected to the boundary with imaginary propagators). This simplification will not change our main conclusions and will highlight the dependence of $\rho(\varphi , N)$ on the field $\varphi$. This step allows us to recognize that the second order $n$-moment has the following general form:
\bea
\langle \varphi^{n} \rangle_{2} & = & \frac{n!}{2!}  \Delta N^2 A^2  \sum_{n_1 n_2}^{n_1 + n_2 = n} \frac{B^{n_1} B^{n_2}}{n_1! n_2!}  \nn \\
&&
\times \Bigg[ ( n_1 \lambda_{n_1 + 1} ) ( n_2 \lambda_{n_2 + 1} ) + ( n_1 \lambda_{n_1 + 1} )  \lambda_{n_2 + 1}  +  \lambda_{n_1 + 1}  ( n_2 \lambda_{n_2 + 1} ) \Bigg], \label{n-moment-V=2}
\eea
where we have identified 
\be
\Delta N = \ln \frac{\tau_0 }{ \tau_f }.
\ee
Let us emphasize that in writing equation~(\ref{n-moment-V=2}) we are only focusing on the dependence of $\langle \varphi^{n} \rangle_{2}$ on $n_1$ and $n_2$. Notice that in the square bracket of the second line we have every possible combination involving the couplings $\lambda_{n_1 + 1}$, $\lambda_{n_2 + 1}$, $n_1 \lambda_{n_1 + 1}$ and $n_2 \lambda_{n_2 + 1}$. The identification allowing us to understand the origin of each term in the second line is simple: If for a given diagram one of the propagators connecting a vertex $\tau_i$ to the boundary is imaginary, then such a diagram will contribute a term proportional to $n_i \lambda_{n_i + 1}$:
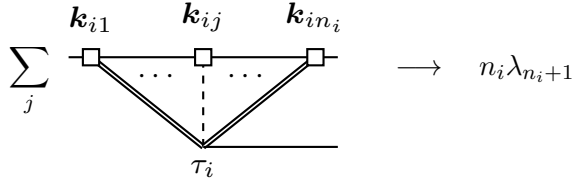
\def\diagramruleone{\tikz[baseline=-1.4ex,scale=1.8, every node/.style={scale=1.4}]{
\coordinate (k1) at (-5ex,0ex);
\coordinate (kj) at (0ex,0ex);
\coordinate (kn) at (5ex,0ex);
\coordinate (tau) at (0,-4ex);
\coordinate (dots1) at (-2.0ex,-1.8ex);
\coordinate (dots2) at (2.0ex,-1.8ex);
\coordinate (tau2) at (6ex,-4ex);
\pgfmathsetmacro{\arista}{0.06}
\draw[thick] (-6ex,0ex) -- (6ex,0ex);
\draw[-,thick,double] (tau) -- (k1);
\draw[-,thick,dashed] (tau) -- (kj);
\draw[-,thick,double] (tau) -- (kn);
\node[anchor=south] at (dots1) {\scriptsize{$\cdots$}};
\node[anchor=south] at (dots2) {\scriptsize{$\cdots$}};
\draw[-,thick] (tau) -- (tau2);
\node[anchor=south] at ($(tau)+(0,-2.0ex)$) {\scriptsize{$\tau_i$}};
\filldraw[color=black, fill=white, thick] ($(k1)-(\arista,\arista)$) rectangle ($(k1)+(\arista,\arista)$);
\node[anchor=south] at ($(k1)+(0ex,0.5ex)$) {\scriptsize{$\k_{i1}$}};
\filldraw[color=black, fill=white, thick] ($(kj)-(\arista,\arista)$) rectangle ($(kj)+(\arista,\arista)$);
\node[anchor=south] at ($(kj)+(0ex,0.5ex)$) {\scriptsize{$\k_{ij}$}};
\filldraw[color=black, fill=white, thick] ($(kn)-(\arista,\arista)$) rectangle ($(kn)+(\arista,\arista)$);
\node[anchor=south] at ($(kn)+(0ex,0.5ex)$) {\scriptsize{$\k_{in_i}$}}
}
}
\be
\sum_j \diagramruleone \quad \longrightarrow  \quad  n_i \lambda_{n_i + 1} \nn
\ee
In the previous diagram, $\tau_i$ is connected to the boundary with $n_i$ legs but one of them is imaginary (the one carrying momentum $k_{ij}$). In addition, there is one extra leg (real or imaginary) connecting $\tau_i$ with another vertex ($\tau_1$ or $\tau_2$ depending on which vertex is $\tau_i$). Similarly, if for a given diagram none of the propagators connecting a vertex $\tau_i$ to the boundary is imaginary, then such a diagram will contribute a term proportional to $\lambda_{n_i + 1}$:
\def\diagramruletwo{\tikz[baseline=-1.4ex,scale=1.8, every node/.style={scale=1.4}]{
\coordinate (k1) at (-5ex,0ex);
\coordinate (kj) at (0ex,0ex);
\coordinate (kn) at (5ex,0ex);
\coordinate (tau) at (0,-4ex);
\coordinate (dots1) at (-2.0ex,-1.8ex);
\coordinate (dots2) at (2.0ex,-1.8ex);
\coordinate (tau2) at (6ex,-4ex);
\pgfmathsetmacro{\arista}{0.06}
\draw[thick] (-6ex,0ex) -- (6ex,0ex);
\draw[-,thick,double] (tau) -- (k1);
\draw[-,thick,double] (tau) -- (kj);
\draw[-,thick,double] (tau) -- (kn);
\node[anchor=south] at (dots1) {\scriptsize{$\cdots$}};
\node[anchor=south] at (dots2) {\scriptsize{$\cdots$}};
\draw[-,thick,dashed] (tau) -- (tau2);
\node[anchor=south] at ($(tau)+(0,-2.0ex)$) {\scriptsize{$\tau_i$}};
\filldraw[color=black, fill=white, thick] ($(k1)-(\arista,\arista)$) rectangle ($(k1)+(\arista,\arista)$);
\node[anchor=south] at ($(k1)+(0ex,0.5ex)$) {\scriptsize{$\k_{i1}$}};
\filldraw[color=black, fill=white, thick] ($(kj)-(\arista,\arista)$) rectangle ($(kj)+(\arista,\arista)$);
\node[anchor=south] at ($(kj)+(0ex,0.5ex)$) {\scriptsize{$\k_{ij}$}};
\filldraw[color=black, fill=white, thick] ($(kn)-(\arista,\arista)$) rectangle ($(kn)+(\arista,\arista)$);
\node[anchor=south] at ($(kn)+(0ex,0.5ex)$) {\scriptsize{$\k_{in_i}$}}
}
}
\be
\sum_j \diagramruletwo \quad \longrightarrow  \quad  \lambda_{n_i + 1} \nn
\ee
Here $\tau_i$ is connected to the boundary with $n_i$ real legs. Necessarily, the extra leg connecting $\tau_i$ with the other vertex is imaginary (recall that every vertex must have at least one imaginary leg connected to them). Coming back to (\ref{n-moment-V=2}), we see that the product $\lambda_{n_1 + 1} \lambda_{n_2 + 1}$ cannot appear without at least one factor $n_1$ or $n_2$.

To continue, given that $\rho_2(\varphi)$ appearing in equation~(\ref{rho_2-sum}) combines both $\langle \varphi^{n} \rangle_{2}$ and $\langle \varphi^{n} \rangle_{1}$ in a single expression, let us write the first order contribution to the $n$-moment as
\bea
\langle \varphi^{n} \rangle_{1} & = & A \,\Delta N \, B^{n}  \, \times \, n \, \lambda_{n} ,  \label{n-moment-V=1-B}
\eea
where, just as for $\langle \varphi^{n} \rangle_{2}$ written in (\ref{n-moment-V=2}), we are only focusing on its dependence on $n$ and $\lambda_n$. Now we are ready to examine the general form of the second order contribution to the PDF. Inserting (\ref{n-moment-V=2}) and (\ref{n-moment-V=1-B}) into (\ref{rho_2-sum}) we obtain:
\bea
\rho_2 (\varphi)  &=&  \frac{1}{2!}  \Delta N^2 A^2 \frac{e^{- \frac{1}{2} \frac{\varphi^2}{\sigma^2}}}{\sqrt{2 \pi} \sigma}  \sum_{n_1=0}^{\infty} \sum_{n_2=0}^{\infty} \frac{1}{n_1! n_2!}  
B^{n_1} B^{n_2}  {\rm He}_{n_1 + n_2} (\varphi / \sigma) \nn \\
&& \hspace{-1.5cm} \times 
\Bigg[  ( n_1 \lambda_{n_1+1} ) ( n_2 \lambda_{n_2+1} ) + ( n_1 \lambda_{n_1+1} ) (  \lambda_{n_2+1} ) + ( \lambda_{n_1+1} ) ( n_2 \lambda_{n_2+1} )+ ( n_1 \lambda_{n_1} ) ( n_2 \lambda_{n_2} )   \Bigg]  .  \label{general-rho-V=2} \qquad 
\eea
To obtain this expression, we made use of the following relation valid for sums:
\be
\sum^{\infty}_{n} \sum_{n_1 n_2}^{n_1 + n_2 = n} = \sum^{\infty}_{n_1} \sum^{\infty}_{n_2} \;.
\ee
A highlight of equation~(\ref{general-rho-V=2}) is that both contributions to the $n$-moments (first order and second order terms) end up having a similar status. The square bracket of the second line contains every possible combination of couplings compatible with the rule whereby at least one vertex is connected to the boundary with an imaginary propagator. The first three terms of the second line come from diagrams with two vertices, which is why the $\lambda$-couplings have the subscript $n_i + 1$ ($n_i$ legs connected to the boundary plus one leg connected to the remaining vertex). The fourth term, instead, comes from the multiplication of two copies of the first order result, which is why the $\lambda$-couplings have the subscript $n_i$ (just $n_i$ legs connected to the boundary). An interesting aspect of this result is that the potentially large factorials coming from the number of channels are cancelled by the factorial appearing in the first line of (\ref{rho_2-sum}). This feature will reappear in a similar manner in the case of arbitrary vertices.

The final step consists in expressing $\lambda_{n_1 + 1}$, $\lambda_{n_2 + 1}$, $n_1 \lambda_{n_1 + 1}$ and $n_2 \lambda_{n_2 + 1}$ in terms of the potential. Using the orthogonality properties of Hermite polynomials and the identity ${\rm He}_n'=n{\rm He}_{n-1}$, we can invert~\eqref{Weir-1} to find
\bea
\lambda_{n_i + 1} &=& \frac{1}{\sigma_L^{n_i}} \int^{\infty}_{-\infty} \!\!\! \de\psi \,   \frac{e^{- \frac{\psi^2}{2 \sigma_L^2}} }{\sqrt{2\pi} \sigma_L}   \, {\rm He}_{n_i} ( \psi / \sigma_L)  \, \mathcal U^{\, '}(\psi) ,
\label{lambda-n+1}
\\
\lambda_{n_i} &=& \frac{1}{\sigma_L^{n_i}} \int^{\infty}_{-\infty} \!\!\! \de\psi \,   \frac{e^{- \frac{\psi^2}{2 \sigma_L^2}} }{\sqrt{2\pi} \sigma_L}   \, {\rm He}_{n_i} ( \psi / \sigma_L)  \, \mathcal U(\psi) ,
\label{lambda-n}
\eea
and using the Hermite operator~\eqref{Herm-Op}, we get
\bea
n_i \, \lambda_{n_i + 1} &=& - \frac{1}{\sigma_L^{n_i}} \int_{-\infty}^{+\infty} \de\psi  \frac{e^{- \frac{\psi^2}{2 \sigma_L^2}}}{\sqrt{2 \pi} \sigma_L} {\rm He}_{n_i}(\psi / \sigma_L)  \mathcal O_{\psi} \, \mathcal U^{\, '}(\psi) , \label{n-lambda-n+1}
\\
n_i \, \lambda_{n_i} &=& - \frac{1}{\sigma_L^{n_i}} \int_{-\infty}^{+\infty} \de\psi  \frac{e^{- \frac{\psi^2}{2 \sigma_L^2}}}{\sqrt{2 \pi} \sigma_L} {\rm He}_{n_i}(\psi / \sigma_L)  \mathcal O_{\psi} \, \mathcal U(\psi) .  \label{n-lambda-n}
\eea
Inserting these results back into (\ref{general-rho-V=2}), one derives an expression which is suitable to perform the sums. 

To simplify the discussion, let us focus on the first term proportional to $( n_1 \lambda_{n_1+1} ) ( n_2 \lambda_{n_2+1} )$. Replacing (\ref{n-lambda-n+1}) in (\ref{general-rho-V=2}) we see that the first term contributes 
\bea
\rho_2 (\varphi)  &\supset& \frac{1}{2!} \Delta N^2 A^2 \frac{e^{- \frac{1}{2} \frac{\varphi^2}{\sigma_L^2}}}{\sqrt{2 \pi} \sigma_L}  
 \int_{-\infty}^{+\infty} \de\psi_1  \frac{e^{- \frac{\psi_1^2}{2 \sigma_L^2}}}{\sqrt{2 \pi} \sigma_L}  \mathcal O_{\psi_1} \, \mathcal U^{\, '} (\psi_1)    
 \int_{-\infty}^{+\infty} \de\psi_2  \frac{e^{- \frac{\psi_2^2}{2 \sigma_L^2}}}{\sqrt{2 \pi} \sigma_L}  \mathcal O_{\psi_2}  \, \mathcal U^{\, '} (\psi_2)  \nn \\
 && \times \sum_{n_1=0}^{\infty} \sum_{n_2=0}^{\infty} \frac{B^{n_1} B^{n_2}}{n_1! n_2!}  
 {\rm He}_{n_1}(\psi_1 / \sigma_L)  {\rm He}_{n_2}(\psi_2 / \sigma_L)  {\rm He}_{n_1 + n_2} (\varphi / \sigma_L)  . \label{rho_2-before_sum}
\eea
As before, here we only focus on the relevant dependence of the expressions on $n_1$ and $n_2$, which is the relevant information to perform the sums. These sums can be achieved thanks to the following formula involving Hermite polynomials:
\bea
 \sum_{n_1 n_2}  \frac{u_1^{n_1}  u_{2}^{n_2} }{n_1!  n_{2} !}  {\rm He}_{n_1} (x_1) {\rm He}_{n_{2}} (x_{2})   e^{-\frac{1}{2} y^2}  {\rm He}_{n_1 + n_{2}} (y)  =  \frac{e^{- \frac{1}{2} \frac{1}{1 - \sum_i^{N} u_i^2 } \left(  y -  \sum_i^{N} u_i x_i    \right)^2  }}{\sqrt{1 - \sum_i^{N} u_i^2 }} .
\eea
Inserting this formula back into (\ref{rho_2-before_sum}) and reorganizing the argument of the exponentials, we finally obtain
\bea
\rho_2 (\varphi)  &\supset& \frac{1}{2!}  \Delta N^2 A^2  \frac{e^{- \frac{1}{2} \frac{\varphi^2}{\sigma_L^2}}}{\sqrt{2 \pi} \sigma_L}    \int_{-\infty}^{+\infty} \de\psi_1    \int_{-\infty}^{+\infty} \de\psi_2     \frac{ 1  }{\sqrt{2\pi} \sqrt{1 - 2 B^2 } \sigma_L }  
\nn \\
&& 
 e^{- \frac{1}{2 \sigma_L^2 (1 - 2 B^2)} \bigg[   ( B \varphi - \psi_1)^2 +  ( B \varphi - \psi_2)^2   -  B^2 ( \psi_2 - \psi_1)^2  \bigg]  } 
 \mathcal O_{\psi_1} \, \mathcal U^{\, '} (\psi_1)  \mathcal O_{\psi_2} \, \mathcal U^{\, '} (\psi_2) . \qquad \label{rho_2-UU}
\eea
This result resembles the first-order PDF structure but in this case we have two convoluted Weierstrass transformations (this, in fact, ensures that $\rho_2$ does not spoil the condition $\int \de\varphi \,\rho(\varphi) = 1$).

To further simplify this expression one would need to perform these integrals in a similar way as we dealt with (\ref{D-exact}) to finally obtain (\ref{main-result-first-order}). The details of how to proceed require us to know precisely the form of $I^{(1)}_{n_1 n_2}$ and $I^{(2)}_{n_1 n_2}$ which, for this discussion, we have disregarded. 
However, equation~(\ref{rho_2-UU}) already captures one of the main points of the present article: To second order, the PDF depends on two powers of ${\mathcal V(\varphi)}$ in addition to field-derivatives acting on it. Schematically, taking into account the remaining terms appearing in (\ref{general-rho-V=2}), we see that the second order contribution to the PDF is given by
\be \label{rho-2}
\rho_2 (\varphi)  \sim \frac{1}{2!} \Delta N^2  \frac{e^{- \frac{1}{2} \frac{\varphi^2}{\sigma_L^2}}}{\sqrt{2 \pi} \sigma_L}   \Bigg[  \mathcal O_{\varphi} \Big( \mathcal V^{\, '} (\varphi)   \mathcal O_{\varphi}  \mathcal V^{\, '} (\varphi)   \Big)   +  \mathcal O_{\varphi} \Big( {\cal V} (\varphi) \mathcal O_{\varphi} {\cal V}(\varphi)  \Big) \Bigg] . 
\ee
The appearance of the Hermite operator in the previous expression is not an accident. By performing a partial integration, it is possible to verify that $\mathcal O_{\varphi} $ satisfies the following general condition:
\be
\int_{-\infty}^{+\infty} \de \varphi   \frac{e^{- \frac{1}{2} \frac{\varphi^2}{\sigma_L^2}}}{\sqrt{2 \pi} \sigma_L}    \mathcal O_{\varphi} F(\varphi) = 0 .
\ee 
Thus, we can see that the dependence of $\rho_2 (\varphi)$ on the potential ${\mathcal V(\varphi)}$ and derivatives with respect to $\varphi$ is such that $\int \de\varphi \,\rho_2 (\varphi) = 0$ ensuring that the full probability density function remains unitary. This, of course, is ensured from the starting relation (\ref{rho-n-moments}).

\subsection{Tree-level case: arbitrary order}

Studying how an arbitrary diagram contributes to the probability density function follows easily from the second order case. As already emphasized, the relevant piece of information allowing us to determine the leading form of a diagram contributing to an $n$-point function, is the way that imaginary propagators are distributed along the diagram. A general tree-level diagram contains $V$ vertices labelled by $\tau_1 , \cdots , \tau_V$, each one connected to the boundary with $n_i$ external legs and connected to other vertices via $s_i$ internal propagators. Thus, the coupling offered by a single vertex $\tau_i$ containing $n_i$ external legs and $s_i$ internal propagators is given by $\lambda_{n_i + s_i}$ and hence a general diagram $D_V$ with $V$ vertices must be proportional to
\be
D_V \propto \lambda_{n_1 + s_1} \lambda_{n_2 + s_2}  \cdots \lambda_{n_V + s_V}.
\ee
Given that we are dealing with connected tree-level diagrams, the number of legs connected by internal propagators is restricted to satisfy
\be
s_1 + s_2 + \cdots + s_V = 2 (V - 1).
\ee
The external momenta flowing through a given vertex $\tau_i$ may be labelled as $\k_{i 1}, \cdots , \k_{i n_i}$, while the internal momenta running from a vertex $\tau_i$ to another vertex $\tau_{j}$ can be simply labelled as $\q_{ij}$. 

The topology of a diagram with $V$ vertices can be labelled by the number of legs $s_i$ of each vertex connected by internal propagators. For instance, for $V=1$ the topology is trivial as $s_1 = 0$. For $V=2$ there is a single possible topology corresponding to two vertices connected by a single propagator, which can be labeled by $(s_1,s_2) = (1,1)$. Similarly, in the case $V=3$ there is a single possible configuration corresponding to $(s_1,s_2,s_3) = (2,1,1)$. Notice that $(s_1,s_2,s_3) = (1,2,1)$ and $(s_1,s_2,s_3) = (1,1,2)$ are equivalent, and in fact correspond to different channels of $(s_1,s_2,s_3) = (2,1,1)$ obtained from permutations of the external momenta. For $V=4$ one encounters two possible topologies given by $(s_1,s_2,s_3,s_4) = (3,1,1,1)$ and $(s_1,s_2,s_3,s_4) = (2,2,1,1)$. The following diagrams are examples of these respective topologies:
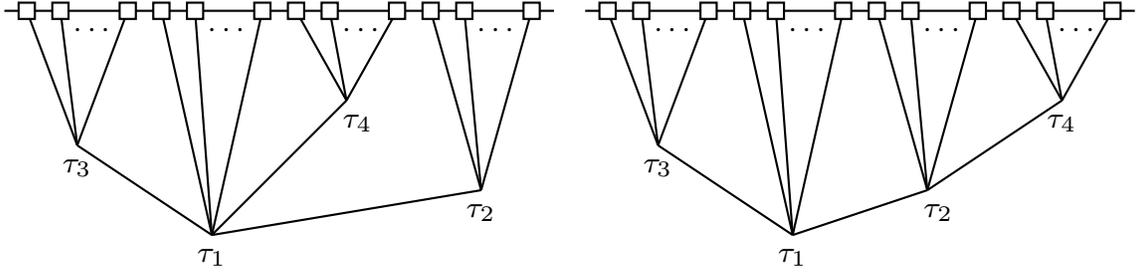
\def\gendiagramone{\tikz[baseline=-1.4ex,scale=1.8, every node/.style={scale=1.4}]{
\coordinate (k1) at (-11ex,0ex);
\coordinate (k2) at (-9.5ex,0ex);
\coordinate (k3) at (-6.5ex,0ex);
\coordinate (k4) at (-5ex,0ex);
\coordinate (k5) at (-3.5ex,0ex);
\coordinate (k6) at (-0.5ex,0ex);
\coordinate (k7) at (1ex,0ex);
\coordinate (k8) at (2.5ex,0ex);
\coordinate (k9) at (5.5ex,0ex);
\coordinate (k10) at (7ex,0ex);
\coordinate (k11) at (8.5ex,0ex);
\coordinate (k12) at (11.5ex,0ex);
\coordinate (t1) at (-8.75ex,-6ex);
\coordinate (t2) at (-2.75ex,-10ex);
\coordinate (t3) at (3.25ex,-4ex);
\coordinate (t4) at (9.25ex,-8ex);
\coordinate (dots1) at (-8.0ex,-1.8ex);
\coordinate (dots2) at (-2.0ex,-1.8ex);
\coordinate (dots3) at (4.0ex,-1.8ex);
\coordinate (dots4) at (10ex,-1.8ex);
\pgfmathsetmacro{\arista}{0.06}
\draw[thick] (-12ex,0ex) -- (12.5ex,0ex);
\draw[-,thick] (t1) -- (k1);
\draw[-,thick] (t1) -- (k2);
\draw[-,thick] (t1) -- (k3);
\draw[-,thick] (t2) -- (k4);
\draw[-,thick] (t2) -- (k5);
\draw[-,thick] (t2) -- (k6);
\draw[-,thick] (t3) -- (k7);
\draw[-,thick] (t3) -- (k8);
\draw[-,thick] (t3) -- (k9);
\draw[-,thick] (t4) -- (k10);
\draw[-,thick] (t4) -- (k11);
\draw[-,thick] (t4) -- (k12);
\draw[-,thick] (t1) -- (t2);
\draw[-,thick] (t2) -- (t3);
\draw[-,thick] (t2) -- (t4);
\node[anchor=south] at (dots1) {\scriptsize{$\cdots$}};
\node[anchor=south] at (dots2) {\scriptsize{$\cdots$}};
\node[anchor=south] at (dots3) {\scriptsize{$\cdots$}};
\node[anchor=south] at (dots4) {\scriptsize{$\cdots$}};
\node[anchor=north] at ($(t1)+(0,0)$) {\scriptsize{$\tau_3$}};
\node[anchor=north] at ($(t2)+(0,0)$) {\scriptsize{$\tau_1$}};
\node[anchor=north] at ($(t3)+(0.5ex,0)$) {\scriptsize{$\tau_4$}};
\node[anchor=north] at ($(t4)+(0,0)$) {\scriptsize{$\tau_2$}};
\filldraw[color=black, fill=white, thick] ($(k1)-(\arista,\arista)$) rectangle ($(k1)+(\arista,\arista)$);
\filldraw[color=black, fill=white, thick] ($(k2)-(\arista,\arista)$) rectangle ($(k2)+(\arista,\arista)$);
\filldraw[color=black, fill=white, thick] ($(k3)-(\arista,\arista)$) rectangle ($(k3)+(\arista,\arista)$);
\filldraw[color=black, fill=white, thick] ($(k4)-(\arista,\arista)$) rectangle ($(k4)+(\arista,\arista)$);
\filldraw[color=black, fill=white, thick] ($(k5)-(\arista,\arista)$) rectangle ($(k5)+(\arista,\arista)$);
\filldraw[color=black, fill=white, thick] ($(k6)-(\arista,\arista)$) rectangle ($(k6)+(\arista,\arista)$);
\filldraw[color=black, fill=white, thick] ($(k7)-(\arista,\arista)$) rectangle ($(k7)+(\arista,\arista)$);
\filldraw[color=black, fill=white, thick] ($(k8)-(\arista,\arista)$) rectangle ($(k8)+(\arista,\arista)$);
\filldraw[color=black, fill=white, thick] ($(k9)-(\arista,\arista)$) rectangle ($(k9)+(\arista,\arista)$);
\filldraw[color=black, fill=white, thick] ($(k10)-(\arista,\arista)$) rectangle ($(k10)+(\arista,\arista)$);
\filldraw[color=black, fill=white, thick] ($(k11)-(\arista,\arista)$) rectangle ($(k11)+(\arista,\arista)$);
\filldraw[color=black, fill=white, thick] ($(k12)-(\arista,\arista)$) rectangle ($(k12)+(\arista,\arista)$);
}
}
\def\gendiagramtwo{\tikz[baseline=-1.4ex,scale=1.8, every node/.style={scale=1.4}]{
\coordinate (k1) at (-11ex,0ex);
\coordinate (k2) at (-9.5ex,0ex);
\coordinate (k3) at (-6.5ex,0ex);
\coordinate (k4) at (-5ex,0ex);
\coordinate (k5) at (-3.5ex,0ex);
\coordinate (k6) at (-0.5ex,0ex);
\coordinate (k7) at (1ex,0ex);
\coordinate (k8) at (2.5ex,0ex);
\coordinate (k9) at (5.5ex,0ex);
\coordinate (k10) at (7ex,0ex);
\coordinate (k11) at (8.5ex,0ex);
\coordinate (k12) at (11.5ex,0ex);
\coordinate (t1) at (-8.75ex,-6ex);
\coordinate (t2) at (-2.75ex,-10ex);
\coordinate (t3) at (3.25ex,-8ex);
\coordinate (t4) at (9.25ex,-4ex);
\coordinate (dots1) at (-8.0ex,-1.8ex);
\coordinate (dots2) at (-2.0ex,-1.8ex);
\coordinate (dots3) at (4.0ex,-1.8ex);
\coordinate (dots4) at (10ex,-1.8ex);
\pgfmathsetmacro{\arista}{0.06}
\draw[thick] (-12ex,0ex) -- (12.5ex,0ex);
\draw[-,thick] (t1) -- (k1);
\draw[-,thick] (t1) -- (k2);
\draw[-,thick] (t1) -- (k3);
\draw[-,thick] (t2) -- (k4);
\draw[-,thick] (t2) -- (k5);
\draw[-,thick] (t2) -- (k6);
\draw[-,thick] (t3) -- (k7);
\draw[-,thick] (t3) -- (k8);
\draw[-,thick] (t3) -- (k9);
\draw[-,thick] (t4) -- (k10);
\draw[-,thick] (t4) -- (k11);
\draw[-,thick] (t4) -- (k12);
\draw[-,thick] (t1) -- (t2);
\draw[-,thick] (t2) -- (t3);
\draw[-,thick] (t3) -- (t4);
\node[anchor=south] at (dots1) {\scriptsize{$\cdots$}};
\node[anchor=south] at (dots2) {\scriptsize{$\cdots$}};
\node[anchor=south] at (dots3) {\scriptsize{$\cdots$}};
\node[anchor=south] at (dots4) {\scriptsize{$\cdots$}};
\node[anchor=north] at ($(t1)+(0,0)$) {\scriptsize{$\tau_3$}};
\node[anchor=north] at ($(t2)+(0,0)$) {\scriptsize{$\tau_1$}};
\node[anchor=north] at ($(t3)+(0.5ex,0)$) {\scriptsize{$\tau_2$}};
\node[anchor=north] at ($(t4)+(0,0)$) {\scriptsize{$\tau_4$}};
\filldraw[color=black, fill=white, thick] ($(k1)-(\arista,\arista)$) rectangle ($(k1)+(\arista,\arista)$);
\filldraw[color=black, fill=white, thick] ($(k2)-(\arista,\arista)$) rectangle ($(k2)+(\arista,\arista)$);
\filldraw[color=black, fill=white, thick] ($(k3)-(\arista,\arista)$) rectangle ($(k3)+(\arista,\arista)$);
\filldraw[color=black, fill=white, thick] ($(k4)-(\arista,\arista)$) rectangle ($(k4)+(\arista,\arista)$);
\filldraw[color=black, fill=white, thick] ($(k5)-(\arista,\arista)$) rectangle ($(k5)+(\arista,\arista)$);
\filldraw[color=black, fill=white, thick] ($(k6)-(\arista,\arista)$) rectangle ($(k6)+(\arista,\arista)$);
\filldraw[color=black, fill=white, thick] ($(k7)-(\arista,\arista)$) rectangle ($(k7)+(\arista,\arista)$);
\filldraw[color=black, fill=white, thick] ($(k8)-(\arista,\arista)$) rectangle ($(k8)+(\arista,\arista)$);
\filldraw[color=black, fill=white, thick] ($(k9)-(\arista,\arista)$) rectangle ($(k9)+(\arista,\arista)$);
\filldraw[color=black, fill=white, thick] ($(k10)-(\arista,\arista)$) rectangle ($(k10)+(\arista,\arista)$);
\filldraw[color=black, fill=white, thick] ($(k11)-(\arista,\arista)$) rectangle ($(k11)+(\arista,\arista)$);
\filldraw[color=black, fill=white, thick] ($(k12)-(\arista,\arista)$) rectangle ($(k12)+(\arista,\arista)$);
}
}
\be
\gendiagramone \quad \gendiagramtwo \nn
\ee
 It is possible to verify that for $V$ vertices the total number of distinct topologies corresponds to the total number of partitions of the integer $V-1$. There is no explicit formula providing the number of such partitions for an arbitrary integer $V$ but there are asymptotic formulas for large $V$, which bound the growth of the function to be much less than factorial. As we shall see, each topology corresponds to a given configuration of derivatives of the potential ${\mathcal V(\varphi)}$ with respect to the field $\varphi$.
 
To continue, let us consider a diagram contributing to $\langle \varphi^{n} ( \k_{11}, \cdots , \k_{V n_V}  ) \rangle '$ (where $n = n_1 + n_2 + \cdots +n_V$), with $V$ vertices characterized by a specific topology and split its propagators into real and imaginary parts. As we saw in Section~\ref{sec:imaginary_props}, we must focus only on those leading diagrams where every vertex is connected to at least one imaginary propagator and at least one imaginary propagator is attached to the boundary. But given that a leading diagram has exactly $V$ imaginary propagators, a single vertex cannot have two or more imaginary propagators connecting it to the boundary. This implies the existence of diagrams where every vertex is attached to the boundary with one imaginary propagator, diagrams where all but one vertex are attached to the boundary with one imaginary propagator and so on. Now comes a critical point: A given vertex $\tau_i$ can have any of its $n_i$ legs attached to the boundary with imaginary propagators. Therefore, after adding up all the diagrams $D_{\tau_i}^{(1)}$ whereby $\tau_i$ is attached to the boundary with one imaginary leg, they will collectively contribute to $\langle \varphi^{n} ( \k_{11}, \cdots , \k_{V n_V}  ) \rangle '$ the following factor
\be
 \sum D_{\tau_i}^{(1)} =
\cdots   \times \Bigg[ \lambda_{n_i + s_i} \frac{H^{2 n_i}}{2^{n_i}} \frac{k_{i 1}^3 + k_{i 2}^3 + \cdots + k_{i n_i}^3}{k_{i 1}^3 k_{i 2}^3 \cdots  k_{i n_i}^3} \int^{\tau_f}_{\tau_0} \frac{\de\tau_i}{\tau_i^4} (\tau_i^3 - \tau_f^3) \Bigg] \times \cdots . \label{general-n-moment-1}
\ee
Recall that to obtain the contribution of a diagram with $V$ vertices to the connected moment $\langle \varphi^n \rangle$ we must multiply (\ref{general-n-moment-1}) by $(2 \pi)^3 \delta^{(3)} (\k_{11} + \cdots + \k_{V n_V})$, integrate every external momentum, and then sum over every channel. In the case where $n_1 \neq n_2 \neq \cdots n_V$ the total number of different channels $N_{\rm ch}$ is given by
\be
N_{\rm ch} = \frac{(n_1 + n_2 + \cdots + n_V)!}{n_1! n_2! \cdots n_V!} . \label{number-of-channels-V}
\ee
However, if two or more $n_i$'s are equal, we must take into account the symmetry factor of the diagram. For example, if all but a total of $n_{\rm eq}$ out of the $n_i$'s are different, then the number of channels will contain an additional symmetry factor $1/n_{\rm eq}!$\,:
\be
N_{\rm ch} =  \frac{1}{n_{\rm eq}!} \frac{(n_1 + n_2 + \cdots + n_V)!}{n_1! n_2! \cdots n_V!} .  \label{number-of-channels-V-eq}
\ee
After adding up all the channels for a particular set of external legs $\{ n_{1} , n_{2} , \cdots , n_{V} \}$, we must add up all possible combinations of these sets such that $n_1 + \cdots + n_V = n$. To not over-count combinations we can perform an ordered sum such that $n_1 < n_2 < \cdots < n_V$ except when two or more sets of external legs are equal. In performing this sum we will encounter the same quantity $N_{\rm ch}$ as in equation~(\ref{number-of-channels-V}) but adjusted by the corresponding symmetry factors indicating that two or more sets of external legs are equal. But the entire sum can be simplified to an unordered sum divided by $V!$. That is:
\bea
\sum^{n_1 + \cdots + n_V = n}_{n_1 < \cdots < n_V}  (\cdots )  && \nn \\
   + \!\!\!\!\!\!   \sum^{n_1 + \cdots + n_V = n}_{n_1 = n_2 < \cdots < n_V} \frac{1}{2!}  (\cdots ) + \cdots +  \sum^{n_1 + \cdots + n_V = n}_{n_1 < n_{V-1} = n_V} \frac{1}{2!}  (\cdots )  &&
\nn \\
 + \!\!\!\!\!\!  \sum^{n_1 + \cdots + n_V = n}_{n_1 = n_2 = n_3 < n_4 \cdots < n_V} \frac{1}{3!}  (\cdots ) + \cdots +  \sum^{n_1 + \cdots + n_V = n}_{n_1 < n_{V-2} = n_{V-1} = n_V} \frac{1}{3!}  (\cdots )  &&
\nn \\
 + \cdots  + \sum^{n_1 + \cdots + n_V = n}_{n_1 = \cdots = n_V} \frac{1}{V!}  (\cdots ) &=& \frac{1}{V!} \sum^{n_1 + \cdots + n_V = n}_{n_1 \cdots n_V}  (\cdots ) . \qquad
\eea
After this step, we can readily write down the form of every type of contribution to the $n$-moment coming from particular types of diagrams. For instance, the contribution $\langle \varphi^{n} \rangle_{V}^{(1)}$ to the $n$-moment emerging from those diagrams where only one vertex is connected to the boundary through a single imaginary propagator is found to have the form:
\bea
\langle \varphi^{n} \rangle_{V}^{(1)} &=& \frac{2 }{3^V H^{2(V + 1)}}  \Big[ \ln (\tau_0 / \tau_f) \Big]^V  \frac{n!}{V!} \sum_{n_1 \cdots n_V}^{n_1 + \cdots + n_V = n} \frac{1}{n_1! \cdots n_V!} 
\left( \frac{ H^{2}}{ 2 } \right)^{n_1} \cdots \left( \frac{ H^{2}}{ 2 } \right)^{n_V} \nn \\
&& 
\times \sum_{s_1 \cdots s_V} \bigg[ ( n_1 \lambda_{n_1 + s_1} )  \lambda_{n_2 + s_2}  \times \cdots \times \lambda_{n_V + s_V} I^{(1)}_{n_1  n_2 \cdots n_V} + \cdots \bigg]  , \label{eq:varphi-n-1-V}
\eea
where the sum $\sum_{s_1 \cdots s_V}$ runs through every possible topology. In the previous expression, the ellipses go through all possible combinations containing only one factor of the form $n_i \lambda_{n_i + s_i}$. 

On the other hand, the quantity $I^{(1)}_{n_1  n_2 \cdots n_V}$ corresponds to a momentum integral that depends on the set $\{ n_1 , \cdots , n_V \}$ only as powers. Just as we did in the case of second-order diagrams, we can simplify our discussion by writing 
\be
I_{n_1  n_2 \cdots n_V} = A^V B^{n_1} \cdots B^{n_V} .
\ee
This will be valid as long as we only want to focus on the dependence of various expressions on $\{ n_1 , \cdots , n_V \}$. Just as in the case of two vertices studied in Section~\ref{sec:tree-level-second-order}, we see that the rule allowing us to understand the origin of each term in the second line of equation~(\ref{eq:varphi-n-1-V}) can be summarized by the following assignments: If a vertex $\tau_i$ contains a single imaginary propagator connected to the boundary (with any other propagator real), then it implies a factor proportional to $n_i \lambda_{n_i + s_i}$:
\def\diagramrulethree
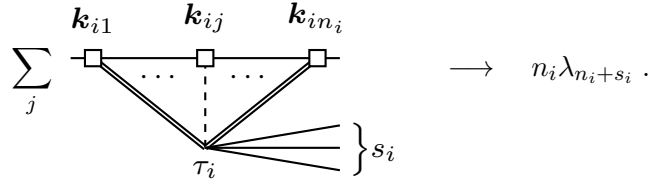
{\tikz[baseline=-1.4ex,scale=1.8, every node/.style={scale=1.4}]{
\coordinate (k1) at (-5ex,0ex);
\coordinate (kj) at (0ex,0ex);
\coordinate (kn) at (5ex,0ex);
\coordinate (tau) at (0,-4ex);
\coordinate (dots1) at (-2.0ex,-1.8ex);
\coordinate (dots2) at (2.0ex,-1.8ex);
\coordinate (tau2) at (6ex,-3ex);
\coordinate (tau3) at (6ex,-4ex);
\coordinate (tau4) at (6ex,-5ex);
\pgfmathsetmacro{\arista}{0.06}
\draw[thick] (-6ex,0ex) -- (6ex,0ex);
\draw[-,thick,double] (tau) -- (k1);
\draw[-,thick,dashed] (tau) -- (kj);
\draw[-,thick,double] (tau) -- (kn);
\node[anchor=south] at (dots1) {\scriptsize{$\cdots$}};
\node[anchor=south] at (dots2) {\scriptsize{$\cdots$}};
\draw[-,thick] (tau) -- (tau2);
\draw[-,thick] (tau) -- (tau3);
\draw[-,thick] (tau) -- (tau4);
\node[anchor=south] at ($(tau3)+(1.5ex,-1.9ex)$) {\scriptsize{$\Big\} s_i$}};
\node[anchor=south] at ($(tau)+(0,-2.0ex)$) {\scriptsize{$\tau_i$}};
\filldraw[color=black, fill=white, thick] ($(k1)-(\arista,\arista)$) rectangle ($(k1)+(\arista,\arista)$);
\node[anchor=south] at ($(k1)+(0ex,0.5ex)$) {\scriptsize{$\k_{i1}$}};
\filldraw[color=black, fill=white, thick] ($(kj)-(\arista,\arista)$) rectangle ($(kj)+(\arista,\arista)$);
\node[anchor=south] at ($(kj)+(0ex,0.5ex)$) {\scriptsize{$\k_{ij}$}};
\filldraw[color=black, fill=white, thick] ($(kn)-(\arista,\arista)$) rectangle ($(kn)+(\arista,\arista)$);
\node[anchor=south] at ($(kn)+(0ex,0.5ex)$) {\scriptsize{$\k_{in_i}$}}
}
}
\be
\sum_j \diagramrulethree \quad \longrightarrow  \quad  n_i \lambda_{n_i + s_i} \; .\nn
\ee
On the contrary, if a vertex $\tau_i$ only contains real propagators connected to the boundary, then it implies a factor proportional to $\lambda_{n_i + s_i}$:
\def\diagramrulefour
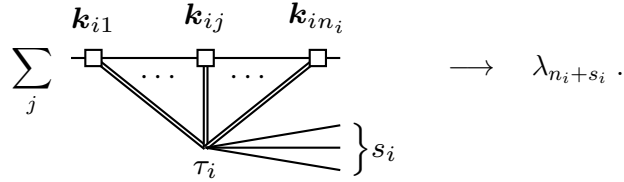
{\tikz[baseline=-1.4ex,scale=1.8, every node/.style={scale=1.4}]{
\coordinate (k1) at (-5ex,0ex);
\coordinate (kj) at (0ex,0ex);
\coordinate (kn) at (5ex,0ex);
\coordinate (tau) at (0,-4ex);
\coordinate (dots1) at (-2.0ex,-1.8ex);
\coordinate (dots2) at (2.0ex,-1.8ex);
\coordinate (tau2) at (6ex,-3ex);
\coordinate (tau3) at (6ex,-4ex);
\coordinate (tau4) at (6ex,-5ex);
\pgfmathsetmacro{\arista}{0.06}
\draw[thick] (-6ex,0ex) -- (6ex,0ex);
\draw[-,thick,double] (tau) -- (k1);
\draw[-,thick,double] (tau) -- (kj);
\draw[-,thick,double] (tau) -- (kn);
\node[anchor=south] at (dots1) {\scriptsize{$\cdots$}};
\node[anchor=south] at (dots2) {\scriptsize{$\cdots$}};
\draw[-,thick] (tau) -- (tau2);
\draw[-,thick] (tau) -- (tau3);
\draw[-,thick] (tau) -- (tau4);
\node[anchor=south] at ($(tau3)+(1.5ex,-1.9ex)$) {\scriptsize{$\Big\} s_i$}};
\node[anchor=south] at ($(tau)+(0,-2.0ex)$) {\scriptsize{$\tau_i$}};
\filldraw[color=black, fill=white, thick] ($(k1)-(\arista,\arista)$) rectangle ($(k1)+(\arista,\arista)$);
\node[anchor=south] at ($(k1)+(0ex,0.5ex)$) {\scriptsize{$\k_{i1}$}};
\filldraw[color=black, fill=white, thick] ($(kj)-(\arista,\arista)$) rectangle ($(kj)+(\arista,\arista)$);
\node[anchor=south] at ($(kj)+(0ex,0.5ex)$) {\scriptsize{$\k_{ij}$}};
\filldraw[color=black, fill=white, thick] ($(kn)-(\arista,\arista)$) rectangle ($(kn)+(\arista,\arista)$);
\node[anchor=south] at ($(kn)+(0ex,0.5ex)$) {\scriptsize{$\k_{in_i}$}}
}
}
\be
\sum_j \diagramrulefour \quad \longrightarrow  \quad  \lambda_{n_i + s_i} \;.\nn
\ee
With these assignments in mind, we can finally add up all types of contributions, ranging from those diagrams where every vertex is connected to the boundary with at least one imaginary propagator, up to those diagrams where only one vertex is connected to the boundary with exactly one imaginary propagator. One finds the general result
\bea
\langle \varphi^{n} \rangle_{V} & = & A^V  \Big[ \ln \frac{\tau_0 }{ \tau_f } \Big]^V  \frac{n!}{V!} \sum_{n_1 \cdots n_V}^{n_1 + \cdots + n_V = n} \frac{B^{n_1} \cdots B^{n_V}}{n_1! \cdots n_V!}   \sum_{s_1 \cdots s_V} \Bigg[ ( n_1 \lambda_{n_1 + s_1} ) \cdots  ( n_V \lambda_{n_V + s_V} ) \nn \\
&&
\hspace{-1cm}
+ ( n_1 \lambda_{n_1 + s_1} ) \cdots ( n_{V-1} \lambda_{n_{V-1} + s_{V-1}} )   \lambda_{n_V + s_V}  + \cdots +  \lambda_{n_1 + s_1}  ( n_2 \lambda_{n_2 + s_2} ) \cdots ( n_V \lambda_{n_V + s_V} ) \nn \\
&&
\hspace{-1cm}
+ \cdots + ( n_1 \lambda_{n_1 + s_1} )  \lambda_{n_2 + s_2}  \cdots   \lambda_{n_V + s_V}  + \cdots +  \lambda_{n_1 + s_1}  \cdots \lambda_{n_{V-1} + s_{V-1}} ( n_V \lambda_{n_V + s_V} ) \Bigg]. \nn \\
\label{n-moment-V=V}
\eea

At this point, we can resort to the general formula (\ref{rho-n-moments}) after expanding each $n$-moment as in equation~(\ref{expansion-n-moments}). This requires us to notice that 
\be
\sum_{n} \sum_{n_1 \cdots n_V}^{n_1 + \cdots + n_V = n} = \sum_{n_1} \cdots \sum_{n_V} . 
\ee
This step gives us back
\bea
\rho_V (\varphi)  &=&  \frac{e^{- \frac{1}{2} \frac{\varphi^2}{\sigma^2}}}{\sqrt{2 \pi} \sigma}  A^V \frac{1}{V!}   \Big[ \ln \frac{\tau_0 }{ \tau_f } \Big]^V  \sum_{n_1...n_V}^{\infty} \frac{B^{n_1} \cdots B^{n_V}}{n_1! \cdots n_V!}  
  {\rm He}_{n_1 + \cdots + n_V} (\varphi / \sigma) \nn \\
&& \hspace{-1.5cm} \times \sum_{s_1 \cdots s_V} \Bigg[ ( n_1 \lambda_{n_1 + s_1} ) \times \cdots \times ( n_V \lambda_{n_V + s_V} ) + \cdots \Bigg]. \label{rho_V-before_sum-0}
\eea
In the previous expression, the ellipses $(+ \cdots)$ inside the square bracket refers to all types of contributions, including those coming from the product of lower $n$-point functions. For example, there must be a term coming from the product $\langle \varphi^{n_1} \rangle_{1} \cdots \langle \varphi^{n_V} \rangle_{1}$. This term will contribute a piece proportional to $(n_1 \lambda_{n_1}) \cdots (n_V \lambda_{n_V})$. To complete this computation, we may express $\lambda_{n_i + s_i}$ and $n_i \lambda_{n_i + s_i}$ in terms of the potential ${\mathcal V(\varphi)}$. These quantities can be respectively written as
\be
\lambda_{n_i + s_i} = \frac{1}{\sigma_L^{n_i}} \int^{\infty}_{-\infty} \!\!\! \de\psi \,   \frac{e^{- \frac{\psi^2}{2 \sigma_L^2}} }{\sqrt{2\pi} \sigma_L}   \, {\rm He}_{n_i} ( \psi / \sigma_L)  \, \mathcal U^{(s_i)}(\psi) ,
\ee
and
\be
n_i \, \lambda_{n_i + s_i} = - \frac{1}{\sigma_L^{n_i}} \int_{-\infty}^{+\infty} \de\psi  \frac{e^{- \frac{\psi^2}{2 \sigma_L^2}}}{\sqrt{2 \pi} \sigma_L} {\rm He}_{n_i}(\psi / \sigma_L)  \mathcal O_{\psi} \, \mathcal U^{(s_i)} (\psi) .  
\ee

Inserting these expressions in equation~(\ref{rho_V-before_sum-0}) one obtains
\bea
\rho_V (\varphi)  &\supset& \frac{1}{V!}   \Big[ \ln \frac{\tau_0 }{ \tau_f } \Big]^V  A^V   \frac{e^{- \frac{1}{2} \frac{\varphi^2}{\sigma^2}}}{\sqrt{2 \pi} \sigma}   \sum_{s_1 \cdots s_V} \Bigg[
  \int_{-\infty}^{+\infty} \de\psi_1  \frac{e^{- \frac{\psi_1^2}{2 \sigma_L^2}}}{\sqrt{2 \pi} \sigma_L}  \mathcal O_{\psi_1} \, \mathcal U^{(s_1 )} (\psi_1)    
 \nn \\
&&
\hspace{-1cm}
\times \cdots  \times
 \int_{-\infty}^{+\infty} \de\psi_V  \frac{e^{- \frac{\psi_V^2}{2 \sigma_L^2}}}{\sqrt{2 \pi} \sigma_L}  \mathcal O_{\psi_V}  \, \mathcal U^{(s_V )} (\psi_V)  + \cdots  \Bigg]  \nn \\
 && 
\hspace{-1cm}
 \times \sum_{n_1=0}^{\infty} \sum_{n_V=0}^{\infty} \frac{B^{n_1} \cdots B^{n_V}}{n_1! \cdots n_V!}  
 {\rm He}_{n_1}(\psi_1 / \sigma_L) \cdots  {\rm He}_{n_V}(\psi_V / \sigma_L)  {\rm He}_{n_1 + \cdots + n_V} (\varphi / \sigma_L)  .  \qquad \label{rho_V-before_sum}
\eea
This expression can be finally summed to yield the desired probability density function. To proceed with the sum we must consider the following general formula involving Hermite polynomials:
\bea
&& \sum^{\infty}_{n_1=0} \cdots \sum^{\infty}_{n_N=0} \frac{u_1^{n_1} \cdots  u_N^{n_N} }{n_1! \cdots n_N !}  {\rm He}_{n_1} (x_1) \cdots {\rm He}_{n_N} (x_N)   e^{-\frac{1}{2} y^2}  {\rm He}_{n_1+ \cdots + n_N} (y)  = \frac{e^{ -\frac{1}{2} \frac{(y - \sum_i u_i x_i )^2}{1 - \sum_i u_i^2 }   } }{\sqrt{1 - \sum_i u_i^2 }} . \qquad  \nn \\
\eea

Using the previous relation one finally arrives at
\bea
\rho_ V(\varphi)  &\supset& \frac{1}{V!}   \Big[ \ln \frac{\tau_0 }{ \tau_f } \Big]^V  A^V  \frac{e^{- \frac{1}{2} \frac{\varphi^2}{\sigma_L^2}}}{\sqrt{2 \pi} \sigma_L}     \int_{-\infty}^{+\infty} \de\psi_1   \cdots   \int_{-\infty}^{+\infty} \de\psi_V  \nn \\ 
&&  \hspace{-1cm} \bigg[ \mathcal O_{\psi_1} \, \mathcal U^{(s_1)} (\psi_1) \cdots  \mathcal O_{\psi_V} \, \mathcal U^{(s_V)}  (\psi_V) + \cdots \bigg]
\\
&& \hspace{-1cm}
 \frac{ 1  }{\sqrt{2\pi} \sqrt{1 - V B^2 } \sigma_L } e^{- \frac{1}{2 \sigma_L^2 (1 - V B^2)} \bigg[   ( B \varphi - \psi_1)^2 + \cdots +  ( B \varphi - \psi_V)^2   -  B^2 ( \psi_1 - \psi_2)^2 - \cdots -  B^2 ( \psi_{V-1} - \psi_V)^2 \bigg]  }  . \nn
\eea
In principle, the $\psi$-integrals can be performed, however, in the present discussion we are not interested in this technical aspect. In terms of $\varphi$, the final result for $\rho_ V (\varphi)$ may be schematically summarized as follows: 
\bea
\rho_V (\varphi) \! &\sim& \! \frac{1}{V!}   \Big[ \ln \frac{\tau_0 }{ \tau_f } \Big]^V  \frac{e^{- \frac{1}{2} \frac{\varphi^2}{\sigma_L^2}}}{\sqrt{2 \pi} \sigma_L} \!\!  \sum_{s_1 \cdots  s_V} \!\! \Bigg[ \nn \\ 
&& \sigma_L^{2 (V-1)}\mathcal O_{\varphi} \bigg(   \mathcal V^{(s_1)} (\varphi)  \times \mathcal O_{\varphi} \mathcal V^{(s_2)} (\varphi) \times\cdots  \times \mathcal O_{\varphi}  \mathcal V^{(s_V)} (\varphi) \bigg) \nn \\
&& + \sigma_L^{2 (V-1)}\mathcal O_{\varphi} \bigg(   \mathcal V^{(s_1)} (\varphi)  \times  \mathcal V^{(s_2)} (\varphi) \times\cdots  \times \mathcal O_{\varphi}  \mathcal V^{(s_V)} (\varphi) \bigg)  +   \cdots \Bigg] . \qquad \label{main-structure-pdf}
\eea
Recall that the $s_i$-labels stand for the number of internal legs connected to a given vertex $\tau_i$, and that the sum $\sum_{s_1 \cdots  s_V}$ runs through all possible topologies allowed for tree-level diagrams with $V$ vertices, restricted to satisfy $s_1 + \cdots + s_V = 2 (V - 1)$. Also, recall that the ellipses $(+   \cdots)$  stand for the sum of every possible term involving the product of $V$ potentials $\mathcal V (\varphi)$ but with a lower number of derivatives.

\subsection{Including loops} 
\label{higher-loops}

To finish, let us comment on the role of loops. Consider an arbitrary tree-level diagram with $V$ vertices and a certain number of real and imaginary propagators. We can now add propagators (real or imaginary) and new vertices to form new diagrams with loops. This procedure would require us to replace the couplings $\lambda_{n_i + s_i}$ (of those vertices $\tau_i$ to which we are attaching new propagators) by new couplings $\lambda_{n_i + s_i + \ell_i}$, where $\ell_i$ is the number of new propagators attached to $\tau_i$. There would be two classes of loops emerging from this procedure: Loops carrying external momentum and those that do not. 

Loops without external momentum would appear by attaching bubble diagrams to a single vertex (say $\tau_i$) of the original diagram. Just as we found in the case of daisy loops analyzed in Section~\ref{loops-V=1}, these type of loops would lead to a new diagram identical to the original but with the coupling $\lambda_{n_i + s_i}$ replaced by 
\be
\lambda_{n_i + s_i} \to \lambda_{n_i + s_i}^{\rm loops} =  \sum_{L=0}^{\infty}   \frac{ \lambda_{n_i + s_i + \ell_i L} }{L!} \left[ D_B \right]^L  ,
\ee
where $\ell_i$ is the number of legs connected to the vertex $\tau_i$ of the original diagram. Here $D_B$ is the value of the bubble diagram, which would necessarily be time-independent. The effect of this type of loop would be to renormalize the potential ${\mathcal V(\varphi)}$ in a highly nontrivial way. Nevertheless, the resulting potential would appear in the PDF in exactly the same way as already anticipated in the previous discussion.

On the other hand, loops with momentum running through them would appear by adding new propagators and vertices in an already existing diagram. It is convenient to notice that forming loops by adding new vertices to a given diagram would be equivalent to add only propagators to another existing tree-level diagram with more vertices. This indicates that we can limit the procedure of introducing loops by only adding new propagators. If we add an imaginary propagator, its effect will be to lower the degree of divergence of the time integrals involved in the vertices. The resulting diagram will contain unbounded momentum-integrals but their degree of divergence with respect to $\tau_f$ will be lower than $\big[ \ln (\tau_0 / \tau_f ) \big]^V$ present in the original diagram. In contrast, the addition of real propagators would keep the degree of divergence $\big[ \ln (\tau_0 / \tau_f ) \big]^V$ of the diagram and introduce unbounded momentum-integrals.  In both cases, the unbounded integrals can be dealt with in the standard way by appropriately redefining both the kinetic term and the potential of the Lagrangian order by order. The key observation is that while this procedure would be somewhat contrived, it would be equivalent to introducing small quantum corrections to the integrals $I_{n_1  n_2 \cdots n_V}$ without modifying the appearance of $\{ n_1 , n_2 , \cdots , n_V \}$ as powers. After all, the dependence of an $n$-moment on the set $\{ n_1 , n_2 , \cdots , n_V \}$ is determined exclusively by the way in which imaginary propagators appear in leading diagrams of order $\big[ \ln (\tau_0 / \tau_f ) \big]^V$, which, as we have seen, is not modified by the introduction of loops. This observation allows us to conclude that the introduction of loops would not modify the overall structure of the PDF shown in equation~(\ref{main-structure-pdf}) as a function of the renormalized potential $\mathcal U_{\rm ren} (\psi)$.

\section{Emergence of Fokker-Planck dynamics}
\label{sec:FP}
\setcounter{equation}{0}

We are finally ready to make contact with the stochastic treatment of fluctuations in de Sitter spacetime. The Fokker-Planck (FP) equation as a probability conservation law for the long-wavelength modes of scalar spectator fields on de Sitter space has been studied extensively. Generalizations and refinements of the original formulation~\cite{Starobinsky:1986fx,Starobinsky:1994bd} have been presented in~\cite{Tokuda:2017fdh,Riotto:2011sf,Moss:2016uix,Burgess:2014eoa,Burgess:2015ajz,Collins:2017haz,Gorbenko:2019rza,Vennin:2015hra,Pinol:2018euk,Pinol:2020cdp,Assadullahi:2016gkk}, while quantum field theory approaches to the FP dynamics have been discussed in~\cite{Tsamis:2005hd,Finelli:2008zg,Burgess:2014eoa,Anninos:2014lwa,Baumgart:2019clc,Baumgart:2020oby,Mirbabayi:2019qtx,Mirbabayi:2020vyt,Cruces:2021iwq}. In what follows, we analyze how our reconstruction of the PDF using QFT tools connects to the stochastic formalism by studying the limit in which it satisfies the FP equation.

Let us first consider the case in which the size of the variance $\sigma_L^2$ of equation~\eqref{variance-0-3} remains fixed over time. Deriving the PDF~(\ref{main-result-first-order}) with respect to the $e$-fold variable $N$, it is straightforward to obtain the following diffusionless Fokker-Planck equation:
\bea \label{Fokker-Planck-no-dif}
\frac{\partial}{\partial N}\rho (\varphi) &=& 
     \frac{1}{3 H^2} \frac{\partial}{\partial \varphi} \left[ \rho (\varphi)  \mathcal V_{\rm eff}^{\, '}(\varphi)  \right]  .
\eea
The term on the right hand side corresponds to the usual drift term encountered in the FP equation derived within the stochastic formalism. Apart from the obvious absence of diffusion ---recall the discussion below equation~\eqref{rho-0}, we notice that the potential appearing in equation~(\ref{Fokker-Planck-no-dif}) is not the potential $\mathcal V (\varphi)$ defined in the original action for the spectator field $\varphi$ but an effective potential $\mathcal V_{\rm eff}(\varphi)$, which takes into account the effect of selecting a finite range of scales on which the statistics is defined (for instance, by excluding subhorizon short wavelength scales at tree level, or by integrating out loops). 

To make contact with the stochastic formalism, recall from our discussion of Section~\ref{sec:variance-t-dep} that our method requires that at any given time $N$ the smallest momentum $p_{\rm IR}$ (limiting the range of scales on which the statistics is defined) must be of order $p_{\rm IR} \sim H e^{- \Delta N}$. Thus, we may choose $p_{L}$ to be a fixed physical scale close to the horizon (\emph{i.e.} $p_{L} \sim H$), while  $p_{\rm IR}$ can be chosen to decay as $p_{\rm IR} = p_{L} e^{- \Delta N}$. In this way, at a given time $N$ our solution $\rho(N,\varphi)$ is always describing the statistics of the maximum possible range of wavenumbers. In terms of comoving scales $k$, these choices would correspond to a growing cutoff $k_{L}(t) = e^{\Delta N} k_{\rm IR}$, while $k_{\rm IR}$ remains constant. With these choices now the parameter $\xi$ appearing in the variance~(\ref{sL}) satisfies:
\be
\ln \xi = \Delta N ,
\ee
implying that it saturates the bound (\ref{bound_DeltaN-xi}). With this choice, the statistics is now describing a range of scales that, with time, is incorporating more and more modes, in consistency with the stochastic approach. Therefore, in this case, the time-dependence of the statistics (\ref{main-result-first-order}) is more contrived since $N$ does not only appear explicitly through $\Delta N$ but also implicitly via 
\be \label{sL-N}
\sigma_L^2 = \frac{H^2}{4 \pi^2} \Delta N,
\ee 
and the cutoff dependence of $\mathcal V_{\rm eff}(\varphi)$ through the parameter $\xi$. 

In the stochastic formalism the only separation of scales is set by $p_{L}$, which divides modes into short and long. This means that $p_{\rm IR} \to 0$ or, equivalently, $\ln \xi = \Delta N \to \infty$. We can formally make sense of this limit by keeping the product $\Delta N \mathcal V$ finite. This is precisely the limit studied in Section~\ref{sec:effective-pot}, where it was shown that 
\be
\mathcal V_{\rm eff} (\varphi) \to \mathcal V_{\rm ren} (\varphi) ,
\ee
where ${\cal V}_{\rm ren} (\varphi)$ is defined in equation~(\ref{Vren}). We notice that ${\cal V}_{\rm ren} (\varphi)$ still depends on the cutoff scale and therefore it contains an implicit dependence on $\Delta N$ such that
\be
\partial_N {\cal V}_{\rm ren} (\varphi) = - \frac{H^2}{8 \pi^2} {\cal V}^{\, ''}_{\rm ren} (\varphi) . \label{V-ren-time}
\ee
With all these considerations in mind, we may now attempt a new derivation of the Fokker-Planck equation incorporating a diffusion term. Notice that with the time-dependent variance \eqref{sL-N}, a derivative with respect to $N$ of the Gaussian part $\rho_0(\varphi , N)$ appearing in~(\ref{main-result-first-order}) is given by:
\be
\frac{\partial }{\partial N} \rho_0(\varphi , N) = \frac{H^2}{8\pi^2} \frac{\partial^2}{\partial \varphi^2} \rho_0(\varphi , N) .
\ee
Then, by replacing $\rho_0 (\varphi , N) = \rho (\varphi , N) \left(1 - \frac{4\pi^2 }{3H^4} \mathcal O_{\varphi} \mathcal V_{\rm eff} \right)$ (valid to first order in the potential), and reordering terms, we finally obtain the following Fokker-Planck equation for $\rho(\varphi,N)$, this time with diffusion:
\bea 
\frac{\partial}{\partial N}\rho (\varphi) &=& \frac{H^2}{8\pi^2} \frac{\partial^2}{\partial \varphi^2} \left[ \rho (\varphi) \left(1 -  \frac{2 \Delta N}{3H^2}   \mathcal V^{\, ''}_{\rm eff}\right)  \right] + 
     \frac{1}{3 H^2} \frac{\partial}{\partial \varphi} \left[ \rho (\varphi)  \mathcal V_{\rm eff}^{\, '}(\varphi)  \right]  . \label{FP-eq}
\eea
This is one of the main results of this article: the Fokker-Planck equation obtained using purely perturbative techniques. It is exact up to first order in the potential $\mathcal V_{\rm eff} (\varphi)$.  

A comment is in order: One might worry that, thanks to (\ref{V-ren-time}), $\mathcal V_{\rm eff} (\varphi)$ is time-dependent and hence our version of the Fokker-Planck equation cannot be compared to the standard Fokker-Planck equation derived using the stochastic formalism. However, it is important to keep in mind that our result is derived in the formal limit $\xi = \Delta N \to \infty$ in which case $\sigma_L^2 \gg H^2$. As a result, in those cases where the potential has a structure such that $\sigma_L^2 \mathcal V'' \sim  \mathcal V$, then the potential would be effectively time-independent but it would still affect the shape of the probability density function $\rho(\varphi , N)$. As an example, consider an axionic potential $\mathcal V(\varphi) = \Lambda^4 (1 - \cos(\varphi / f_a))$. Then, if $\sigma_L^2 > f_a^2 \gg H^2$, the time-derivative (\ref{V-ren-time}) would be highly suppressed using from equation~\eqref{sL-N}, one can deduce that the shape of the axionic potential would produce an imprint of size $\Delta N \mathcal V^{\,''}(\varphi) / H^2 \sim \frac{\sigma_L^2}{f_a^2} \mathcal V(\varphi) / H^4$, which is not necessarily suppressed in the limit examined in this discussion. 

Finally, note that the main difference between our result and the standard Fokker-Planck equation is that~(\ref{FP-eq}) contains a secular piece inside the square bracket of the diffusion term, whose field-theoretic origin we clarify in Ref.~\cite{Palma:2023uwo}. (See also~\cite{Palma:2025oux} for a complementary discussion). Therefore, equation~(\ref{FP-eq}) matches the usual Fokker-Planck equation in the limit $\frac{\Delta N}{H^2}   \mathcal V^{\, ''}_{\rm eff} \to 0$, which is the regime where the first-order solution~\eqref{main-result-first-order} holds. However, taking this correction at face value, we see that as the system approaches equilibrium, the diffusion term is bound to change drastically, invalidating the usual equilibrium solution $\rho (\varphi) \propto e^{ - \frac{8 \pi^2}{3 H^4} {\mathcal V(\varphi)}}$. As time passes, one is forced to resum higher order terms with respect to $\Delta N \mathcal V$ in order to obtain a solution valid for the equilibrium. Thus, our approach, being perturbative in nature, does not allow us to go beyond this limitation.
\subsection{Higher order corrections to the Fokker-Planck equation}
We now turn our focus on corrections to the drift term induced by higher orders in perturbation theory, i.e. entering the nonperturbative regime, which affect the PDF throughout the whole dynamics (including the equilibrium). Corrections of the type discussed below have been argued for in~\cite{Cohen:2021fzf,Cohen:2020php,Green:2022ovz,Cohen:2022clv}. In the case of a polynomial potential, the requirement of keeping the combination $\Delta N {\cal V}^{\, ''}/H^2$ small for finite $\Delta N$ implies that we are describing the statistics for small values of $\varphi$, i.e. away from the tails. In~\cite{Palma:2019lpt}, it was shown that to first order in the potential, the PDF~\eqref{main-result-first-order} can be derived from the Gaussian by means of a change of variables via a generalized local ansatz, i.e. it governs the statistics of a function of a Gaussian variable. The failure of the first-order solution~\eqref{main-result-first-order} towards the tails of the distribution is consistent with the claim of~\cite{Celoria:2021vjw} that in the tails the method of the generalized local ansatz fails and there is a need for a resummation to all orders in the potential.

Let us  explicitly demonstrate the appearance of higher order terms. To this end, let us for simplicity come back to the diffusionless equation~(\ref{Fokker-Planck-no-dif}) and, for the sake of analysis, take it to be valid to all orders with respect to the potential $\mathcal V$. Our wish is to assess the relevance of higher order corrections in (\ref{Fokker-Planck-no-dif}). Integrating once, we obtain
\be
\rho (\varphi, N) =  \rho_0 (\varphi) + \int^{N}_{N_0} \de N \frac{1  }{3 H^2 } \frac{\partial }{\partial \varphi} \bigg[  \rho (\varphi)  {\cal V}' (  \varphi)  \bigg] ,
\ee
where $\rho_0 (\varphi)$ is nothing but the time-independent Gaussian distribution of equation~(\ref{rho-0}), valid for the case ${\cal V} = 0$. Iterating this solution an infinite number of times, and integrating with respect to time (which is possible thanks to the fact that $\rho_0$ is independent of time) we obtain the formal solution:
\bea
\rho (\varphi, N) &=&  \rho_0 (\varphi) \Bigg[ 1 +  \frac{\Delta N }{3 H^2\sigma_L^2 } \mathcal O_{\varphi}  {\cal V}  + \frac{1}{2!} \left(\frac{\Delta N }{3 H^2\sigma_L^2 } \right)^2   \mathcal O_{\varphi}  \int \de \varphi \, {\cal V}^{\, '}     \mathcal O_{\varphi} {\cal V}     \nn \\
&& + \frac{1}{3!} \left(\frac{\Delta N }{3 H^2 \sigma_L^2} \right)^3 \mathcal  O_{\varphi} \int \de \varphi \,  {\cal V}^{\, '}  \mathcal O_\varphi  \int \de \varphi \,   {\cal V}^{\, '}  \mathcal O_\varphi  {\cal V}       + {\cal O}\left(\Delta N^4\right) \Bigg] ,
\eea
where every operator is understood to act on everything to its right. 

Now, to learn something useful, we may observe that every coefficient of order $\Delta N^n$ contains a term $(\mathcal O_{\varphi}  {\cal V} )^n$. For example, to second order, we have
\be \label{ro_2-fi-N}
\rho_2 (\varphi, N) =  \rho_0 (\varphi) \Bigg[ 1 +  \frac{\Delta N }{3 H^2 \sigma_L^2} \mathcal O_{\varphi}  {\cal V}  + \frac{1}{2!} \left(\frac{\Delta N }{3 H^2 \sigma_L^2} \right)^2  \Big[      \left( \mathcal O_{\varphi} {\cal V}  \right)^2     + \sigma_L^2 {\cal V}^{\,'} \left( \mathcal O_{\varphi} {\cal V}\right)^{'}\Big] \Bigg] .
\ee
We thus see that many terms anticipated by our analysis of Section~\ref{sec:general_structure_pdf} are absent. For instance, already at second order, we see that a term proportional to $\left(\mathcal O_\varphi {\cal V}^{\, '}\right)^2$, that appears in equation~\eqref{rho-2}, is absent in this solution. Note that this term captures the genuine second order contribution to the $n$-point functions~\eqref{fi^n_2^2}, while the terms present in~\eqref{ro_2-fi-N} correspond to the square of the first-order contribution~\eqref{fi^n_2^1}. The discussion can be extended to arbitrary order, since this series captures only terms stemming from powers of lower-order ones, which exclude contributions of the form $\left(\mathcal O_\varphi {\cal V}^{\, '}\right)^n$. Thus we conclude that equation~(\ref{Fokker-Planck-no-dif}) must have additional contributions to all orders in the potential ${\cal V}$ accounting for those terms. For example, at second order we expect
\be
\frac{\partial }{\partial N} \rho (\varphi) =   \frac{1  }{3 H^2 } \frac{\partial }{\partial \varphi} \bigg[  \rho (\varphi)  {\cal V}^{\, '} (  \varphi)  \bigg]  + \gamma \Delta N \rho (\varphi) \left( \mathcal O_\varphi {\cal V}^{\, '} \right)^2 + \cdots ,
\ee
where $\gamma$ is some constant parameter. The same reasoning should hold for the diffusive case~\eqref{FP-eq}. 

This type of corrections should arise
if one were to treat the noise appearing in the Langevin equation as a non-Gaussian random variable in the way described in~\cite{Palma:2019lpt}; that is as receiving corrections of the form $\hat\xi \sim \sum [[\hat\xi_{\rm g},\hat {\cal H}],\hat {\cal H}] \cdots ,\hat {\cal H}]]$, where $\hat\xi_{\rm g}$ is the Gaussian random variable and $\hat {\cal H}$ is the interaction Hamiltonian (given in terms of the potential ${\cal V}$). Again, at equilibrium, as long as fluctuations around the vacuum remain small and Gaussian the importance of these corrections should diminish. We leave further investigation along this direction for future work.

\setcounter{equation}{0}
\section{Summary \& Discussion}
\label{sec:conclusions}

We have focused on the computation of the probability density function $\rho (\varphi)$ of a spectator scalar field $\varphi$ in a de Sitter spacetime with its dynamics determined by a potential ${\mathcal V(\varphi)}$. Our main goal was to uncover the general structure of $\rho (\varphi)$ as a function of ${\mathcal V(\varphi)}$ dictated by perturbation theory. Our approach contrasts with other well-known methods such as the Starobinsky-Yokoyama stochastic approach~\cite{Starobinsky:1986fx,Starobinsky:1994bd}. While our procedure is unable to yield an equilibrium solution for $\rho(\varphi)$ valid after a long time, it gives us back a solution valid after the modes have crossed the horizon, at a time where $\rho(\varphi)$ is still evolving as it is adapting dynamically to the shape of the potential ${\mathcal V(\varphi)}$. We posit that our approach is in line with what would be expected in realistic models of inflation, where the primordial curvature perturbation $\zeta$ may interact with $\varphi$ receiving small non-Gaussian deformations due to the evolution of $\rho (\varphi)$. To finish, we would like to discuss and emphasize a few issues encountered along the previous sections.

\subsection{Structure of the PDF}

Our perturbative approach forced us to reconstruct $\rho(\varphi)$ as an expansion of the form $\rho (\varphi) = \rho_0 (\varphi) + \rho_1 (\varphi) + \rho_2 (\varphi) + \cdots$ where the subscript reveals the order of $\rho (\varphi)$ in terms of powers of ${\mathcal V(\varphi)}$. The form of an arbitrary term $\rho_{V}(\varphi)$ is summarized in equation~(\ref{main-structure-pdf}). We have not derived in detail the exact form of coefficients entering this expansion except for the case $V=1$ [recall that there are also integrals such as those appearing in equation~(\ref{main-result-first-order})]. Despite of this technical shortcoming, we can see that terms appearing in $\rho(\varphi)$ at lower order reappear as powers at higher orders. For instance, the term $\mathcal O_\varphi {\cal V} $ appearing at $\rho_1 (\varphi)$ reappears in the form of $\frac{1}{V!}(\mathcal O_\varphi {\cal V} )^V$ in $\rho_{V}(\varphi)$. This suggests that a more detailed computation could give us back a resummation of the form
\be
\rho(\varphi) = \frac{1}{\sqrt{2 \pi} \sigma_L} \exp \bigg[ - \frac{1}{2} \frac{\varphi^2}{\sigma_L^2} + \alpha_1 \mathcal O_{\varphi} {\cal V} + \alpha_2  ( \mathcal O_{\varphi} {\cal V}^{\, '} )^2 + \alpha_3 ( \mathcal O_{\varphi} {\cal V}^{\, ''} ) ( \mathcal O_{\varphi} {\cal V}^{\, '} )^2   + \cdots  \bigg] , \label{guess-general-rho-V}
\ee
where the $\alpha_V$'s are coefficients of order $\ln [\tau_0 / \tau_f]^V$ to be determined, and the ellipses denote every other terms allowed by the topology of diagrams. In fact, if this is true, given that we know $\alpha_1$ from the computation to first order, and given that $\int \de \varphi \rho(\varphi) = 1$ is preserved, it should be possible to reconstruct the form of every $\alpha_V$. In the wavefunction of the universe approach the probability density function is given as $\rho(\varphi) \propto |\Psi (\varphi)|^2$, where $\Psi (\varphi)$ (the wavefunction) is explicitly an exponential of diagrams. Then, given that the vertex content of these diagrams is also determined by ${\cal V}$, it is reasonable to expect the form~(\ref{guess-general-rho-V}) as a general result. We leave this for future examination.

\subsection{Fokker-Planck from Schwinger-Keldysh without Langevin}
In a QFT context, the FP dynamics (with higher order $\cal V$-corrections) can be thought of as a resummation of Feynman diagrams: to first order in the potential, the PDF of a spectator field filtered in comoving-momentum space, is an exact solution to a Fokker-Planck equation with a modified diffusion; that is, equation~\eqref{main-result-first-order} is a solution of~\eqref{FP-eq}. In addition, we have shown that diffusion depends on the choice of the cutoff used to separate long and short modes. Finally, we argued that higher orders in perturbation theory (organized as a series in the interaction potential) induce corrections to both the diffusion- and the drift-dynamics of the PDF, which become relevant in equilibrium. This opens a way to systematically compute such distributions, which are relevant in a variety of contexts such as CMB/LSS statistics~\cite{Palma:2019lpt,Munchmeyer:2019wlh} and primordial black hole production~\cite{Ezquiaga:2019ftu,Celoria:2021vjw,Hooshangi:2023kss,Cai:2022erk,Cai:2021zsp,Hooshangi:2021ubn,Pi:2022ysn,Vennin:2020kng, Achucarro:2021pdh, Gow:2022jfb}.

\subsection{Large $n$-factorials}

The counting of diagrams contributing to amplitudes in particle processes (e.g., decays in a $\lambda \varphi^4$ theory) has been a matter of concern regarding the behavior of observables~\cite{Voloshin:1992qn}. The number of diagrams involved in our computations is consistent with that found in previous works on amplitudes~\cite{Brown:1992ay, Son:1995wz, Libanov:1994ug, Khoze:2017ifq, Ghosh:2016fvm} (for instance, for $\lambda \varphi^4$, the number of channels we obtain via (\ref{number-of-channels-V-eq}) goes as $\sim n! \lambda^{(n-2)/2}$ for large $n$). In our case, the resummation of $\rho (\varphi)$ ensures that any large $n$-factorial will be cancelled by the  factor $1 / n!$ multiplying any $n$-moment $\langle \varphi^n \rangle$ appearing in equation~(\ref{rho-n-moments}). Other works have examined this issue within the computation of cosmological $n$-point correlation functions (see for example \cite{Panagopoulos:2020sxp}).

\subsection{Infrared divergences and time-dependence of the PDF}

The leading logarithmic contributions to $n$-point correlation functions can be traced back to the fact that at zeroth order the theory consists of a massless spectator field in a de Sitter spacetime. This behavior has been dubbed an infrared divergence, and its existence has been treated as an obstruction to deal with massless or nearly-massless scalars in a de Sitter spacetime.  
There is however an alternative point of view. First of all, it should be clear that if the scalar potential ${\mathcal V(\varphi)}$ consisted only of the quadratic term $\frac{1}{2} m^2 \varphi^2$, the first order result for the $2$-point function $\langle \varphi^2 (\k_1 , \k_2 ) \rangle$ would be proportional to $m^2 \ln (\tau_0 / \tau_f )$. But before this result could diverge other higher order terms of order  $\left[ m^2  \ln (\tau_0 / \tau_f ) \right]^V$ would step in and eliminate the divergence, simply because the resummation of the full result $\langle \varphi^2 (\k_1 , \k_2 ) \rangle$ must coincide with the well-known analytic solution written in terms of Hankel functions~\cite{Gorbenko:2019rza}. This, of course, must be true for any $n$-point function and any type of potential ${\mathcal V(\varphi)}$. The logarithmic time-dependence of our $n$-point functions only emphasizes the perturbative nature of our computations and the need for a nonperturbative scheme to obtain the full time-dependence of $n$-point functions (or the probability density function) revealing the form of $\rho(\varphi)$ for large times. In this regard, our solution for $\rho (\varphi)$ displays the initial phase of the evolution of $\rho(\varphi)$ enforced by ${\mathcal V(\varphi)}$, and can be trusted as long as $\Delta N \ll H^2 / \cal V^{\, ''}$.

\subsection{Beyond Gaussian resummation}
Let us close with a possible extension of our results. In using Gaussian random fields we assume that any nonlinear configuration admits a weak-coupling/small-amplitude convergent expansion, which can be resummed and extrapolated to strong coupling/large amplitudes. In nonlinear theories there might exist however configurations that admit no linear expansion (e.g. instantons), which in turn forbids the use of free-field propagators. In such a case one would have to compute the nonlinear solution and infer the PDF from it. (Such a method is employed in~\cite{Celoria:2021vjw}, however, there it does correspond to resummation of diagrams.) It would be interesting to explore the distribution of such purely nonlinear configurations.

\section*{Acknowledgments}
We are grateful to Sebasti\'an C\'espedes, Xingang Chen, Francisco Colipi, Ian Crosby, Jinn-Ouk Gong, Javier Huenupi, Ellie Hughes, Gabriel Mar\'in, Enrico Pajer, S\'ebastien Renaux-Petel, Bruno Scheihing Hitschfeld and Vicharit Yingcharoenrat for useful discussions and comments. We would also like to thank the organizers of the Workshop ``Revisiting cosmological non-linearities in the era of precision surveys" YITP-T-23-03 for inviting us to present this work. GAP acknowledges support from the Fondecyt Regular project number 1210876 (ANID). SS is supported by Thailand NSRF via PMU-B [grant number B05F650021].

\end{document}